\documentclass[numberedappendix]{emulateapj}
\usepackage{natbib}
\usepackage{graphicx}
\usepackage{dcolumn}
\usepackage{bm}
\usepackage{amsmath}
\usepackage{amstext}
\usepackage{pstricks}
\usepackage{color}
\renewcommand{\baselinestretch}{1.18}

\newcommand\bp{\begin{figure}}
\newcommand\ep{\end{figure}}
\newcommand\bpm{\begin{figure*}}
\newcommand\epm{\end{figure*}}
\newcommand\reffig[1]{Figure \ref{fig:#1}}

\newcommand\refsec[1]{\S \ref{sec:#1}}
\newcommand\Refsec[1]{Section \ref{sec:#1}}
\newcommand\reftbl[1]{Table \ref{tbl:#1}}

\newcommand{\HI}{H\,{\scriptsize I}}

\newcommand{\gsim}{ \mathop{}_{\textstyle \sim}^{\textstyle >} }

\newcommand{\degree}{^\circ}

\def\rosat{{\it ROSAT}}
\def\ROSAT{{\it ROSAT}}
\def\asca{{\it ASCA}}

\def\sgras{Sgr~A$^{*}$}
\newcommand{\nughz}{\nu_{\rm GHz}}  
\newcommand{\be}{\begin{equation}}
\newcommand{\ee}{\end{equation}}
\newcommand{\bea}{\begin{eqnarray}}
\newcommand{\eea}{\end{eqnarray}}

\newcommand{\gev}{\rm \, GeV}
\newcommand{\gevflux}{\rm \, GeV\,cm^{-2} s^{-1} sr^{-1}}

\newcommand{\yr}{\rm \, yr}
\newcommand{\kms}{\ensuremath {\rm km/s}}
\newcommand{\cm}{\rm \, cm}
\newcommand{\Fermi}{\emph{Fermi}}
\newcommand{\WMAP}{\emph{WMAP}}

\begin{document}

\title{Giant gamma-ray bubbles from \Fermi-LAT: AGN activity or bipolar Galactic wind?}

\author{Meng Su\altaffilmark{1,3}, Tracy R. Slatyer\altaffilmark{1,2}, Douglas P. Finkbeiner\altaffilmark{1,2}}

\altaffiltext{1}{ 
  Institute for Theory and Computation,
  Harvard-Smithsonian Center for Astrophysics, 
  60 Garden Street, MS-51, Cambridge, MA 02138 USA } 

\altaffiltext{2}{ 
  Physics Department, 
  Harvard University, 
  Cambridge, MA 02138 USA }
\altaffiltext{3}{mengsu@cfa.harvard.edu}

\begin{abstract}

Data from the \Fermi-LAT reveal two large gamma-ray bubbles, extending 50 degrees above and below the Galactic center, with a width of about 40 degrees in longitude. The gamma-ray emission associated with these bubbles has a significantly harder spectrum ($dN/dE \sim E^{-2}$) than the IC emission from electrons in the Galactic disk, or the gamma-rays produced by decay of pions from proton-ISM collisions. There is no significant spatial variation in the spectrum or gamma-ray intensity within the bubbles, or between the north and south bubbles. The bubbles are spatially correlated with the hard-spectrum microwave excess known as the \WMAP\ haze; the edges of the bubbles also line up with features in the \ROSAT\ X-ray maps at $1.5-2$ keV.

We argue that these Galactic gamma-ray bubbles were most likely created by some large episode of energy injection in the Galactic center, such as past accretion events onto the central massive black hole, or a nuclear starburst in the last $\sim$ 10 Myr. Dark matter annihilation/decay seems unlikely to generate all the features of the bubbles and the associated signals in \WMAP\ and \ROSAT; the bubbles must be understood in order to use measurements of the diffuse gamma-ray emission in the inner Galaxy as a probe of dark matter physics. Study of the origin and evolution of the bubbles also has the potential to improve our understanding of recent energetic events in the inner Galaxy and the high-latitude cosmic ray population.
\end{abstract}


\section{Introduction}
\label{sec:introduction}

The inner Milky Way is home to a massive black hole (MBH), surrounded
by clusters of young stars and giant molecular clouds \citep[see e.g.][for a review]{Morris:1996}.  The nuclear
star cluster has a half-light radius of $\sim 5$pc.  Although there are indications of past activity, the BH is quiescent today.

Fe K$\alpha$ echoes from molecular clouds around \sgras\ have been understood as relics of activity in the past few hundred years \citep{Sunyaev:1993,Koyama:1996}. On a longer timescale, one might expect relics of past
activity in high energy CRs and hot gas, perhaps far off the disk.
The most obvious observables would be $e^-$ CR (visible in inverse
Compton gammas and microwave synchrotron) and thermal emission
(X-rays).

This work presents a multiwavelength study of the inner Galaxy and
identifies several large-scale (tens of degrees) gamma-ray features,
most notably 2 large (spanning $-50^\circ < b < 50^\circ$) structures that
we refer to as the ``\Fermi\ bubbles''.  We suggest that these bubbles are associated
with previously discovered structures in the X-rays and
microwaves, and possibly with analogous smaller-scale structures visible in the FIR.

\subsection{Previous High-Energy Excesses}
Observations of gamma-ray emission in the inner Galaxy at $E \la 1\gev$ go
back decades to \emph{COS-B} \citep{Strong:1984cb,COS-B}, and \emph{SAS-2}
\citep{Fichtel:1975,Kniffen:1981kn} \citep[see][for a
review]{Bloemen:1989}. Later data from the \emph{EGRET} experiment aboard the
\emph{Compton Gamma-ray Observatory} extended to the high-energy side
of the $\pi^0$ bump\citep{Smialkowski:1997sm,Dixon:1998di}. However, EGRET lacked the sensitivity and
angular resolution to reveal the detailed structure of gamma-ray
emission toward the inner Galaxy. The
\emph{Fermi Gamma-ray Space Telescope} provides greatly improved data
up to $\sim100\gev$, with
sufficient angular resolution to map out interesting
structures.\footnote{See
\texttt{http://fermi.gsfc.nasa.gov/ssc/data/}}

At lower energies,
the \emph{ROSAT} All-Sky Survey at 1.5 keV \citep{Snowden:ROSAT}
revealed a biconical X-ray structure over the inner tens of degrees around
the Galactic center (GC), later interpreted as a superwind bubble
(SWB) with energetics of the order of $10^{54-55}$ ergs
\citep{Sofue:2000, Bland:2003}. On smaller scales, the \emph{Midcourse Space
Experiment} combined with \emph{IRAS} data also confirms the existence of a
limb-brightened bipolar structure, the so called Galactic center lobe (GCL), with origin at the GC on the
\emph{degree} scale \citep[see e.g.][for a summary of multiwavelength observations of GCL]{Law:2010}.
The inferred energy injection of both these
bipolar structures, despite their different scales, is $\sim 10^{54-55}$ ergs, with
an estimated age of $\sim 10^6$ yr for the GCL  
and $\sim 10^7$ yr for the SWB. Several Galactic Center Shells (GCS),
tens of pc in size, have been found with total energy of order
$\sim10^{51}$ ergs \citep{Sofue:2003}. These shells and filaments are
claimed to originate from one or more episodes of rapid energy release.

\subsection{Microwave Excess: the WMAP Haze}
Beyond direct evidence of shell structures, microwave observations
also provide intriguing indications of energy release toward the
GC. 

At tens of GHz, the {\it Wilkinson Microwave Anisotropy Probe} (\WMAP) \footnote{\texttt{http://map.gsfc.nasa.gov/}} provides sensitive degree resolution full sky maps of diffuse microwave emission. By subtracting templates including Galactic H$\alpha$, Haslam 408 MHz soft synchrotron, and thermal dust emission to remove the different known emission mechanisms in these maps, a microwave residual excess (named ``the microwave haze'') with spherical (non-disklike) morphology about $\sim4$ kpc in radius toward the GC (visible up to at least $|b| \approx 30\degree$) has been recognized \citep{Finkbeiner:2003im}.  It has a spectrum of about $I_\nu \sim \nu^{-0.5}$, harder than typical synchrotron, but softer than free-free. The microwave haze was later interpreted as synchrotron emission from a hard spectrum of $e^-$ cosmic rays. Other hypotheses such as free-free, spinning dust, or thermal dust have failed to explain its morphology, spectrum, or both \citep{Finkbeiner:2004us,Dobler:2008ww}. 
However, the most recent \WMAP\ 7-year data have not detected the haze polarization predicted by the synchrotron hypothesis \citep{Gold:2010fm}, implying either heavily tangled magnetic fields, or an alternative emission mechanism.  With that caveat in mind, we will assume the \WMAP\ haze is synchrotron and consider the implications. 

\subsection{A Hard Electron CR Spectrum}
A simple model, in which the electron CRs that form the haze have diffused from supernova shocks in the disk, cannot fully explain the data for standard diffusion assumptions. The $23-33$ GHz spectrum of the haze synchrotron is as hard as that generated from shocks, and it seems extremely unlikely that these electrons can diffuse several kpc from the disk without significant softening of the spectrum. The synchrotron cooling timescale for cosmic ray electrons emitting at frequency $\nu$ is $\tau_{\rm syn}\approx10^{6}\,\,B_{100}^{-3/2}\,\nughz^{-1/2}\,\,{\rm yr}$, where $B_{100} = B/100 \mu G$ \citep{Thompson:2006}. Besides the hard spectrum, it is difficult to form the distinctly non-disklike morphology of the haze with any population of sources concentrated in the disk (as is believed to be true of supernovae).

The presence of a distinct component of diffuse hard $e^-$ CR far off the plane is intriguing in itself, and has motivated proposals where the haze is generated by pulsars, other astrophysical processes, or the annihilation of dark matter \citep{Hooper:2007kb, Cholis:0811, Zhang:2008tb, Harding:2009ye, Kaplinghat:2009ix, McQuinn:2010, Malyshev:2010xc}. Other indications of excess electronic activity in the Milky Way may be found in recent measurements of local electron and positron cosmic rays. The ATIC, \emph{Fermi} and H.E.S.S experiments have observed a hardening in the $e^+ + e^-$ spectrum at $20-1000$ GeV \citep{aticlatest, Aharonian:2009ah, Abdo:2009zk}, with an apparent steepening at $\sim 1$ TeV, and the PAMELA experiment has measured a rising positron fraction above 10 GeV. Taken together, these measurements imply a new source of hard electrons and positrons, which may be related to the \WMAP\ haze. The coexistence of \rosat\ X-ray bipolar features and the \WMAP\ haze toward the inner Galaxy also suggests the interesting possibility of a common physical origin for these signals. 

\subsection{Inverse Compton Excess from Fermi-LAT}

Fortunately, if the \WMAP\ haze is synchrotron radiation from a hard electron population located around the GC, the same CRs would also produce IC scattered gammas, allowing an independent probe of the CR population. IC photons provide valuable complementary information about the spatial distribution of the $e^-$ CR (given a model for the ISRF), which in turn can constrain hypotheses about their origin. The unprecedented angular resolution and sensitivity of \emph{Fermi}-LAT allows us to probe the gamma-ray counterpart to the microwave haze in detail for the first time.

Previous work employing the first year \Fermi-LAT data isolated a spectrally hard ``gamma-ray haze'' with similar morphology to the \WMAP\ microwave haze \citep{fermihaze}. In this work, we show that the \Fermi-LAT sky maps constructed from 1.6 yr data (600 days) reveal two large gamma-ray lobes, extending 50 degrees above and below the GC, with a width of about 40 degrees in longitude. These two ``bubble''-like structures have relatively sharp edges and are symmetric with respect to the galactic plane and the minor axis of the galactic disk. The gamma-ray signal reveals similar morphology to the \WMAP\ haze, and is also suggestive of a common origin with features in the \rosat\ X-ray maps at 1.5 keV towards the GC.

As we will discuss, the sharp edges, bilobular shape, and apparent centering on the GC of these structures suggest that they were created by some large episode of energy injection in the GC, such as a past accretion onto the central black hole, or a nuclear starburst in the last $\sim$10 Myr. It is well known that the GC hosts a massive black hole and massive clusters of recently formed stars \citep{Paumard:2006}. Either of these could potentially provide the necessary energy injection by driving large-scale galactic winds or producing energetic jets; we will outline some of the advantages and disadvantages of each scenario.

\subsection{Structure of This Paper}
In \Refsec{data} we briefly review the \emph{Fermi}-LAT data and our data analysis procedure. In \Refsec{bubbles}, we show the 1.6 yr \Fermi\ data maps, reveal the \Fermi\ bubble features and show that they are robust when different models for the expected Galactic diffuse emission are subtracted. We characterize the morphology of the bubbles in some detail and employ regression template fitting to reveal a hard, spatially uniform spectrum for the gamma-ray emission associated with the bubbles. In \Refsec{othermaps}, we show that features in the \rosat\ soft X-rays and the \WMAP\ microwave haze are spatially correlated with the \Fermi\ bubbles, and the \WMAP\ haze and \Fermi\ bubbles are consistent with being produced from the same electron CR population (by synchrotron and IC respectively). \Refsec{Interp} presents our understanding based on the analysis in \Refsec{bubbles} and \Refsec{othermaps}. \Refsec{explanation} discusses possible scenarios to produce the \Fermi\ gamma-ray bubbles. \Refsec{CR} focuses on the origin of the electron CRs and the challenges in explaining the spectral and spatial profiles of the gamma-ray emission from the bubbles. We discuss the implications of the \Fermi\ bubbles for several topics of interest in \Refsec{discussion}, and give our conclusions and suggest future work in \Refsec{conclusion}.


\begin{figure*}[ht]
\begin{center}
\includegraphics[width=1.0\textwidth]{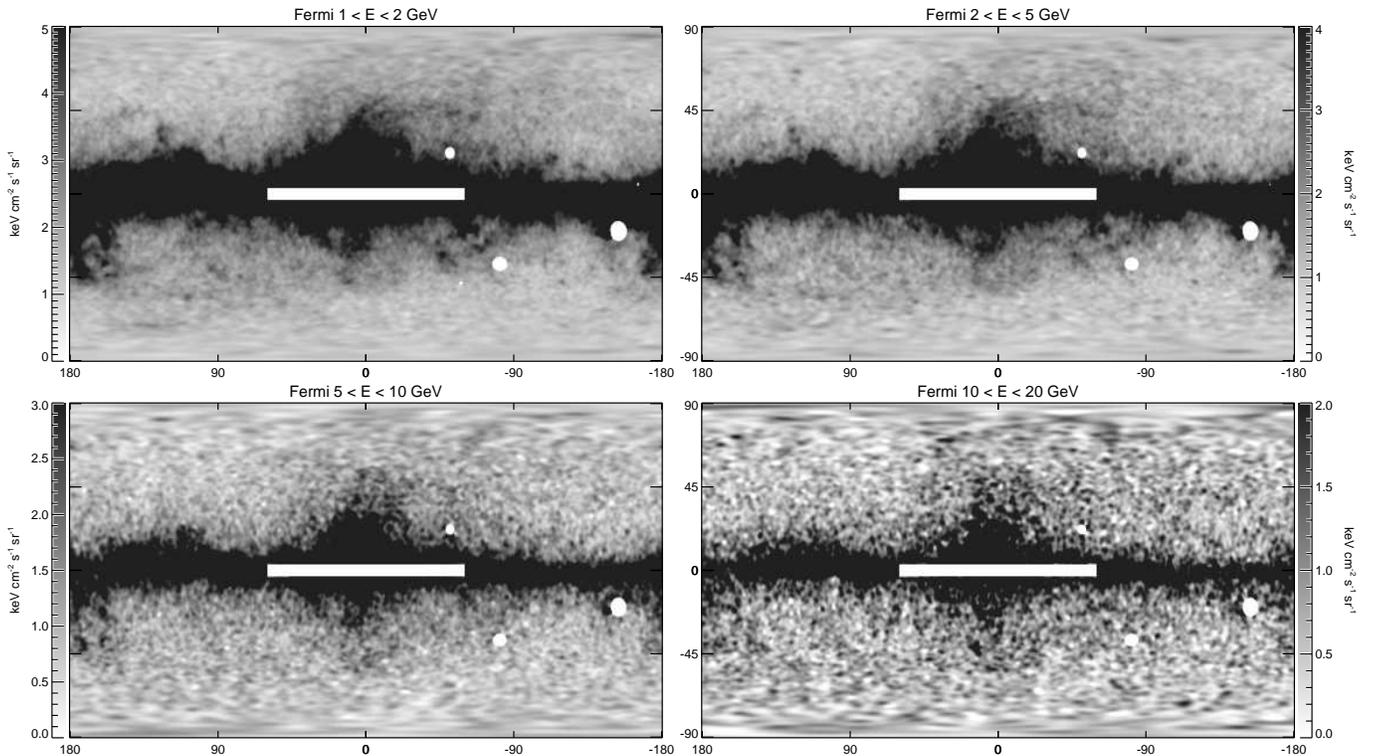}
\end{center}
\caption{All-sky \Fermi-LAT 1.6 year maps in 4 energy bins. Point sources have been subtracted, and large sources, including the inner disk ($-2\degree < b < 2\degree, -60\degree < \ell < 60\degree$), have been masked. 
}
\label{fig:fermimaps}
\end{figure*}


\begin{figure*}[ht]
\begin{center}
\includegraphics[width=1.0\textwidth]{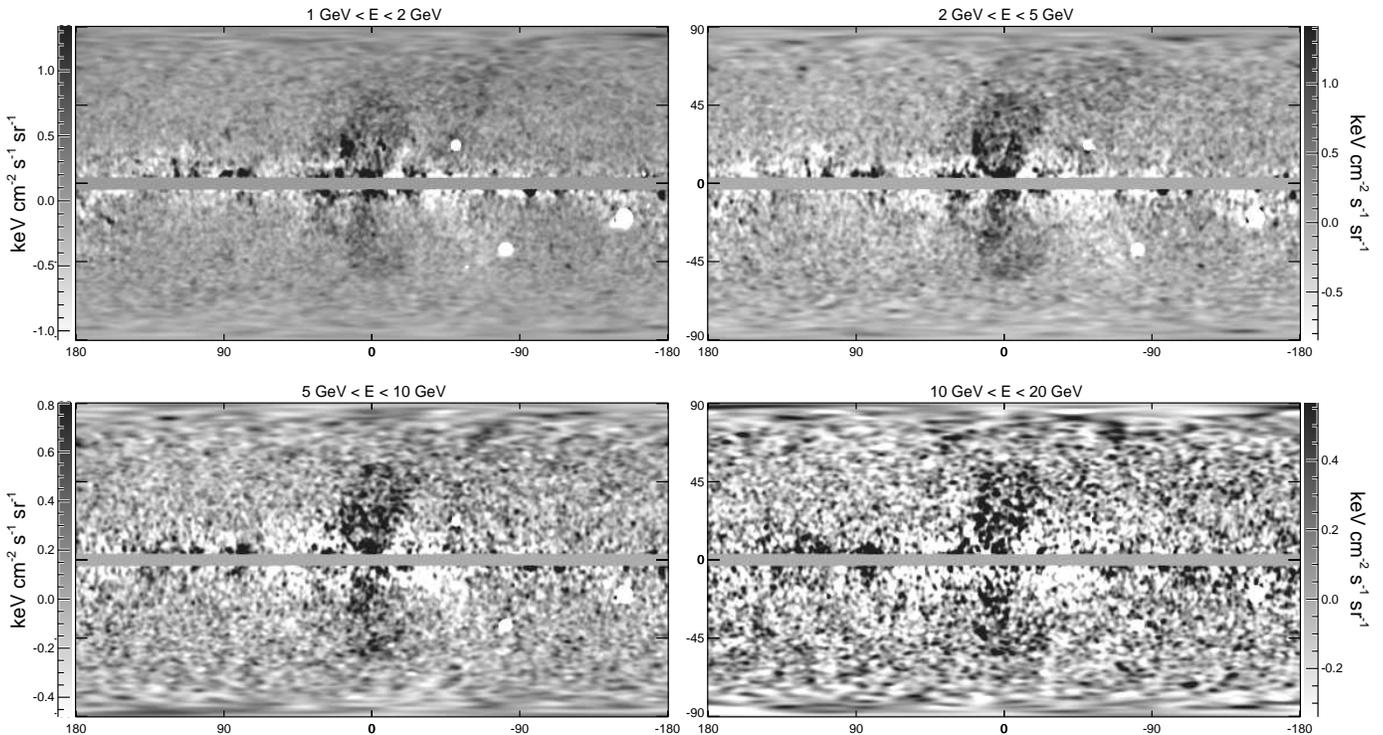}
\end{center}
\caption{All-sky residual maps after subtracting the \Fermi\ diffuse Galactic model from the LAT 1.6 year maps in 4 energy bins (see \refsec{fermidiffuse}). Two bubble structures extending to $b\pm 50\degree$ appear above and below the GC, symmetric about the Galactic plane.
}
\label{fig:diffusemodel}
\end{figure*}

\begin{figure*}[ht]
\begin{center}
\includegraphics[width=1.0\textwidth]{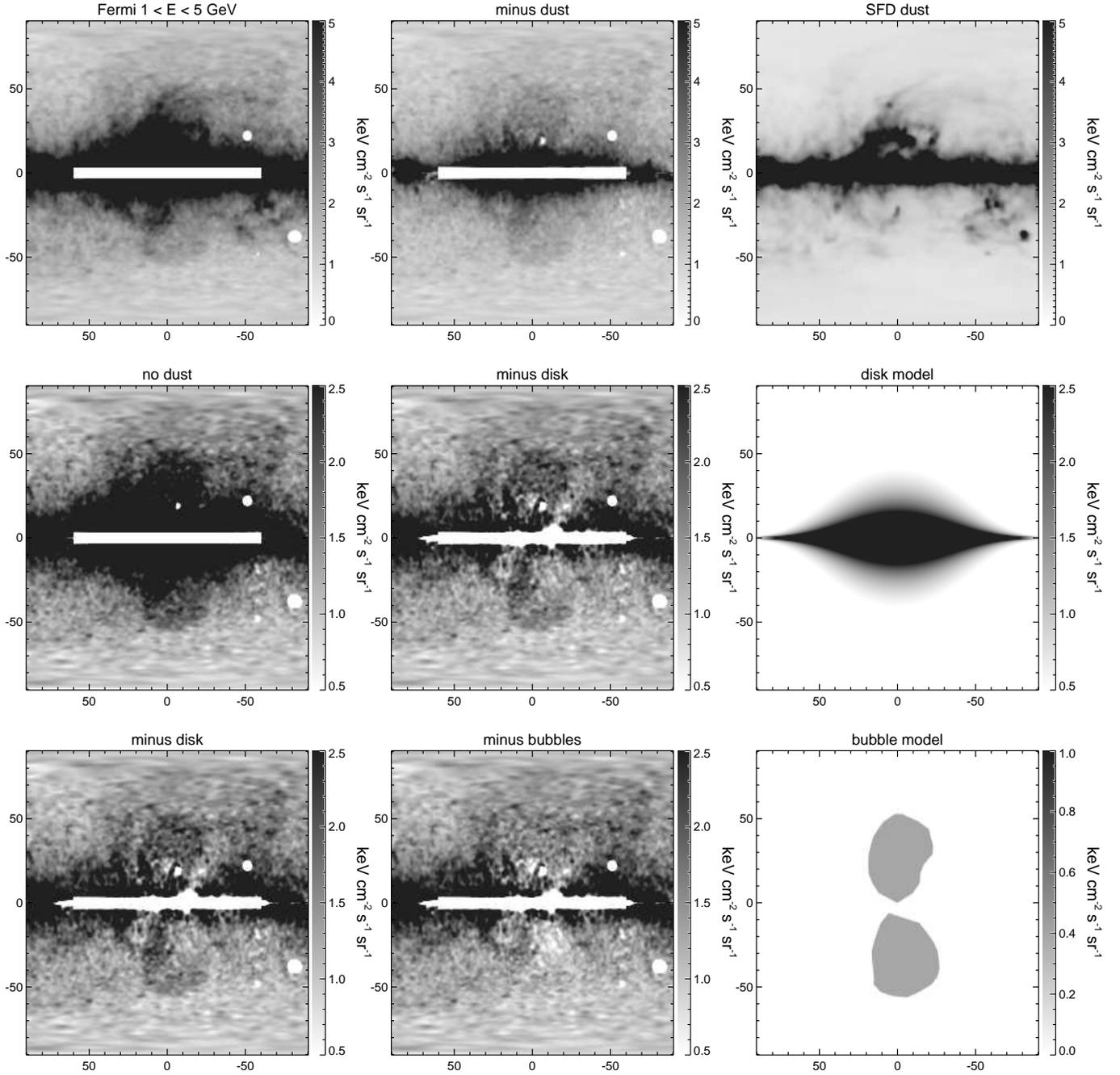}
\end{center}
\caption{
Template decomposition of the \Fermi-LAT 1.6 year $1-5$ GeV map (see \refsec{simplediffuse}). \emph{Top left: }Point source subtracted $1-5$ GeV map, and large sources, including the inner disk $(-2\degree < b < 2\degree, -60\degree < \ell < 60\degree)$, have been masked. \emph{Top middle:} The $1-5$ GeV map minus SFD dust map (\emph{top right} panel) which is used as a template of $\pi^0$ gammas. \emph{Middle row:} The \emph{left} panel is the same as the \emph{top middle} panel but stretched $2\times$ harder. The \emph{middle} panel subtracts a simple geometric disk template (shown in the \emph{right} panel), representing mostly inverse Compton emission, to reveal features close to the Galactic center. Two large bubbles are apparent (spanning $-50\degree < b < 50\degree$). \emph{Bottom row:} The \emph{left} panel is the same as the \emph{middle} panel of the \emph{second row}. Finally we subtract a simple bubble template (\emph{right} panel), with a shape derived from the edges visible in the maps, and uniform projected intensity. After subtracting the bubble template, the two bubbles features have nearly vanished (\emph{bottom middle} panel), indicating a nearly flat intensity for the \Fermi\ bubbles.
}
\label{fig:bubbles}
\end{figure*}


\begin{figure*}[ht]
\begin{center}
\includegraphics[width=1.0\textwidth]{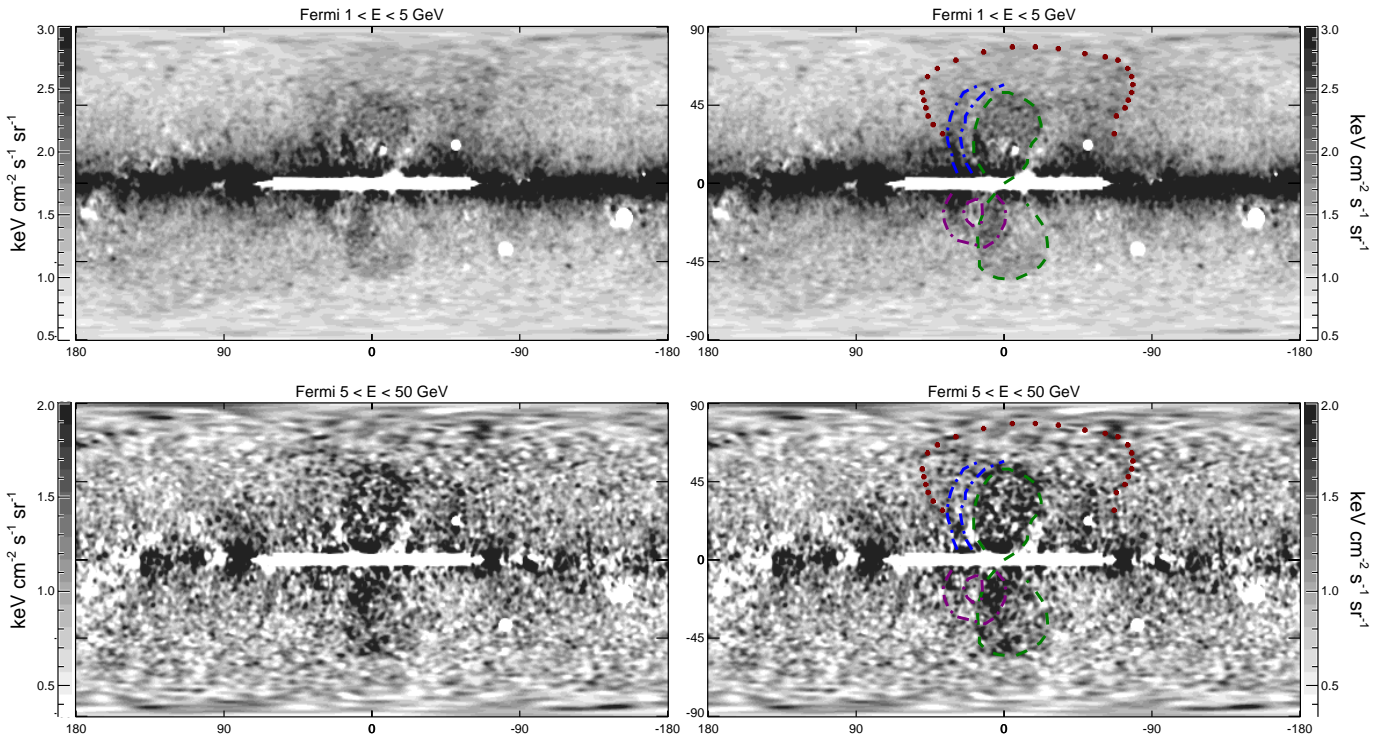}
\end{center}
\caption{Full sky residual maps after subtracting the SFD dust and disk templates from the \Fermi-LAT 1.6 year gamma-ray maps in two energy bins. Point sources are subtracted, and large sources, including the inner disk ($-2\degree < b < 2\degree, -60\degree < \ell < 60\degree$), have been masked. Two large bubbles are seen (spanning $-50\degree < b < 50\degree$) in both cases. \emph{Right panels:} Apparent \Fermi\ bubble features marked in color lines, overplotted on the maps displayed in the \emph{left panels}. Green dashed circles above and below the Galactic plane indicate the approximate edges of the north and south \Fermi\ bubbles respectively. Two blue dashed arcs mark the inner (dimmer) and outer (brighter) edges of the \emph{northern arc} -- a feature in the northern sky outside the north bubble. The red dotted line approximately marks the edge of \emph{Loop I}. The purple dot-dashed line indicates a tentatively identified ``donut'' structure. }
\label{fig:fullsky}
\end{figure*}


\begin{figure*}[ht]
\begin{center}
\includegraphics[width=0.4\textwidth]{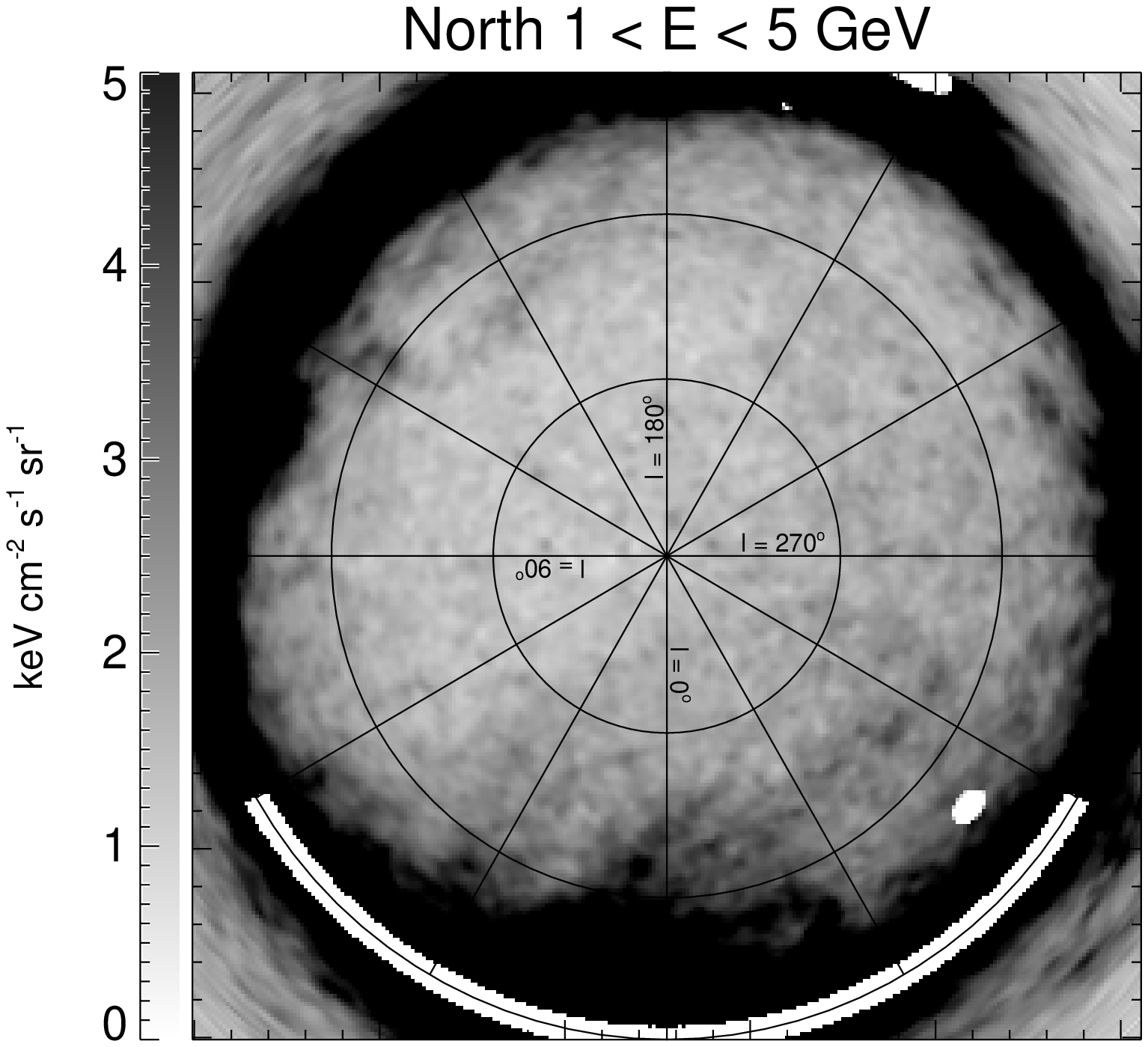}
\includegraphics[width=0.4\textwidth]{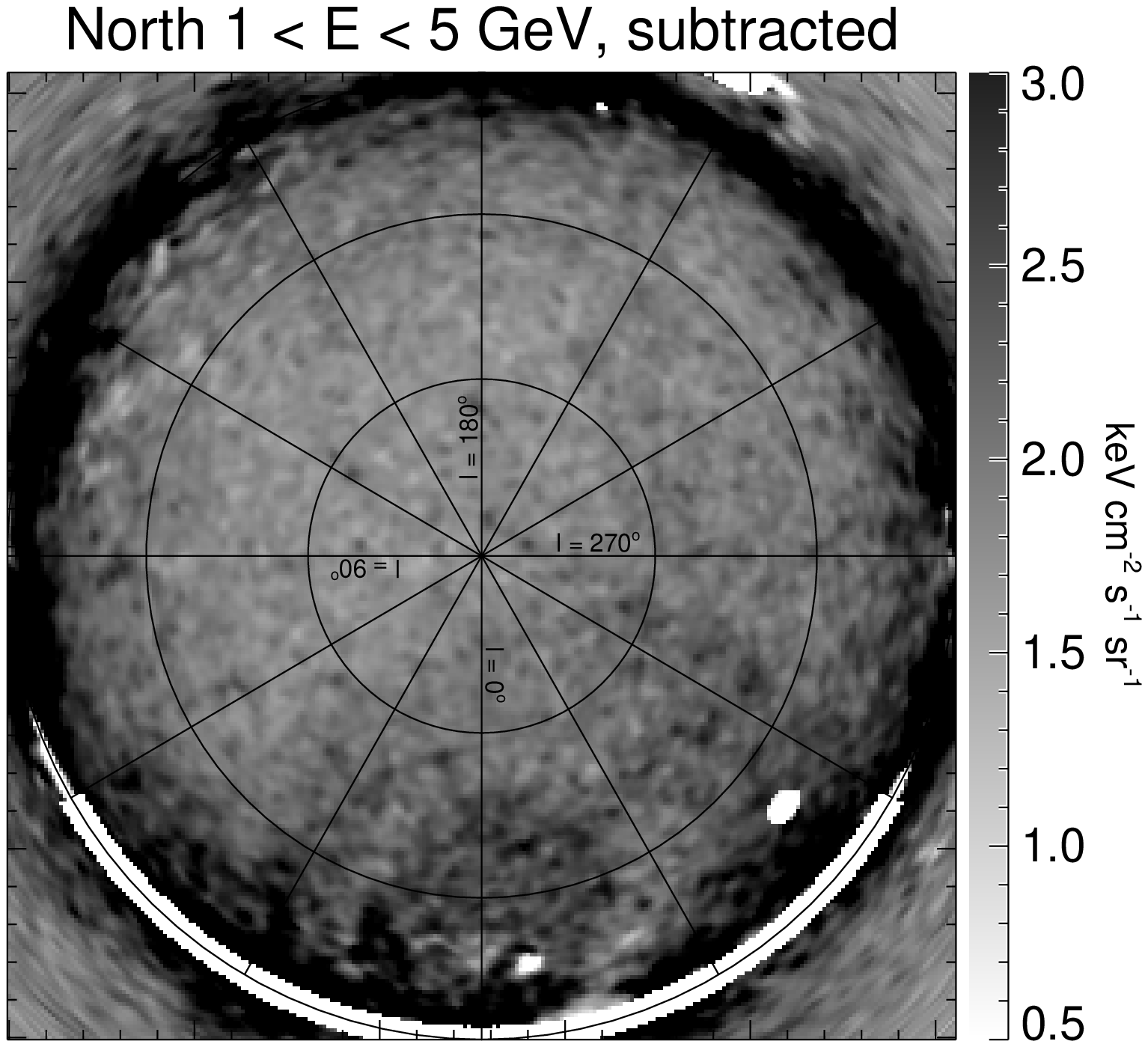}
\includegraphics[width=0.4\textwidth]{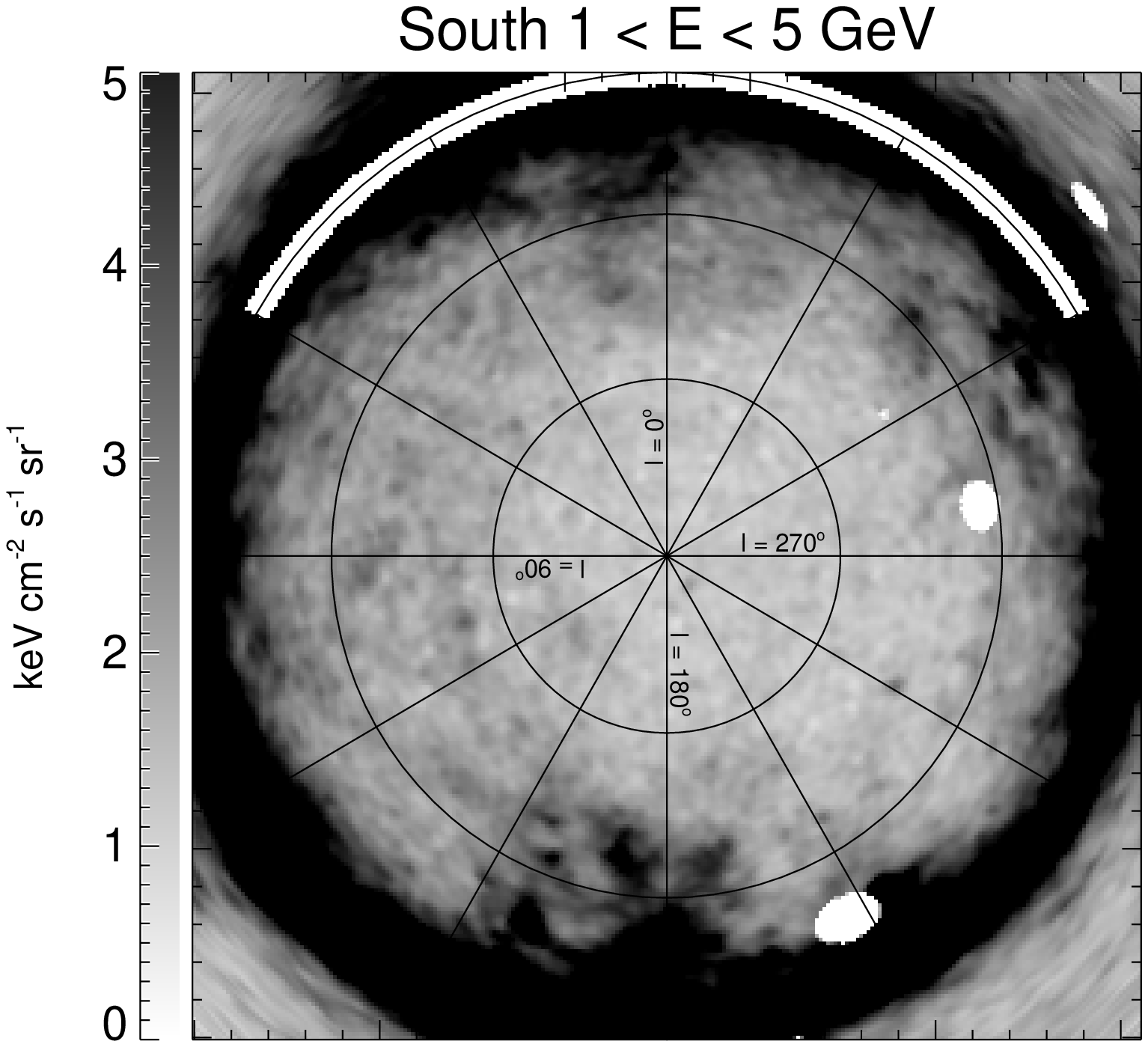}
\includegraphics[width=0.4\textwidth]{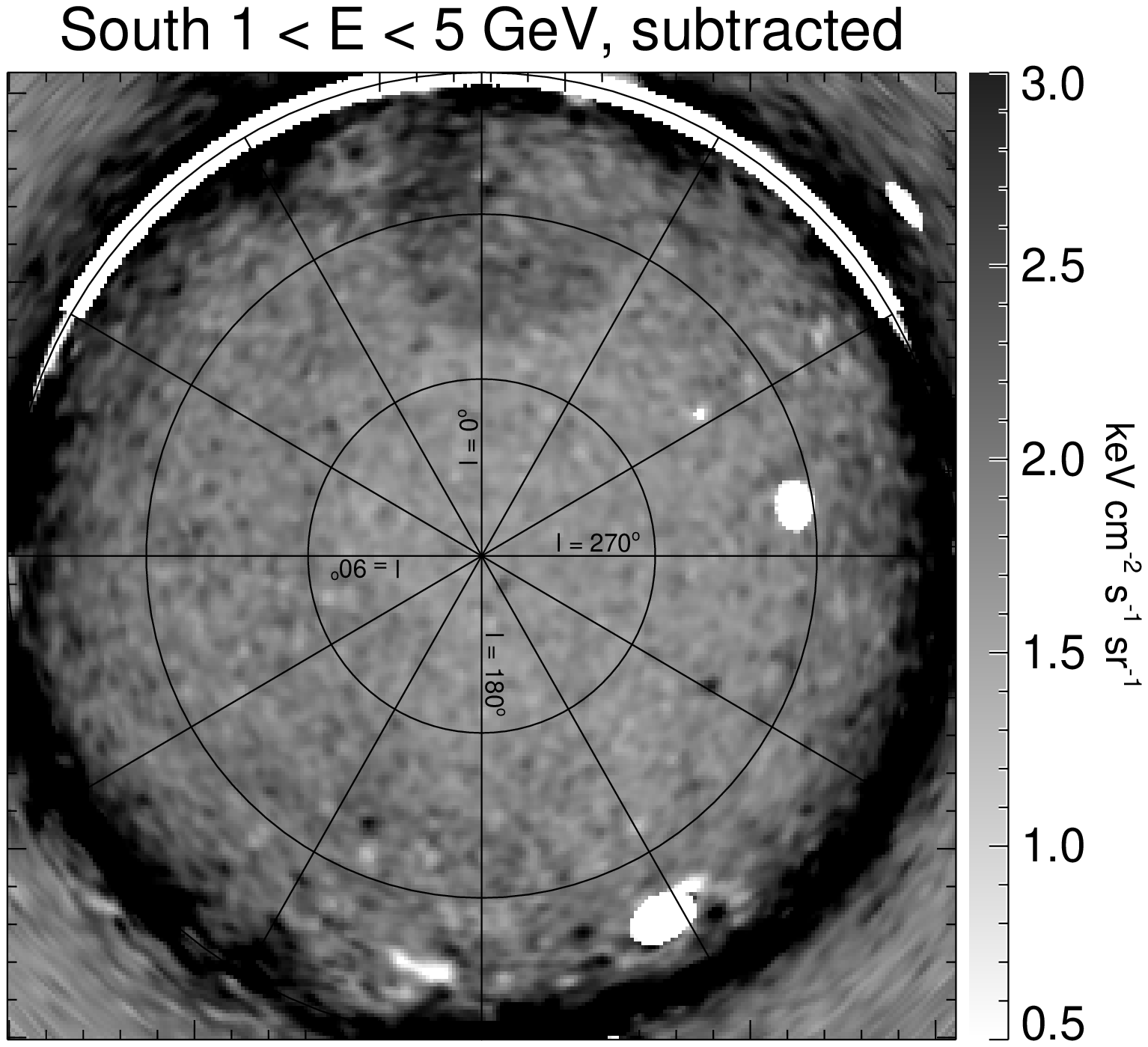}
\end{center}
\caption{The zenithal equal area (ZEA) projection with respect to both north pole (\emph{upper panels}) and south pole (\emph{lower panels}) for the $1-5$ GeV energy band before (\emph{left panels}) and after (\emph{right panels}) subtracting the SFD dust and simple disk templates to reveal the \Fermi\ bubbles. }
\label{fig:zea}
\end{figure*}


\begin{figure*}[ht]
\begin{center}
\includegraphics[width=1.0\textwidth]{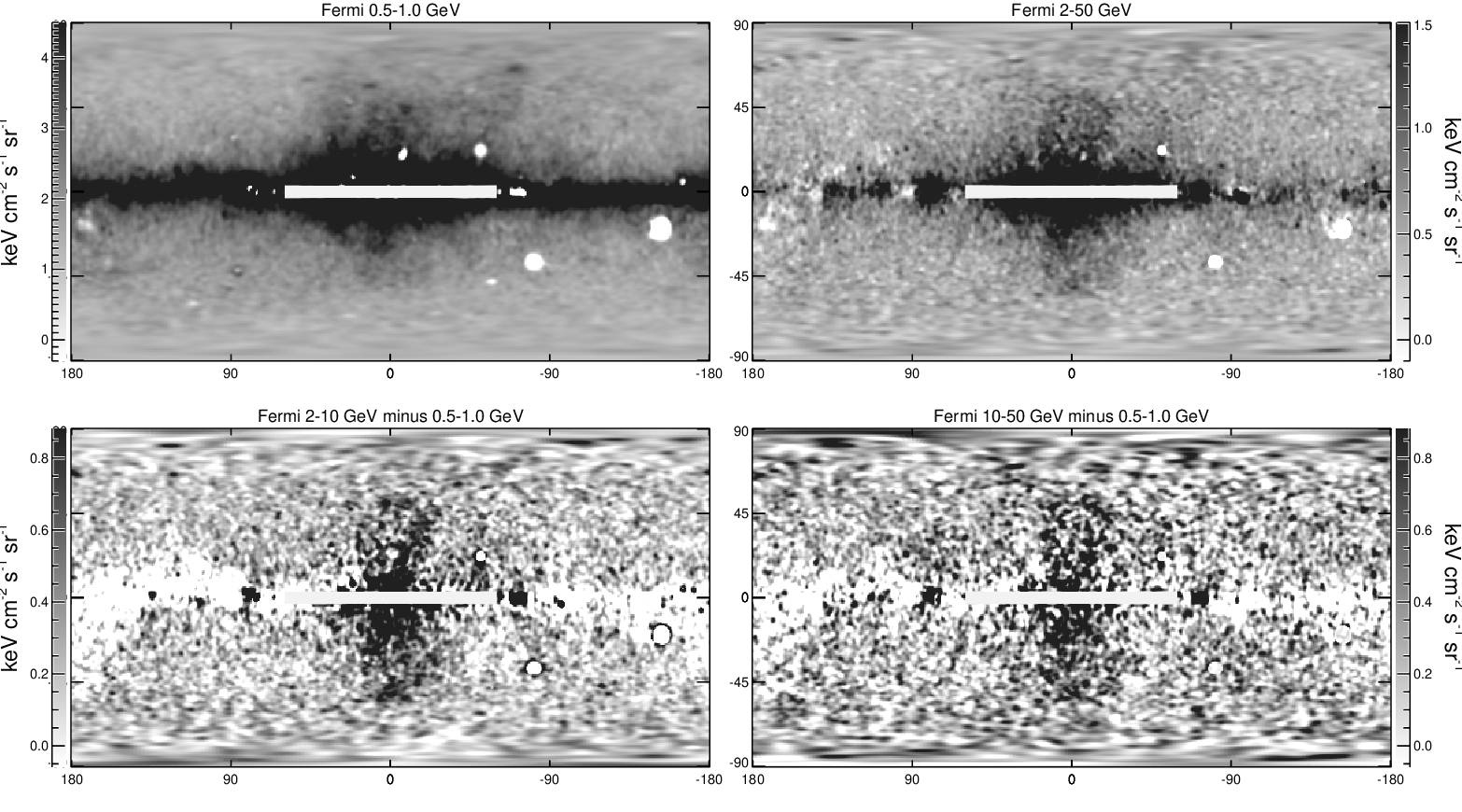}
\end{center}

\caption{\emph{Top left: } Full sky \Fermi-LAT 1.6 year $0.5-1.0$ GeV map subtracts the SFD dust map as a template of $\pi^0$ gammas. \emph{Top right: } The same as \emph{top left} panel, but for energy range $2-50$ GeV (note the different gray scale for the two panels). \emph{Bottom left: }The $2-10$ GeV \Fermi\ gamma-ray map subtracting the \emph{top left} $0.5-1.0$ GeV residual map which is used as a template of ICS of \emph{starlight}. \emph{Bottom right:} The same as \emph{bottom left} panel but for $10-50$ GeV map subtracting the \emph{top left} $0.5-1.0$ GeV residual map. The \Fermi\ bubble structures are better revealed after subtracting the lower energy $0.5-1.0$ GeV residual map with extended disk-like emission. }
\label{fig:fermiselfclean}
\end{figure*}

\section{Fermi Data and Map Making}
\label{sec:data}

The Large Area Telescope (LAT; see \citealt{Gehrels:1999ri};
\citealt{Atwood:2009}; as well as the \emph{Fermi}
homepage\footnote{\texttt{http://fermi.gsfc.nasa.gov/}}) 
is a pair-conversion telescope consisting of 16
modules of tungsten and silicon-strip trackers, on top of a calorimeter with a thickness of 7
radiation lengths.  It has a scintillating anti-coincidence detector
that covers the tracker array, and a programmable trigger and data
acquisition system.  The LAT provides a wide field of view, and covers
the energy range from about 30 MeV to 300 GeV.

The spacecraft occupies a low Earth orbit with an inclination of
$25.6\degree$. The field of view is so wide that the entire sky may
be covered in two orbits by rocking the spacecraft north of zenith on
one orbit and south of zenith on the next.  This scan strategy exposes
the LAT to atmospheric gammas at high zenith angles.  We make use of
only events designated ``Class 3'' (P6\_V3 diffuse class) by the LAT pipeline
with a zenith angle cut of $105\degree$.

The events are binned into a full sky map using HEALPix, a
convenient iso-latitude equal-area full-sky pixelization widely used
in the CMB community.\footnote{HEALPix software and documentation can be found at
  \texttt{http://healpix.jpl.nasa.gov}, and the IDL routines used in
  this analysis are available as part of the IDLUTILS product at
  \texttt{http://sdss3data.lbl.gov/software/idlutils}.} Spherical
harmonic smoothing is straightforward in this pixelization, and we
smooth each map by the appropriate kernel to obtain a Gaussian PSF of
$2\degree$ FWHM.  Because the PSF of the initial map must be
smaller than this, at energies below 1 GeV we use only
front-converting events (which have a smaller PSF).  A larger
smoothing scale would help improve S/N, but a relatively small
smoothing scale is necessary to see sharp features (such as the bubble
edges).  Furthermore, for the comparisons and linear combination
analysis described in e.g. \refsec{lowEdiffuse}. it is necessary to
smooth the maps at each energy to a common PSF. We generate maps from
$1-4\degree$, and find that a FWHM of $2\degree$ works well for our
purposes.  See \cite{fermihaze} for details on map construction,
smoothing, masking, and for instructions on how to download the maps.

Our current gamma-ray maps (v2\_3) constructed from the \Fermi\ data
have greater signal/noise compared to the previously released v1\_0
maps. They contain photon events from mission times 239557417.494176
to 291965661.204593, for about 606 days or 1.66 years of data (rather
than 1 year).  We refer to these as the ``1.6 year maps.''  
As in \cite{fermihaze}, we construct maps of
front-converting and back-converting events separately, smooth to a
common PSF, and then combine them.  The point source subtraction has
improved: instead of interpolating over every source in the 3-month
catalog, we use the 1-year catalog,\footnote{Available from
  \texttt{http://fermi.gsfc.nasa.gov/ssc/data}} and subtract each
point source from the maps in each energy bin, using estimates of the
PSF from the \Fermi\ science tools.\footnote{See
  \texttt{http://fermi.gsfc.nasa.gov/ssc/data/analysis/documentation/}}
For the 200 brightest and 200 most variable sources, the subtraction
is noticeably imperfect, so we interpolate over the core of the PSF
after subtracting the best estimate.  We take care to expand the mask
for very bright sources (Geminga, 3C 454.3, and LAT PSR J1836+5925).
The resulting map is appropriate for diffuse work at $|b| >
3\degree$.  At $|b| < 3\degree$ the maps are severely compromised by
the poor subtraction and interpolation over a large number of point
sources.  Further details of the map processing may be found in
Appendix B of \cite{fermihaze}. The v2\_3 maps used in this work and color versions of the map are
available for download.\footnote{Available at \texttt{http://fermi.skymaps.info}}



\begin{figure*}[ht]
\begin{center}
\includegraphics[width=1.0\textwidth]{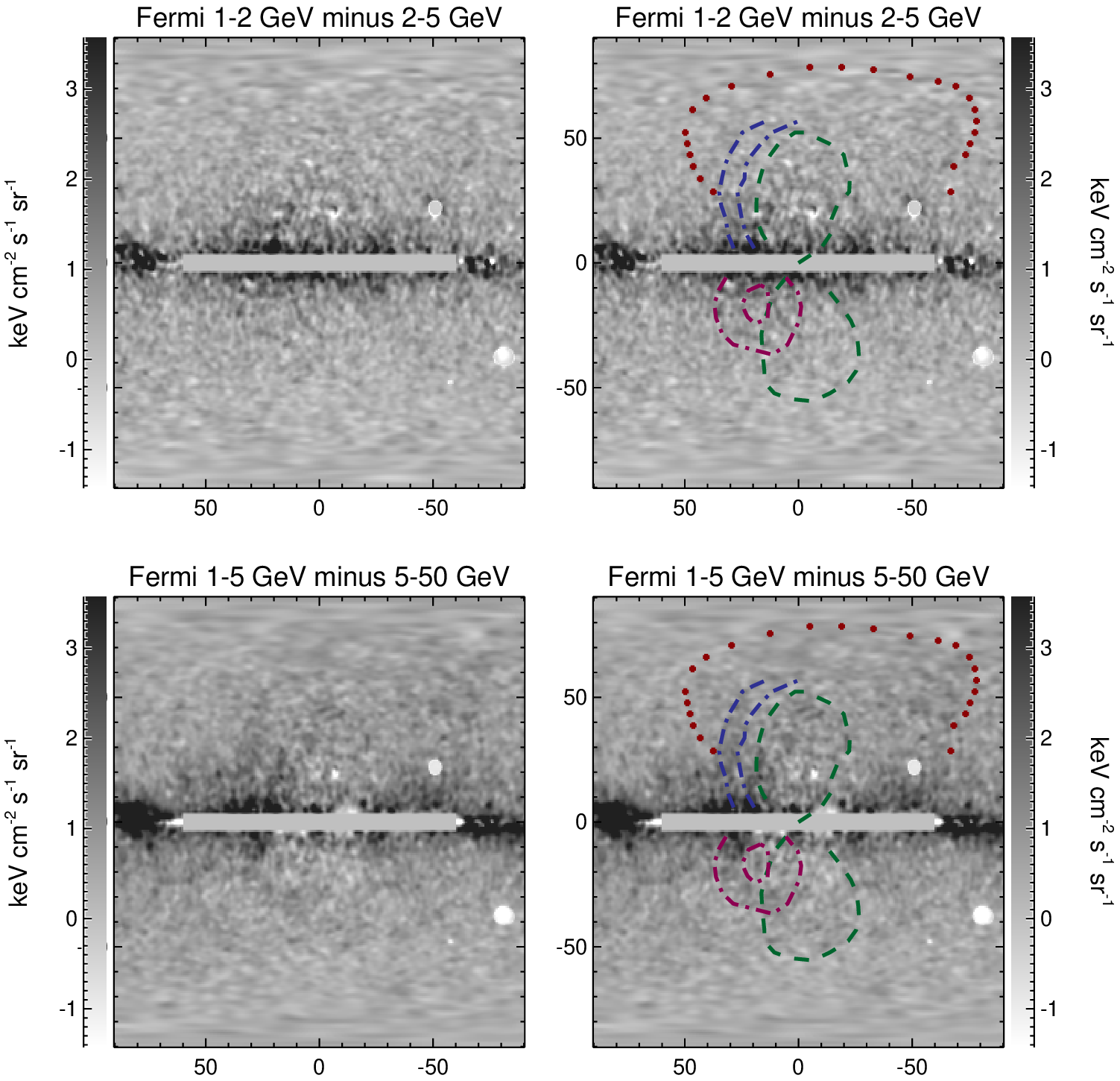}
\end{center}
\caption{Comparison of the \Fermi\ bubbles between different energy bins. \emph{Top left:} Subtraction of the $2-5$ GeV residual map from the $1-2$ GeV residual map; each residual map is constructed from the data by regressing out the SFD dust and disk templates to best reveal the \Fermi\ bubbles. The difference map is consistent with Poisson noise away from the masked region, and the bubble features can hardly be recognized, indicating that different spatial regions of the \Fermi\ bubbles have the same spectrum. \emph{Top right:} The same map as \emph{left} panel, but with the \Fermi\ bubble features overplotted for comparison. The marked features are the same as those plotted in \reffig{fullsky}, and are listed in \reftbl{bubblecoords}. \emph{Bottom row:} Same as the \emph{upper panels}, but subtracting the $5-50$ GeV residual map from the $1-5$ GeV residual map.}
\label{fig:energysplit}
\end{figure*}


\begin{figure*}[ht]
\begin{center}
\includegraphics[width=1.0\textwidth]{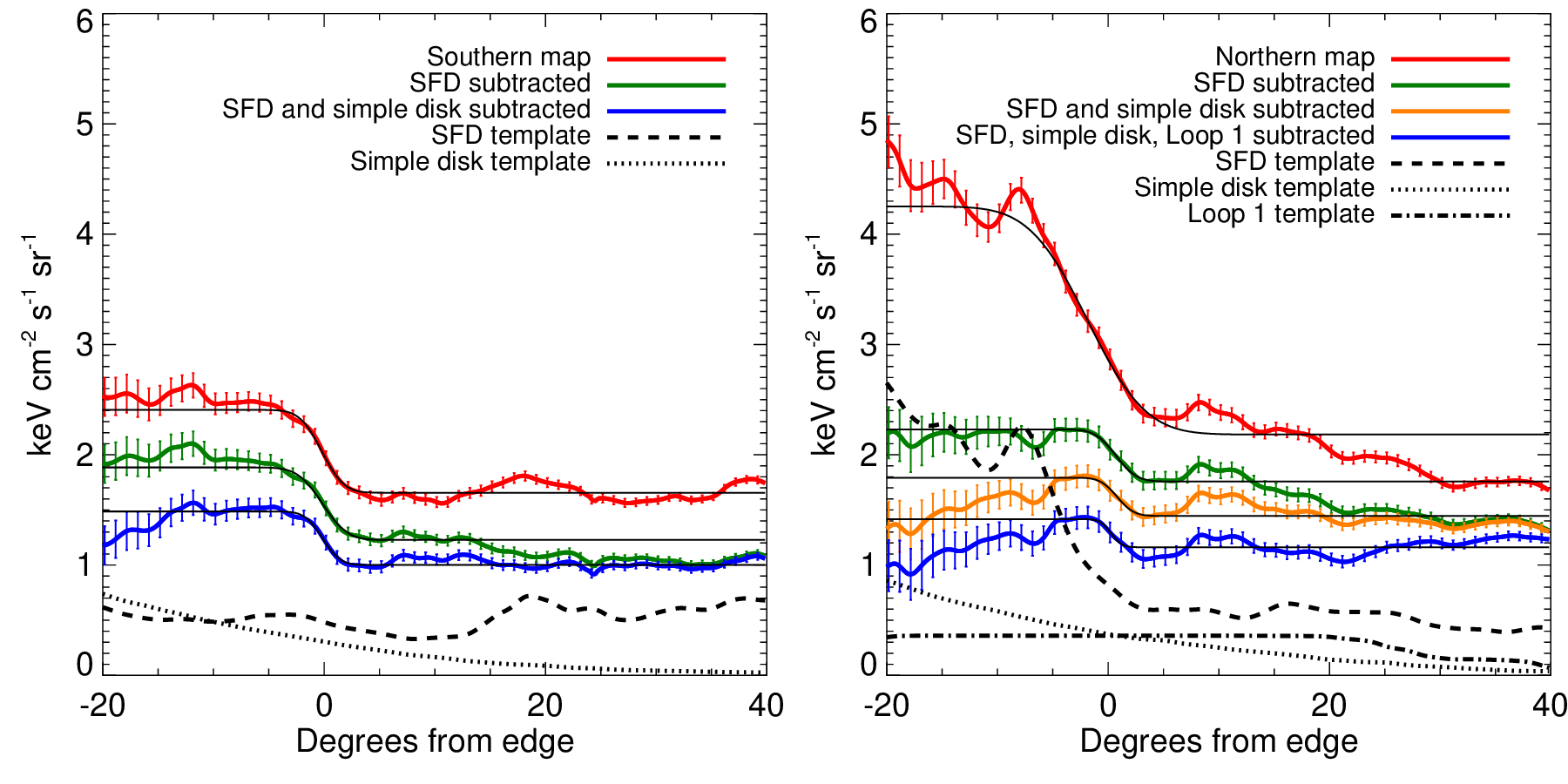} \\
\includegraphics[width=1.0\textwidth]{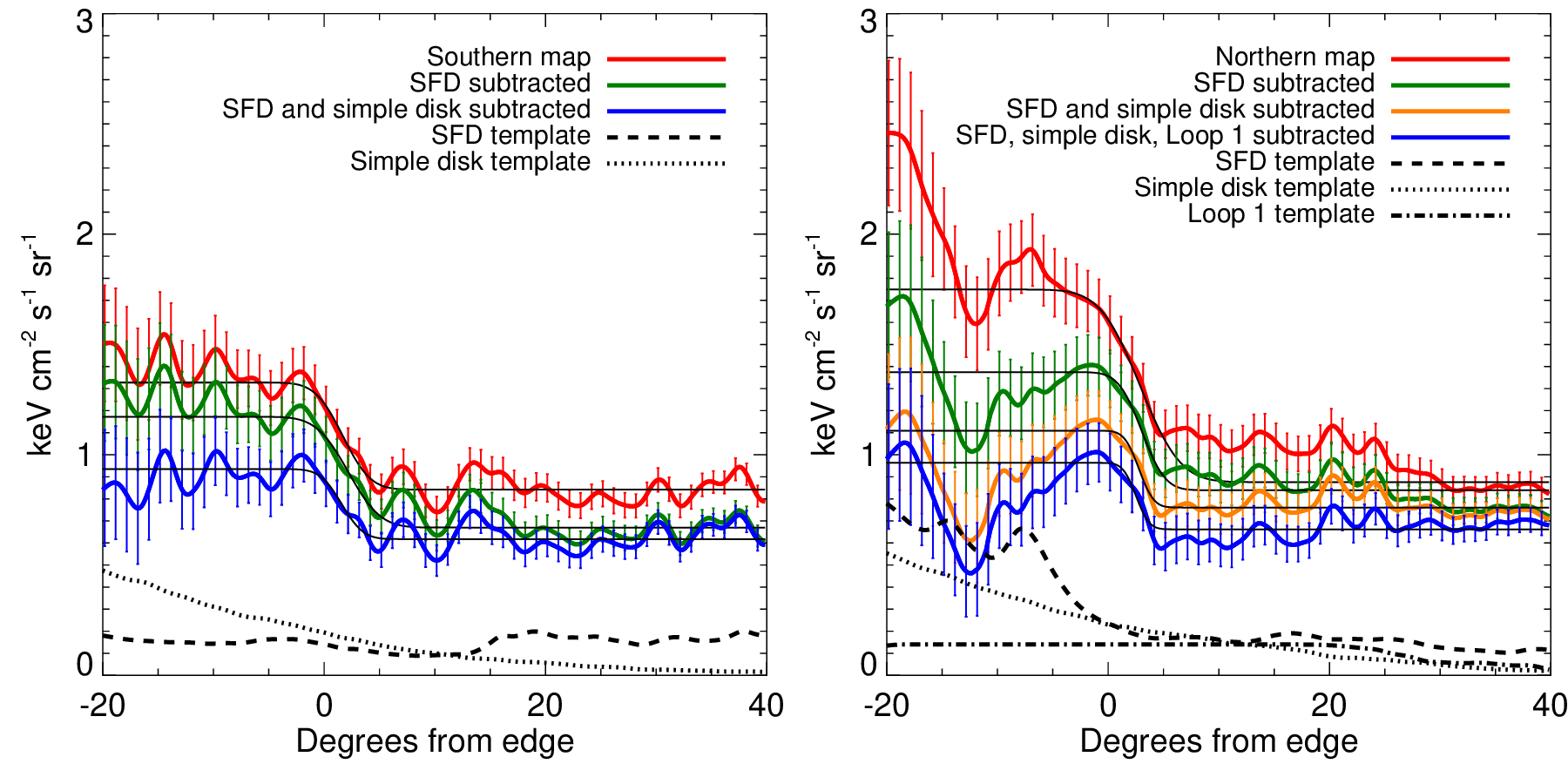}
\end{center}
\caption{Intensity as a function of radial distance from the bubble edge, averaged over great circle arcs intersecting the bubble center and lying at $|b| > 28^\circ$. Results are shown for (\emph{left}) the southern bubble, and (\emph{right}) the northern bubble, for (\emph{top}) the averaged $1-2$ and $2-5$ GeV maps, and (\emph{bottom}) the averaged $5-10$ and $10-20$ GeV maps. Different lines show the results at different stages of the template regression procedure and the corresponding errors are plotted (see text for an outline of the error analysis).}
\label{fig:edgeplot}
\end{figure*}


\begin{figure*}[ht]
\begin{center}
\includegraphics[width=1.0\textwidth]{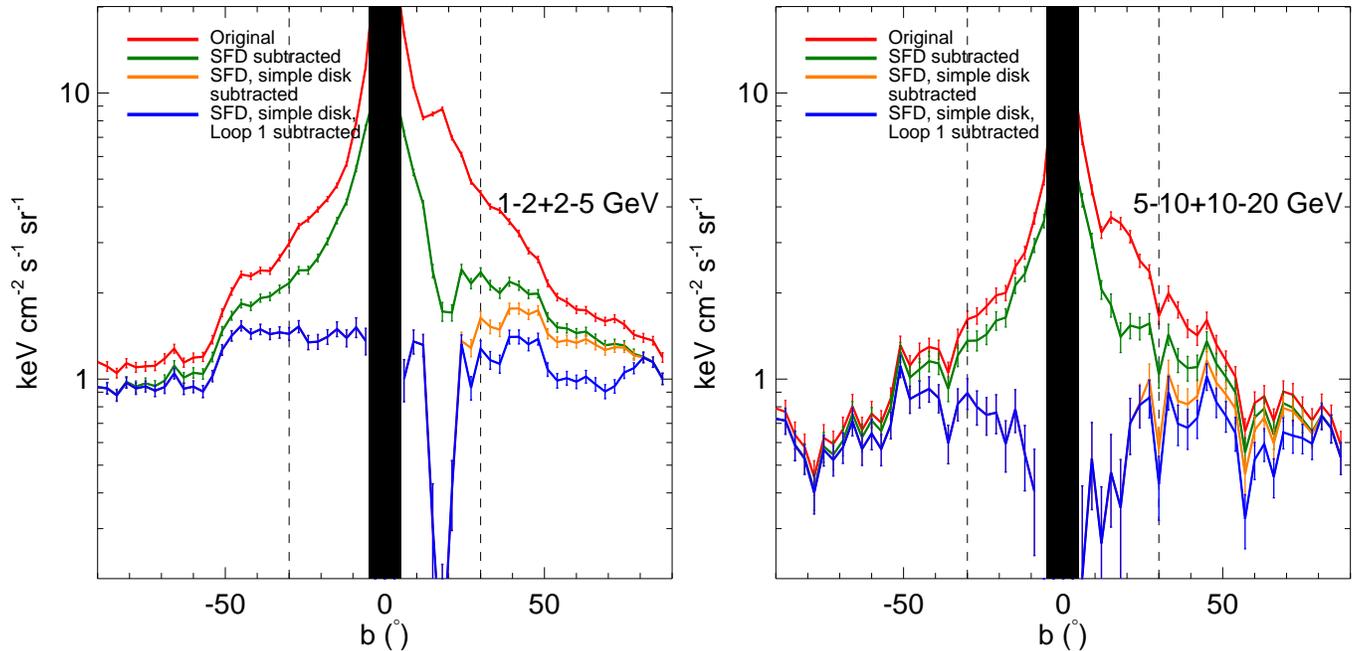}
\end{center}
\caption{Intensity averaged over the central 20 degrees in longitude, as a function of latitude, for (\emph{left}) the averaged $1-2$ and $2-5$ GeV maps, and (\emph{right}) the averaged 5-10 and 10-20 GeV maps. We construct great circle arcs perpendicular to the $l=0$ great circle, extending $10^\circ$ in each direction (east and west), and average the emission over each such arc; the ``$b$'' label corresponding to each arc, and the $x$-axis of the plot, refers to the value of $b$ at $l=0$. Different lines show the results at different stages of the template subtraction process. The large oversubtraction at $b \sim 15^\circ$ in the north, especially pronounced in the low-energy data, is associated with a bright feature in the SFD dust map.}
\label{fig:latitudeprofile}
\end{figure*}


\begin{figure*}[ht]
\begin{center}
\includegraphics[width=.8\textwidth]{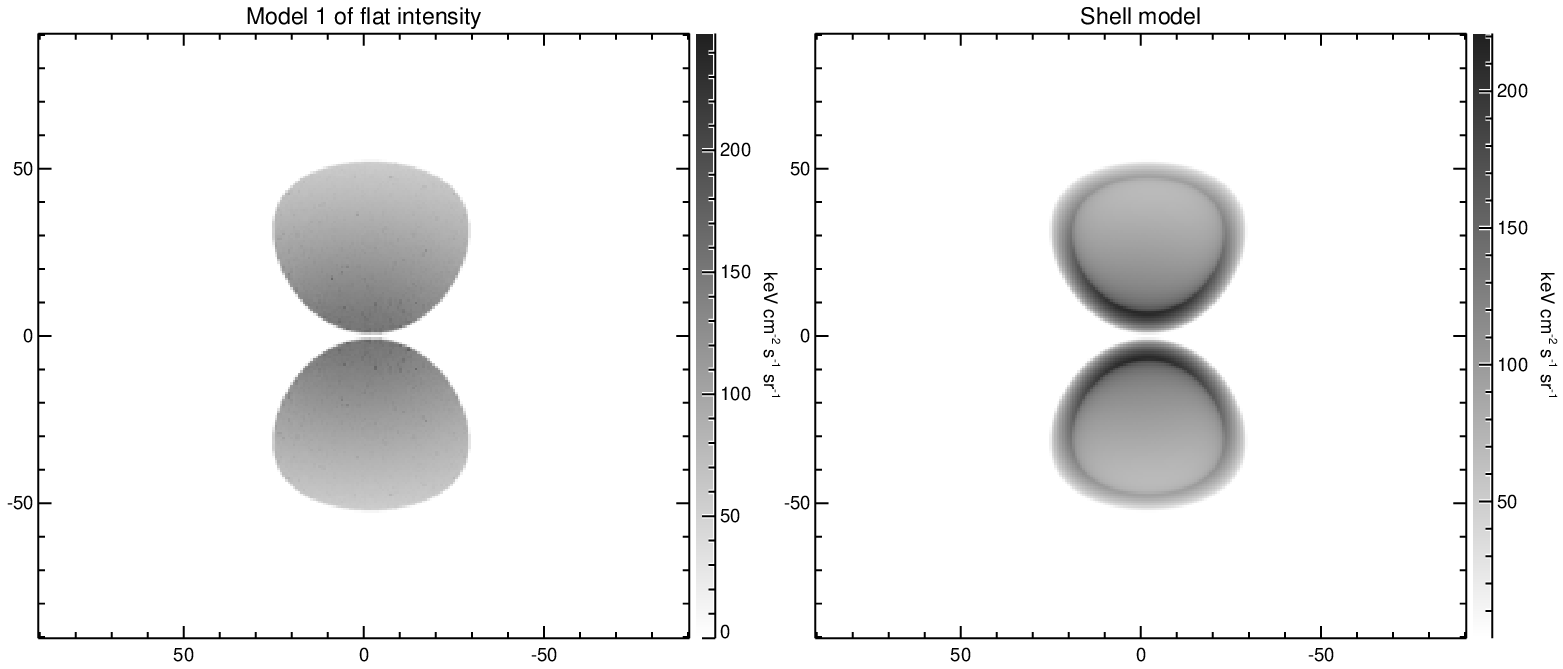}
\includegraphics[width=.8\textwidth]{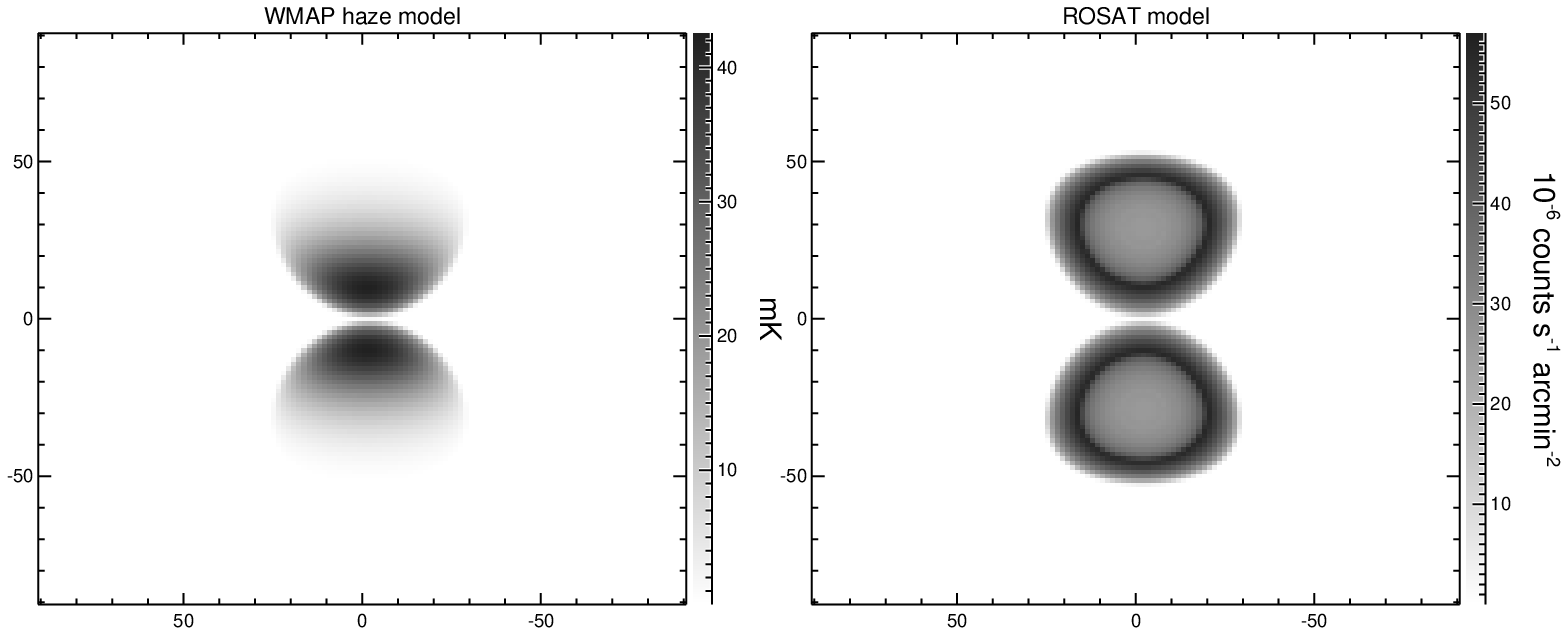}
\end{center}
\caption{
Projected emission from four emissivity distributions to illustrate the qualitative features of each. 
\emph{Top left:} A toy model of the \Fermi\ bubbles with flat projected intensity. In this model, the volume emissivity (assuming the ISM is optically thin to gamma rays) is proportional to $(R^2-r^2)^{-1/2}$ for $r < R$, and zero otherwise, where $R$=3.5 kpc is the approximate radius of the two bubbles, and $r$ is the distance to the center of the north or south bubble. \emph{Top right:} A bubble model with compressed gas shells with a thickness of 0.5 kpc; the electron CR density in the shell is a factor of 5 higher than in the interior of the bubbles. For this model, a limb-brightened edge of the bubbles is clearly visible, a feature which is not seen in the \Fermi\ data. \emph{Bottom left:} An illustrative toy model for the \WMAP\ haze.  The haze synchrotron emissivity depends on the electron CR density and the magnetic field; here we take $B = B_{0} e^{-z/z_{0}}$, where $z_{0} = 2$ kpc, and show the line of sight integral of this B-field through bubble volume.  Even though the synchrotron emissivity is not simply the product of CR density times field strength, this panel suggests that the decreasing intensity of the \WMAP\ haze at high latitudes is due to the decay of the Galactic magnetic field away from the Galactic plane. \emph{Bottom right:} A toy model for the \rosat\ X-ray features. The observed soft X-rays are limb-brightened and we assume all gas is uniformly distributed within a compressed shell, with no contribution from the interior, and X-ray emission is proportional to the gas density squared. The thickness of the shell is 1 kpc.}
\label{fig:template}
\end{figure*}

\section{\emph{FERMI} bubbles}
\label{sec:bubbles}

\subsection{Diffuse Galactic Emission Models}
\label{sec:diffusemodel}

At low ($\sim$1 GeV) energies, and close to the Galactic plane, the gamma-rays observed by \emph{Fermi} are dominated by photons from the decay of $\pi^0$ particles, produced by the collisions of CR \emph{protons} with ambient ionized gas and dust in the ISM. Collisions of high energy CR \emph{electrons} with the ISM (primarily protons, but also heavier nuclei) produce bremsstrahlung radiation. The \Fermi\ all-sky gamma-ray maps in different energy bands are shown in \reffig{fermimaps}. In order to uncover the \Fermi\ bubble features better, significant $\pi^0$ emission, bremsstrahlung, and IC emission from the Galactic disk must be removed.  We take three approaches for the foreground removal. One is to use the \Fermi\ Diffuse
Galactic Model provided by the \Fermi\ team\footnote{See \texttt{http://fermi.gsfc.nasa.gov/ssc/data/access/lat/BackgroundModels.html}}(\refsec{fermidiffuse}). The second approach employs a linear combination of templates of known emission mechanisms (\refsec{simplediffuse}), using existing maps from multiwavelength observations and/or constructed geometric templates. The third approach is taking advantage of the lower energy band $0.5-1.0$ GeV \Fermi\ map to form a template of a diffusion emission model (\refsec{lowEdiffuse}).

\subsubsection{\emph{Fermi} Diffuse Galactic Model}
\label{sec:fermidiffuse}

The \Fermi\ diffuse Galactic model\footnote{Available from \texttt{http://fermi.gsfc.nasa.gov/ssc/data}. The version of the diffuse model we use is $gll.iem.v02$.} is a comprehensive model of Galactic gamma-ray emission from the ISM, and serves as a background
estimate for point source removal. This model is based on template fits to the gamma-ray data, and includes an IC component generated by the \texttt{GALPROP} cosmic ray
propagation code. \texttt{GALPROP} calculates the steady state solution to the
diffusion-energy-loss equation, given the 3D gas distribution,
interstellar radiation field, B-field model, CR diffusion
assumptions, and many other input parameters
\citep{Strong:1999sv,galprop09,Strong:2007}.  The model is constrained by gamma-ray
and microwave observations, locally measured CR spectra, etc.  By
using a well motivated physical model, one can solve for the spectral
and spatial dependence of the injection function, i.e. the $e^-$ and
$p$ CR primary source spectra, as a function of position and
energy. The diffuse model is the key connection between the input
assumptions and the observables, and is essential for interpretation
of the \Fermi-LAT data.  It is important to make it as complete as possible.

In this model, the $\pi^0$ emission is modeled with maps of
interstellar gas: \HI\ from the Leiden/Argentine/Bonn (LAB) Galactic Survey \citep{Kalberla:2005} and CO
from the CfA composite CO survey \citep{Dame:2001}.  Because the
$\pi^0$ emission is a function of both the gas density and the proton
CR density, which varies with Galactocentric radius, it is desirable
to allow the emissivity of the gas to vary.  Both the \HI\ and CO
surveys contain velocity information, which allows separation into six
Galactocentric annuli (rings) with boundaries at 4.0, 5.5, 7.0, 10.0,
16.5, and 50 kpc.  The spectrum of each is allowed to float, with the
constraint that the sum of the rings along each line of sight
approximates the observed signal.  This freedom also allows for
varying amounts of bremsstrahlung (with varying spectrum) which also
scales with the ISM density.  The contribution from IC is modeled with
\texttt{GALPROP} as described above, and included in the ring fit\footnote{A description of this model is available at \texttt{fermi.gsfc.nasa.gov/ssc/data/access/lat/ring\_for\_FSSC\_final4.pdf}}.

This procedure provides a diffuse model that faithfully reproduces
most of the features of the diffuse Galactic emission.  One
shortcoming is the existence of ``dark gas'' \citep{Grenier:2005}, clouds
with gamma-ray emission that do not appear in the \HI\ and CO surveys.
These features are seen in dust maps \citep{Schlegel:1997yv} and may simply be
molecular H clouds underabundant in CO.

The Fermi diffuse model is primarily intended as a background for point source detection, and comes with a number of caveats. However these caveats apply mainly near the Galactic plane, and at $E > 50 \gev$.  It is nevertheless useful for qualitatively revealing features in the diffuse emission at high latitude. 
In \reffig{diffusemodel}, we show the residual maps after subtracting the \Fermi\ diffuse Galactic model in different energy bins. A double-lobed bubble structure is clearly revealed, with similar morphology in the different energy bins. We note that the bubble is neither limb brightened nor centrally brightened, consistent with a flat \emph{projected} intensity distribution.

\subsubsection{Simple Template-Based Diffuse Galactic Model}
\label{sec:simplediffuse}

Since the dominant foreground gamma-rays originate from $\pi^0$ gammas produced by CR protons interacting with the ISM, the resulting gamma-ray distribution should be morphologically correlated with other maps of spatial tracers of the ISM. A good candidate is the Schlegel, Finkbeiner, \& Davis (SFD) map of Galactic dust, based on $100\,\micron$ far IR data \citep{Schlegel:1997yv}. The $\pi^0$/bremsstrahlung gamma-ray intensity is proportional to the ISM density $\times$ the CR proton/electron density integrated along the line of sight. As long as the CR proton/electron spectrum and density are approximately spatially uniform,  the ISM column density is a good tracer of $\pi^0$/bremsstrahlung emission. The dust map has some advantages over gas maps: there are no problems with self absorption, no concerns about ``dark gas'' \citep{Grenier:2005}, and the SFD dust map has sufficient spatial resolution (SFD has spatial resolution of 6', and LAB is 36'). On the other hand, SFD contains no velocity information, so it is impossible to break the map into Galactocentric rings. Nevertheless, it is instructive to employ the SFD map to build a very simple foreground model. The goal is to remove
foregrounds in a fashion that reveals the underlying structure
with as few physical assumptions as possible. We will compare the resulting residuals using this simple diffuse model with those using the \Fermi\ diffuse Galactic model. 

As an example, we reveal the \Fermi\ bubble structure from $1-5$ GeV \Fermi-LAT 1.6 yr data in \reffig{bubbles}. We use the SFD dust map as a template of the $\pi^0$ gamma foreground. The correlation between \Fermi\ and
SFD dust is striking, and the most obvious features are removed by this
subtraction (\emph{top row} in \reffig{bubbles}).  This step makes the bubbles above and below the GC easily visible.  The revealed bubbles are not aligned with any structures in the
dust map, and cannot plausibly be an artifact of that subtraction.

Next, a simple disk model is subtracted (\reffig{bubbles}, \emph{middle row}).  The
purpose of this subtraction is to reveal the structure deeper into the
plane, and allow a harder color stretch.  The functional form is
$(\csc|b|)-1$ in latitude and a Gaussian ($\sigma_\ell=30\degree$) in
longitude. The disk model mostly removes the IC gamma-rays produced by cosmic ray electrons interacting with the ISRF including CMB, infrared, and optical photons; as discussed previously, such electrons are thought to be mostly injected in the Galactic disk by supernova shock acceleration before diffusing outward.

Finally, we fit a simple double-lobed geometric \emph{bubble} model with flat gamma-ray intensity to the data, to remove the remaining large-scale residuals towards the GC (\reffig{bubbles}, \emph{bottom row}).  In this model, we identify the approximate edges of the two bubble-like structures towards the GC in the \emph{bottom left} panel (shown with \emph{dashed green} line in \emph{right} panels of \reffig{fullsky}). We then fill the identified double-lobed bubble structure with uniform gamma ray intensity, as a template for the ``\Fermi\ bubbles'' (\emph{bottom right} panel of \reffig{bubbles}). If the \Fermi\ bubbles constitute the projection of a three dimensional two-bubble structure symmetric to the Galactic plane and the minor axis of the Galactic disk, taking the distance to the GC $R_\odot$ = 8.5 kpc, the bubble centers are approximately 10 kpc away from us and 5 kpc above and below the Galactic center, extending up to roughly 10 kpc as the most distant edge from GC has $|b|\sim50\degree$.  No structures like this appear in \texttt{GALPROP} models, and in fact \texttt{GALPROP} is often run with a box-height smaller than this. Because the structures are so well centered on the GC, they are unlikely to be
local.

In \reffig{fullsky}, we show the full sky residual maps at $1-5$ GeV and $5-50$ GeV after subtracting the SFD dust and the disk model to best reveal the \Fermi\ bubble features. Although photon Poisson noise is much greater in the $5-50$ GeV map, we identify a \Fermi\ bubble structure morphologically similar to the structure in the $1-5$ GeV map, present both above and below the Galactic plane.

In \reffig{zea}, we show the full sky maps at $1-5$ GeV with the zenithal equal area (ZEA) projection with respect to both north pole and south pole. We found no interesting features appear near the poles.

\subsubsection{Low Energy Fermi Map as a Diffuse Galactic Model}
\label{sec:lowEdiffuse}

In \reffig{fermiselfclean}, we show the $0.5 - 1$ GeV and $2 - 50$ GeV residual maps after subtracting only the SFD dust map as a template of foreground $\pi^0$ gammas. The residual maps should be dominated by IC emission from CR electrons interacting with the ISRF. We use the $0.5 - 1$ GeV maps as a template of IC emission from high energy electrons scattering \emph{starlight}, and subtract the template from higher energy maps (the \emph{lower panels} of \reffig{fermiselfclean}). The \Fermi\ bubble structures are clearly revealed. We thus conclude that the \Fermi\ bubbles are mostly from high energy electron CRs IC scattering on CMB photons, and IR photons at higher energies (see \refsec{Interp} for more discussion).
 By comparing the \Fermi\ diffuse Galactic model subtraction (\reffig{diffusemodel}) and our simple template model subtraction (\reffig{fullsky}), we find that the bubble structures are robust to quite different foreground subtractions.  It is difficult to see how such emission could arise --
especially with sharp edges \Refsec{feature} -- as an artifact of these subtractions.

\subsection{\emph{Fermi} Bubbles: Morphology}
\label{sec:morphology}

\subsubsection{Morphological Features}
\label{sec:feature}

In the \emph{right} panels of \reffig{fullsky}, we illustrate the edges of the \Fermi\ bubbles and some other features. We find that the \Fermi\ bubbles have distinct sharp edges, rather than smoothly falling off as modeled in \cite{fermihaze}. Besides the two bubbles symmetric with respect to the Galactic plane, we find one giant \emph{northern arc} that embraces half of the north bubble, that extends from the disk up to $b\sim50$, with $\ell$ ranging from roughly $-40\degree$ to $0\degree$. It has a brighter and sharper outer edge in the $1-5$ GeV map. On a even larger scale, we identify a fainter structure extended up to $b\sim80\degree$, with $\ell$ ranging from roughly $-80\degree$ to $50\degree$. We will show in \refsec{othermaps} that this large extended structure corresponds to the \emph{North Polar Spur} emission associated with \emph{Loop I} \citep[as seen for example in the Haslam 408 MHz map][]{1982A&AS...47....1H}. We will discuss the possible relation of the \Fermi\ bubble, the \emph{northern arc}, and the \emph{Loop I} feature in \refsec{discussion}. In the $1-5$ GeV map, we also identify a ``donut-like'' structure in the south sky with $b$ ranging from roughly $-35\degree$ to $0\degree$ and $\ell$ from roughly $0\degree$ to $40\degree$. The coordinates of the \Fermi\ bubble edges, \emph{northern arc}, \emph{Loop I}, and the ``donut'' feature identified from the $1-5$ GeV map are listed in \reftbl{bubblecoords}.

In \reffig{energysplit} we compare the \Fermi\ bubble morphology in different energy bins. We show the difference of the $1 - 2$ and $2 - 5$ GeV residual maps in the \emph{upper panels}; each residual map is the result of subtracting the SFD dust map and the simple disk model to best reveal the \Fermi\ bubbles. The difference maps between the $1 - 5$ and $5 - 50$ GeV maps are shown in the \emph{lower panels}. The bubble features almost disappear in the difference maps, indicating that different parts of the \Fermi\ bubbles have similar spectra.

To study the sharp edges of the bubbles at high latitude more carefully, we examine the (projected) intensity profiles along arcs of great circles passing through the estimated centers of the north and south bubbles, and intersecting the bubble edge (as defined in \reffig{fullsky}) at $|b| > 28^\circ$. Along each such ray, we define the intersection of the arc with the bubble edge to be the origin of the coordinate system; we then perform an inverse-variance-weighted average of the intensity profile along the rays (as a function of distance from the bubble edge). We subtract a constant offset from the profile along each ray, prior to averaging the rays together, to minimize aliasing of point sources onto the averaged profile, and then add the averaged offset back in at the end. The inverse variance for each data point is obtained from the Poisson errors in the original photon data, prior to any subtraction of point sources or templates (however, the smoothing of the map is taken into account). When the rays are averaged together, the naive inverse variance in the result is multiplied by a factor of the annulus radius (for the points being averaged together) divided by $4 \pi \sigma^2$, where $\sigma$ is the $1\sigma$ value of the PSF, and the annulus width is taken to be $1^\circ$ (the spacing between the points along the rays; this is comparable to the smoothing scale, so there may still be unaccounted-for correlations between the displayed errors); this is done to take into account that the number of independent measurements being sampled by the rays can be far less than the number of rays, especially close to the center of the bubbles. This procedure is repeated for all the stages of the template subtraction, using the simple disk template for inverse Compton scattering (ICS) for illustration (our conclusions do not depend on this choice).

The results are shown in \reffig{edgeplot} for the averaged $(1-2)+(2-5)$ GeV maps, and the averaged $(5-10)+(10-20)$ GeV maps. In both energy ranges the edges are clearly visible; in the south, this is true even before any templates are subtracted. The intensity profile of the north bubble is strikingly similar to profile of the south bubble. For both of the north and south bubbles, no significant edge-brightening or limb-brightening of the bubbles is apparent from the profiles, the flux is fairly uniform inside the bubbles.

In \reffig{latitudeprofile}, we plot the intensity profile as a function of latitude from the south to the north pole. We construct great circle arcs perpendicular to the $l=0$ great circle, extending $10^\circ$ in each direction (east and west), and average the emission over each such arc. The flatness of the bubbles with latitude (except possibly close to the Galactic plane), and the sharp edges at high latitude, are also apparent here.

We note that the flat intensity of the bubbles is striking. As we show in \reffig{template}, if we assume that the \Fermi\ bubbles are projected from spherically symmetric three-dimensional bubbles centered above and below the GC, a non-trivial emissivity distribution in the bubble interior is required to produce a flat projected intensity distribution (\emph{upper left} panel of \reffig{template}). If the ``bubbles'' originate from IC scattering, this suggests a rather non-uniform density distribution for the electron CRs, which -- combined with a nearly uniform spectral index -- presents challenges for many models for the electron injection. The expected intensity profile for a shock generated bubble with a compressed gas shell is shown in the \emph{upper right} panel; it is noticeably limb-brightened, in contrast to observations. 

\subsubsection{Spectrum of Gamma-ray Emission}
\label{sec:templatefit}

\begin{figure*}[ht]
\begin{center}
\includegraphics[width=.8\textwidth]{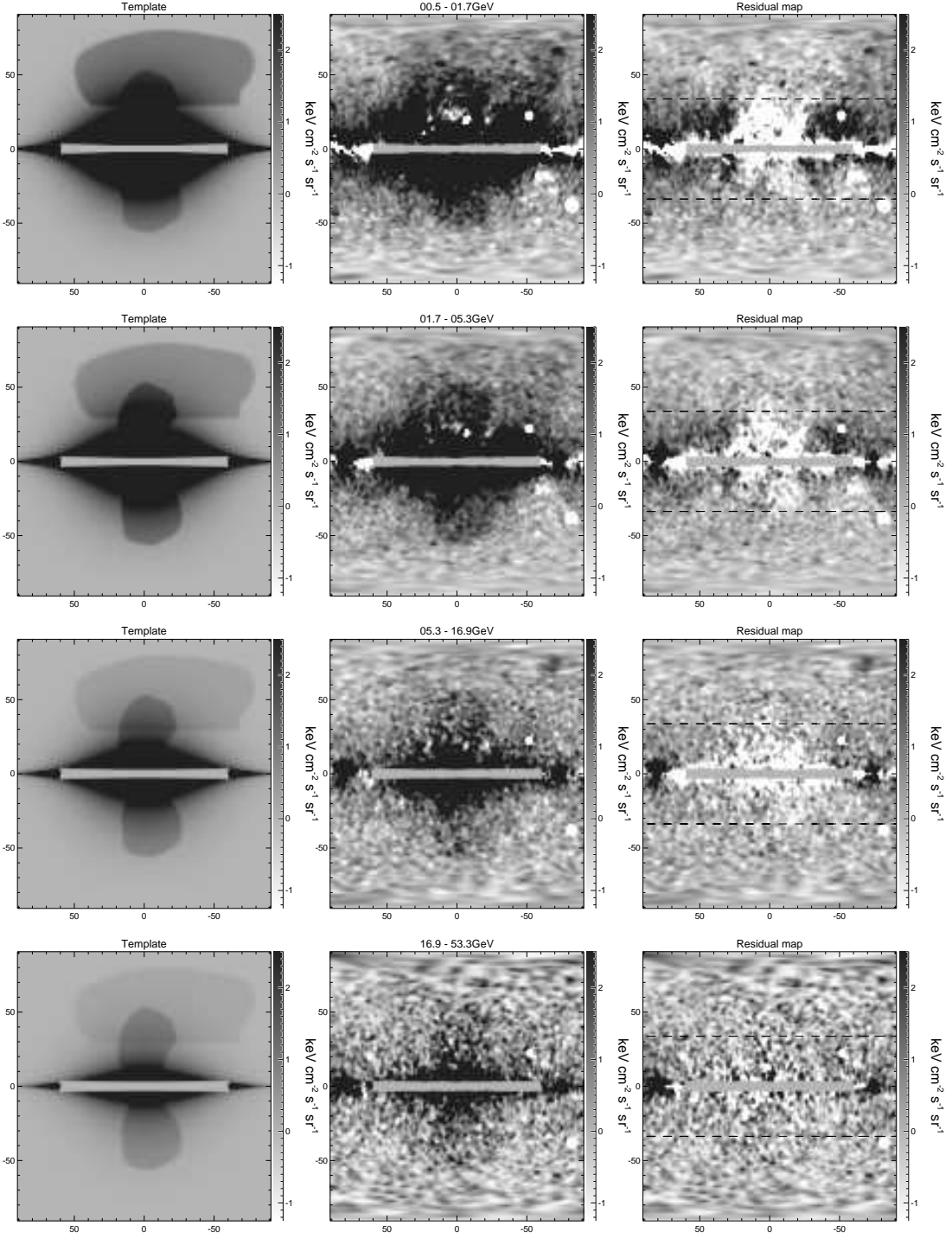}
\end{center}
\caption{
The models obtained from the multi-template fits, compared with the \Fermi\ maps, in different energy bins. The \emph{left} column shows the linear combination of the disk, \emph{Loop I}, uniform, and bubble templates that provide the best fit to the \Fermi\ maps after subtracting the best fit SFD dust template (shown in the \emph{middle} column).  The difference maps between the combined template and the data are shown in the \emph{right} column. The template fitting is done for the region with $|b| > 30\degree$ to avoid contaminations from the Galactic disk (shown with black dashed line in the \emph{right} column residual maps). The subtraction of the model largely removes the features seen in the \Fermi\ maps with  $|b| > 30\degree$. We use the same gray scale for all the panels. We find that both the disk IC template and \emph{Loop I} features fade off with increasing energy, but the bubble template is required for all the energy bands and does not fade off with increasing energy. The oversubtraction in the residual maps, especially at lower energy bins, is due to the simple disk IC model, which is not a good template across the entire disk. However, outside the masked region, the residual maps are consistent with Poisson noise without obvious large scale features.
}
\label{fig:bubbledisk}
\end{figure*}

We now attempt to estimate the spectrum of the gamma rays associated with the \Fermi\ bubbles, and the spatial variation of the spectrum. In order to reveal the hardness of the spectrum of the \Fermi\ bubbles, and quantitatively study the intensity flatness of the bubble interiors, we do a careful regression template fitting. First, we maximize the Poisson likelihood of a simple diffuse emission model involving 5-templates (see \refsec{simplediffuse}). In this model, we include the SFD dust map (\reffig{bubbles}, \emph{right panel} of the top row) as a tracer of $\pi^0$ emission which is dominant (or nearly so) at most energies on the disk and significant even at high latitudes, the simple disk model (\reffig{bubbles}, \emph{right panel} of the \emph{second row}), the bubble template (\reffig{bubbles}, \emph{right panel} of the bottom row), the \emph{Loop I} template (see e.g. \reffig{bubbledisk}, \emph{left panel} of the \emph{top row}), and a uniform background as templates to weight the \Fermi\ data properly. 

The significant isotropic background is due to extra-galactic emission and charged particle contamination, including heavy nuclei at high energies. The \Fermi\ collaboration has measured the extragalactic diffuse emission using additional cuts to reduce charged particle contamination \cite{Abdo:2010nz}: below $\sim 20$ GeV the isotropic contribution in our fits is roughly a factor of 2 larger than the extragalactic diffuse emission, but has a similar spectral slope. At energies above $\sim 20$ GeV the isotropic contribution becomes much harder, which we attribute to charged particle contamination.

For each set of model parameters, we compute the Poisson log likelihood,
\be
  \ln {\mathcal L} = \sum_i k_i\ln\mu_i - \mu_i - \ln(k_i!),
\ee
where $\mu_i$ is the synthetic map (i.e., linear combination of templates) at pixel $i$, and $k$ is the map of observed data. The last term is a function of only the observed maps. The $1\sigma$ Gaussian error is calculated from the likelihood by $\Delta\ln {\mathcal L} = 1/2$.  The error bars are simply the square root of the diagonals of the covariance matrix. We refer to Appendix B of \cite{fermihaze} for more details of the likelihood analysis. Maps of the models constructed from linear combinations of these five templates, and the residual maps between the \Fermi\ data and the combined templates at different energy bins, are shown in \reffig{bubbledisk}. In this fit, we mask out all pixels with Galactic latitude $|b| < 30\degree$ (the dashed black line in the residual maps).

\begin{figure*}[ht]
\begin{center}
\includegraphics[width=.8\textwidth]{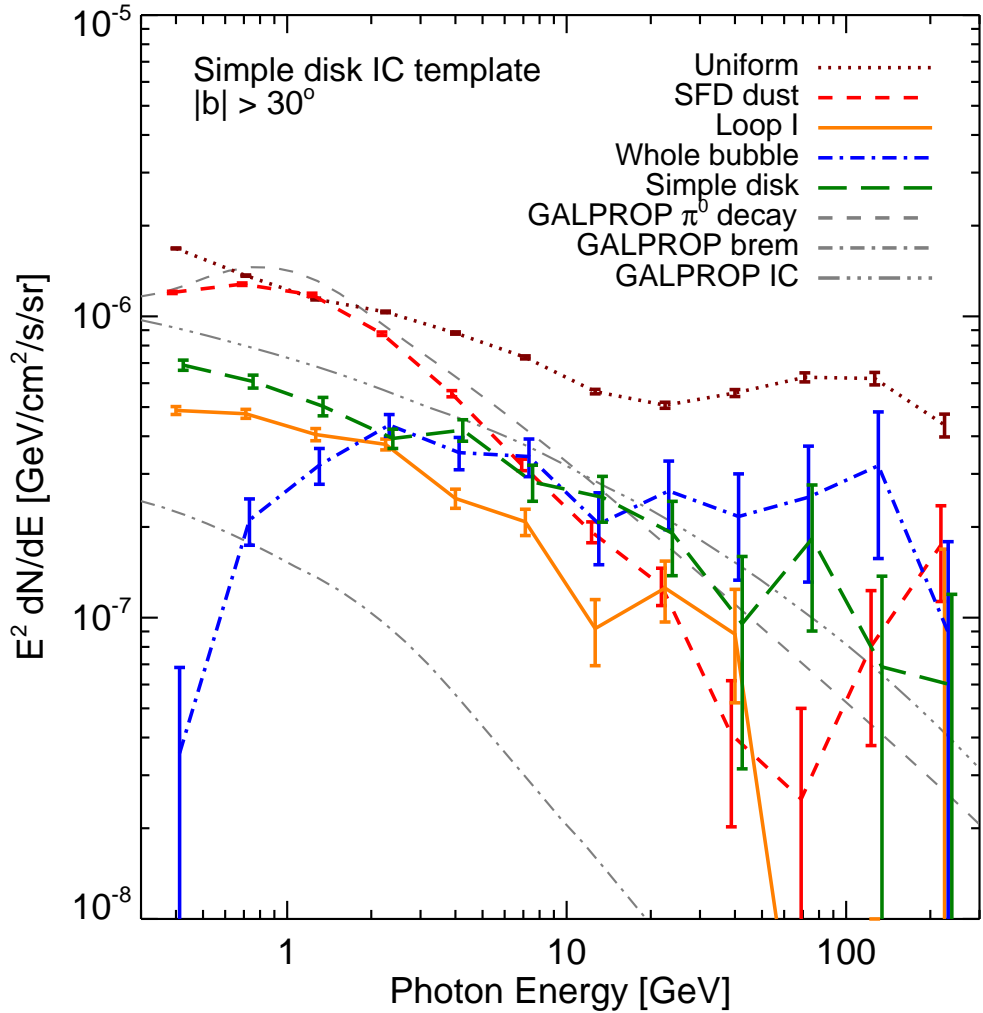}
\end{center}
\caption{
  Correlation spectra for the 5-template fit employing a simple disk model for the IC (and to a lesser degree bremsstrahlung) emission from supernova-shock-accelerated electrons (see \refsec{templatefit}). The SFD-correlated spectrum is shown by the red short-dashed line which roughly traces $\pi^0$ emission (the gray dashed line indicates a \texttt{GALPROP} prediction for $\pi^0$ emission). The disk-correlated emission is shown by the green dashed line, which traces the soft IC (gray triple-dot-dashed line) and bremsstrahlung (gray dot-dashed line) component. The spectrum of the uniform emission, which traces the isotropic background (including possible cosmic-ray contamination), is shown as a dotted brown line. The solid orange line indicates the spectrum of emission correlated with \emph{Loop I}, which has a similar spectrum to the disk-correlated emission. Finally, the blue dot-dashed line shows the spectrum correlated with the \Fermi\ bubble template. The bubble component has a notably harder (consistent with flat) spectrum than the other template-correlated spectra, and the models for the various emission mechanism generated from \texttt{GALPROP}, indicating that the \Fermi\ bubbles constitute a distinct component with a hard spectrum. The fitting is done over the $|b| > 30\degree$ region. Note that these \texttt{GALPROP} ``predictions'' are intended only to indicate the expected spectral shape for these emission components, for reference. The correlation coefficients for the SFD map and simple disk model are multiplied by the average value of these maps in the bubble region (defined by the \emph{bottom right} panel of \reffig{bubbles}, with a $|b| > 30^\circ$ cut) to obtain the associated gamma-ray emission; see \refsec{templatefit} for details, and \reftbl{normfac} for a summary of the normalization factors.
}
\label{fig:bubblespecdisk}
\end{figure*}

\begin{figure*}[ht]
\begin{center}
\includegraphics[width=.8\textwidth]{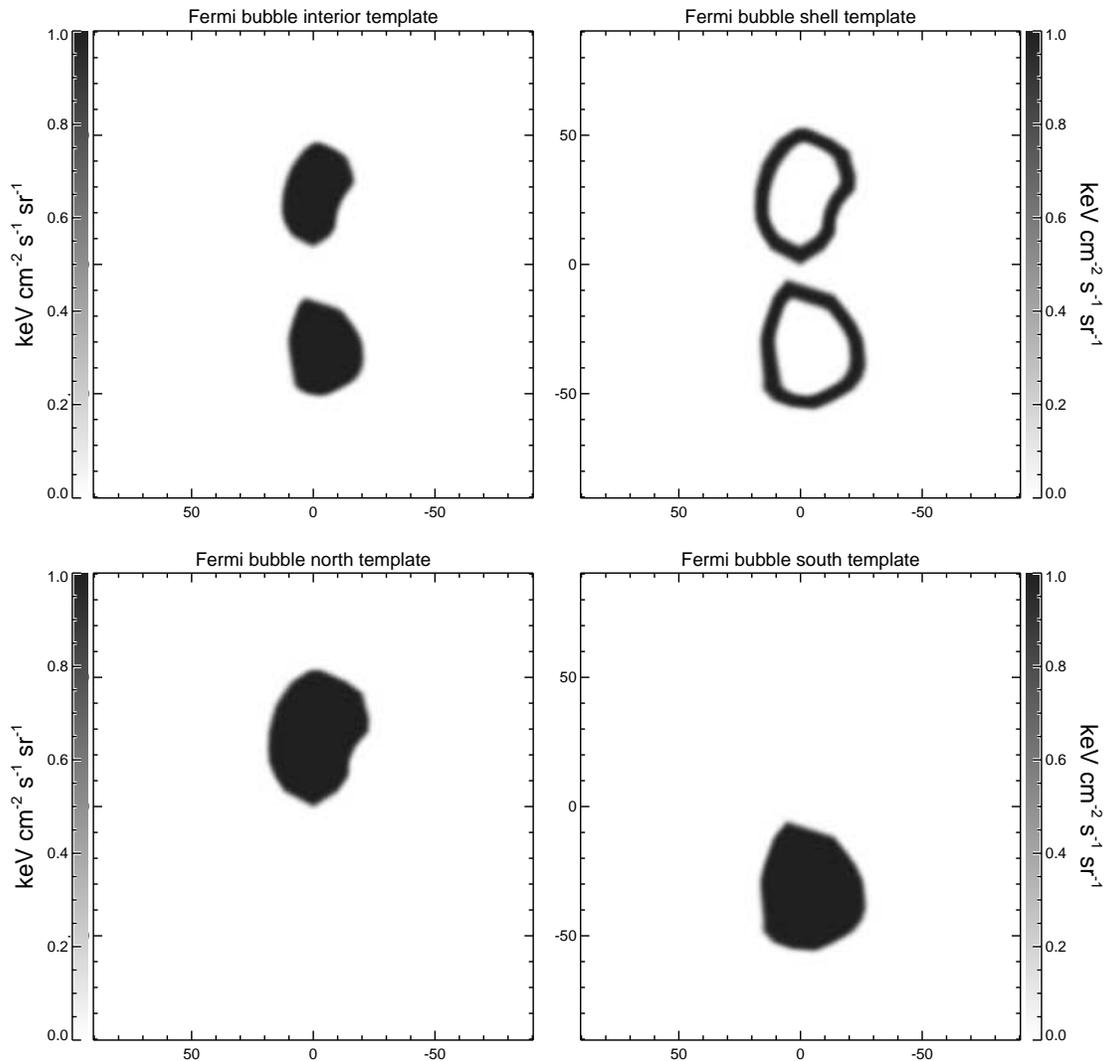}
\end{center}
\caption{\emph{Top row:}
We split our \Fermi\ bubble template (shown in the \emph{bottom right} panel of \reffig{bubbles}) into two components for template fitting: an interior template (\emph{top left}) and a shell template ({\emph {top right}}) with uniform intensity, in order to reveal any potential spectrum difference with the template fitting technique. \emph{Bottom row:} We split the \Fermi\ bubble template into north bubble template (\emph{bottom left}) and south bubble template (\emph{bottom right}). If the two bubbles have the same origin, they should not only have similar morphologies but also consistent spectra.}
\label{fig:splittemplate}
\end{figure*}

Template-correlated spectra for the 5-template fit are shown in \reffig{bubblespecdisk}. The fitting is done with regions of $|b| > 30\degree$.  For a template that has units (e.g., the SFD dust map is in $E_{B-V}$ magnitudes) the correlation spectrum has obscure units (e.g. gamma-ray emission per magnitude).  In such a case we multiply the correlation spectrum by the average SFD value in the bubble region, defined by the \emph{bottom right} panel of \reffig{bubbles}, masking out the $|b| < 30\degree$ region. For the uniform, \emph{Loop I}, and bubble templates (including inner, outer, north, and south), no renormalization is done.  These templates are simply ones and zeros (smoothed to the appropriate PSF), so the outer bubble spectrum is simply the spectrum of the bubble shell template shown in \reffig{splittemplate}, \emph{not} the mean of this template over the whole bubble region. The normalization factors for different templates are listed in \reftbl{normfac}.

In \reffig{bubblespecdisk}, we show spectra for $\pi^0$ emission, bremsstrahlung and inverse Compton scattering calculated using a sample \texttt{GALPROP} model (tuned to match locally measured protons and anti-protons as well as locally measured electrons at $\sim20-30$ GeV), as an indication of the expected spectral shapes. The spectra for the SFD and the simple disk template reasonably match the model expectations. The dust map mostly traces the $\pi^0$ emission, and the simple disk model resembles a combination of IC and bremsstrahlung emission. The spectrum for emission correlated with the \Fermi\ bubbles is clearly significantly harder than either of these components, consistent with a \emph{flat} spectrum in $E^2dN/dE$. This fact coupled with the distinct spatial morphology of the \Fermi\ bubbles indicates that the IC bubbles are generated by a \emph{separate} electron component. We also note that the spectrum of the bubble template falls off significantly at energy $\lesssim$1 GeV. This feature is robust with respect to the choice of templates. The fitting coefficients and corresponding errors of each template are listed in \reftbl{bubdisk}. We will discuss some implications of the falling spectrum in \refsec{atic}.

\begin{figure*}[ht]
\begin{center}
\includegraphics[width=.8\textwidth]{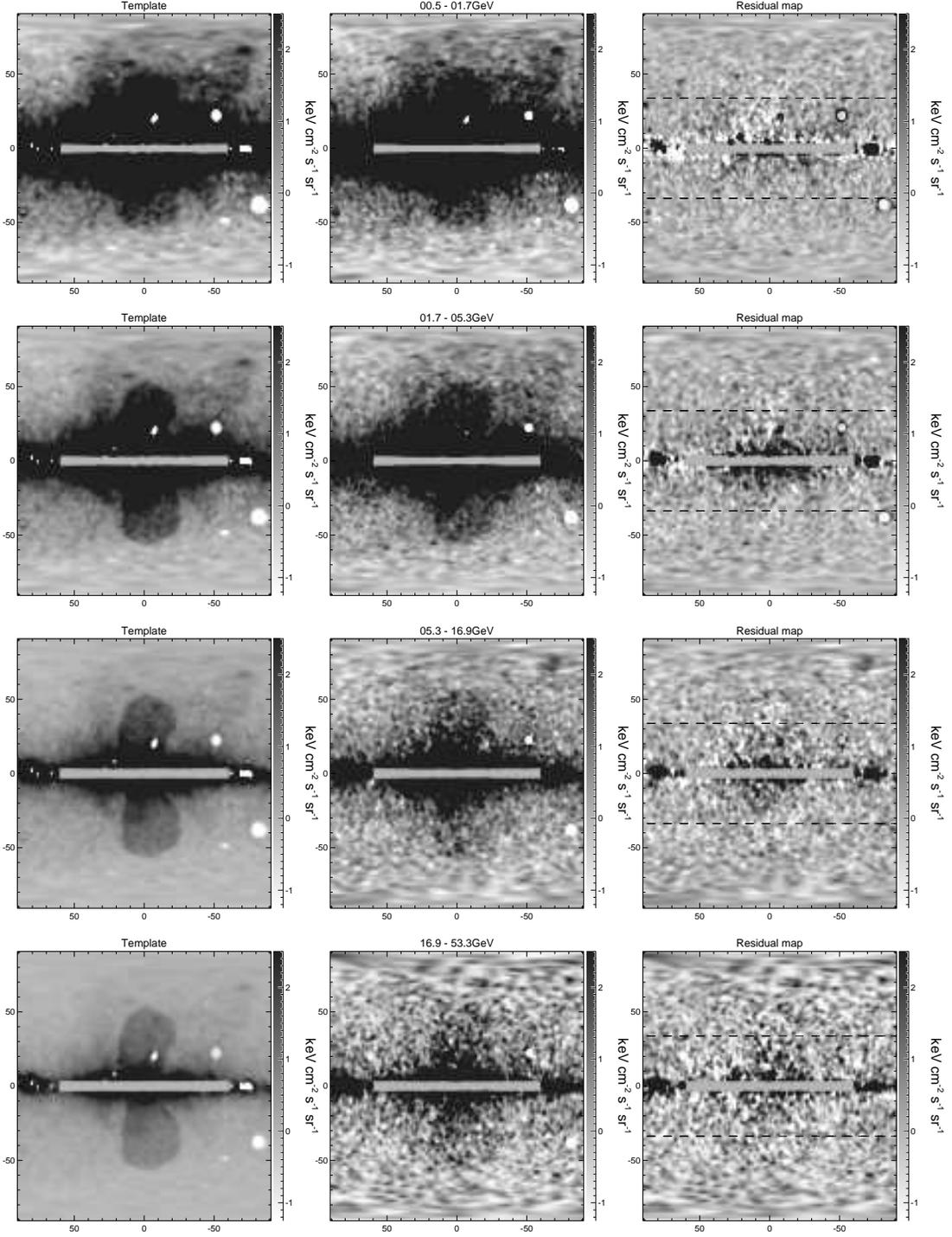}
\end{center}
\caption{
The same as \reffig{bubbledisk}, but using the \Fermi\ $0.5-1$ GeV residual map after subtracting emission correlated with the SFD dust map (to remove the $\pi^0$ gammas) as a template for low-latitude IC emission (originating primarily from scatterings on \emph{starlight}, and likely involving softer supernova-shock-accelerated electrons). The \emph{left} column shows the best fit combined template including the residual $0.5 - 1$ GeV map, uniform, SFD dust, and bubble templates. The \emph{middle} column shows the \Fermi-LAT data in different energy bins.  The difference maps between the combined template and the data are shown in the \emph{right} column. The template fitting is done over the region with $|b| > 30\degree$ (shown with black dashed line in the \emph{right} column residual maps). We use the same grey scale for all the panels. We find that the bubble template does not fade away with increasing energy. The 4-template fit works extremely well, at $|b| > 30\degree$ the residual maps are consistent with Poisson noise without obvious large scale features, and there are no obvious sharp features in the data closer to the disk. 
}
\label{fig:bubblefermilowE}
\end{figure*}

\begin{figure*}[ht]
\begin{center}
\includegraphics[width=.8\textwidth]{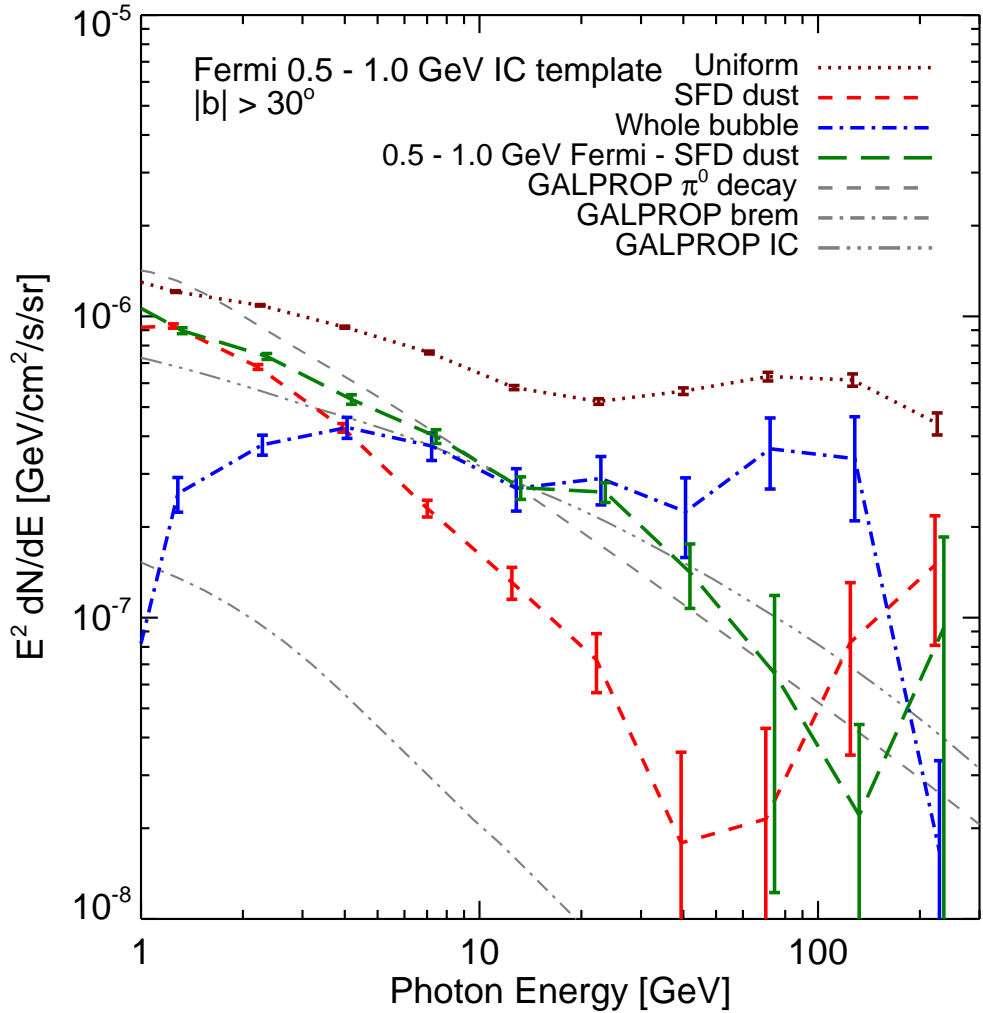}
\end{center}
\caption{ Same as \reffig{bubblespecdisk}, but correlation spectra
  for the 4-template fit employing the \Fermi\ $0.5 - 1$ GeV residual
  map (after subtracting the SFD dust) as a template for the
  \emph{starlight} IC. The line style is the same as
  \reffig{bubblespecdisk}. Again, we find that the spectrum correlated
  with the \Fermi\ bubble template (blue dot-dashed line) is harder
  (consistent with flat in $E^2 dN/dE$) than the spectra correlated with
  the other templates, and the models for the various emission mechanism
  generated from \texttt{GALPROP}, indicating that the \Fermi\ bubbles
  constitute a distinct gamma-ray component with a hard spectrum. The
  fitting is done for $|b| > 30\degree$.  As in \reffig{bubblespecdisk},
  the correlation spectra have been normalized to a reference region;
  see \refsec{templatefit} for details, and \reftbl{normfac} for a
  summary of the normalization factors.}
\label{fig:bubblespecfermilowE}
\end{figure*}

To demonstrate the robustness of the spectrum we have derived for the
\Fermi\ bubbles, we make use of the \Fermi\ $0.5 - 1$ GeV residual map
(after subtracting the SFD dust map to largely remove the $\pi^0$
gammas) as a template of IC emission, and perform a 4-template fit
(\refsec{lowEdiffuse}). These gamma rays mostly originate from IC scattering of a relatively soft population of electrons in the disk, but might also contain gammas from IC scattering on \emph{starlight} by a latitudinally extended electron population.  We use the SFD dust map as a template for $\pi^0$ gammas as previously, and include the uniform background and the bubble template as in the previous 5-template fit. The fitting is done with regions of $|b| > 30\degree$. For the SFD dust map and the \Fermi\ $0.5 - 1$ GeV IC template, the correlation coefficients are weighted by the mean of each template in the ``bubble'' region. The resulting model and the difference maps with respect to the \Fermi\ data, at different energy bands, are shown in \reffig{bubblefermilowE}. The residuals are remarkably small. The spectrum is shown in \reffig{bubblespecfermilowE}. The \Fermi\ $0.5 - 1$ GeV IC template appears to contain a small fraction of $\pi^0$ gammas, but the spectral index is consistent with the predicted \texttt{GALPROP} IC component. The fitting coefficients and corresponding errors of each template are listed in \reftbl{resdisk}.

\begin{figure*}[ht]
\begin{center}
\includegraphics[width=.48\textwidth]{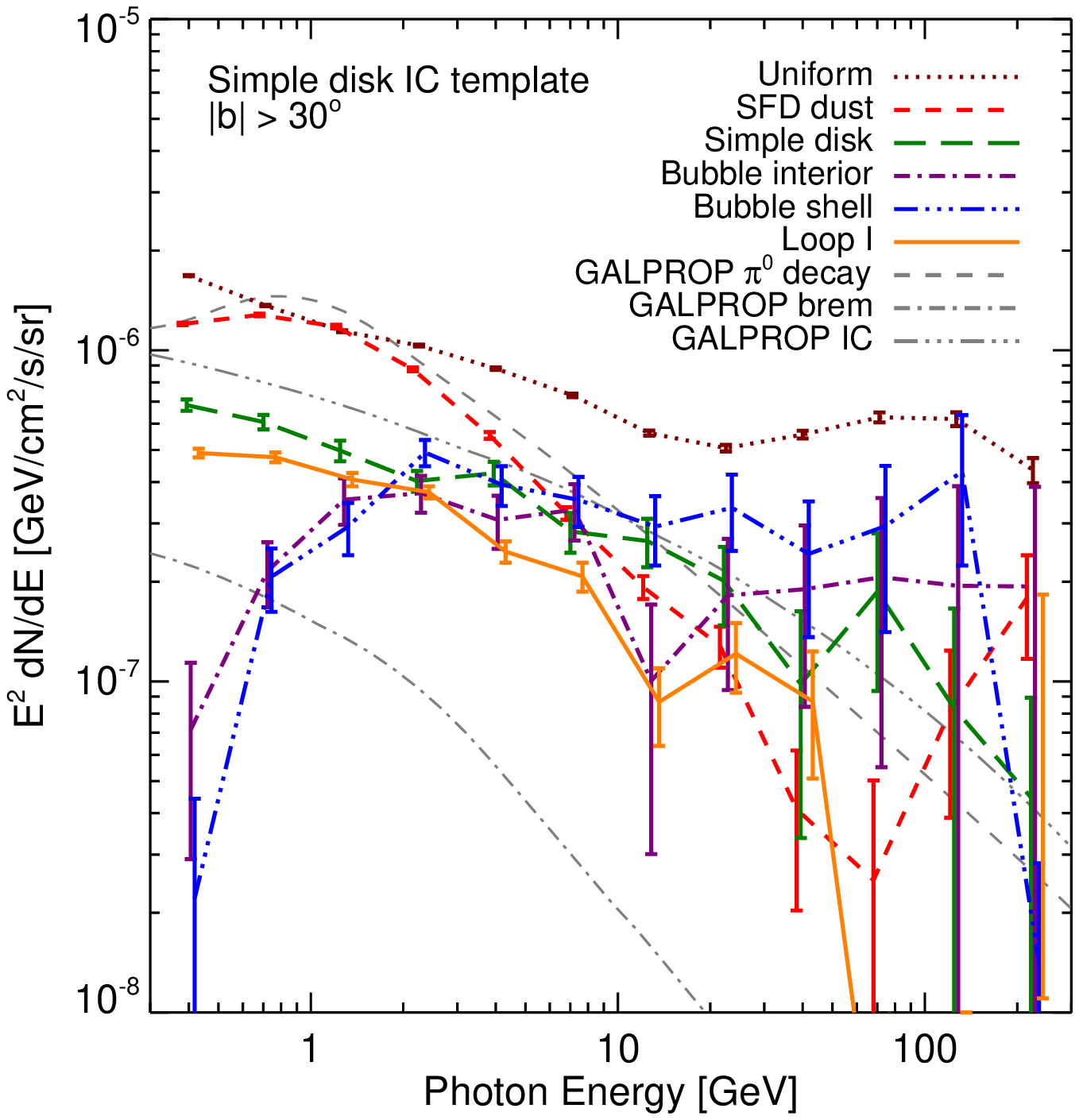}
\includegraphics[width=.48\textwidth]{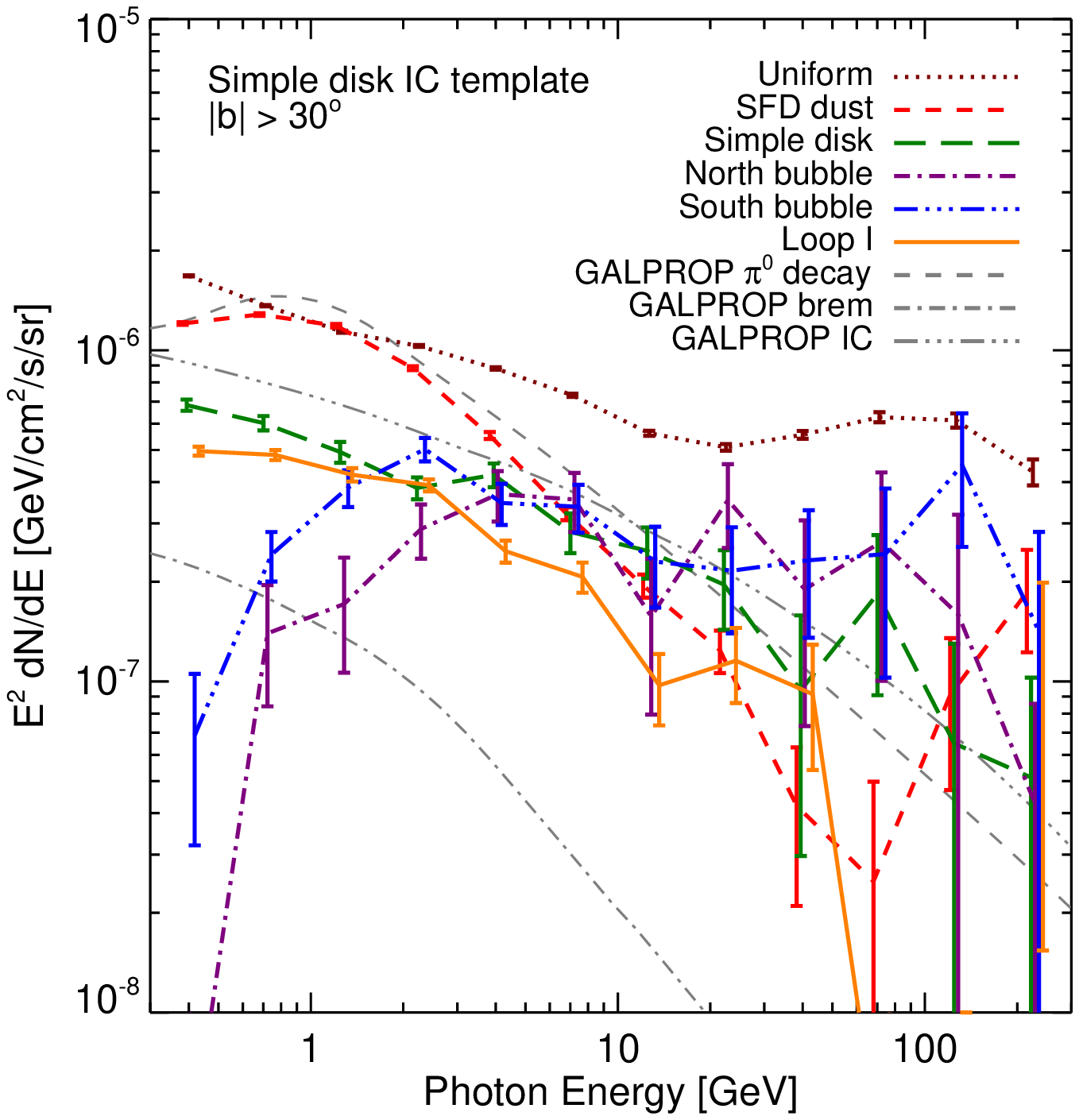}
\includegraphics[width=.48\textwidth]{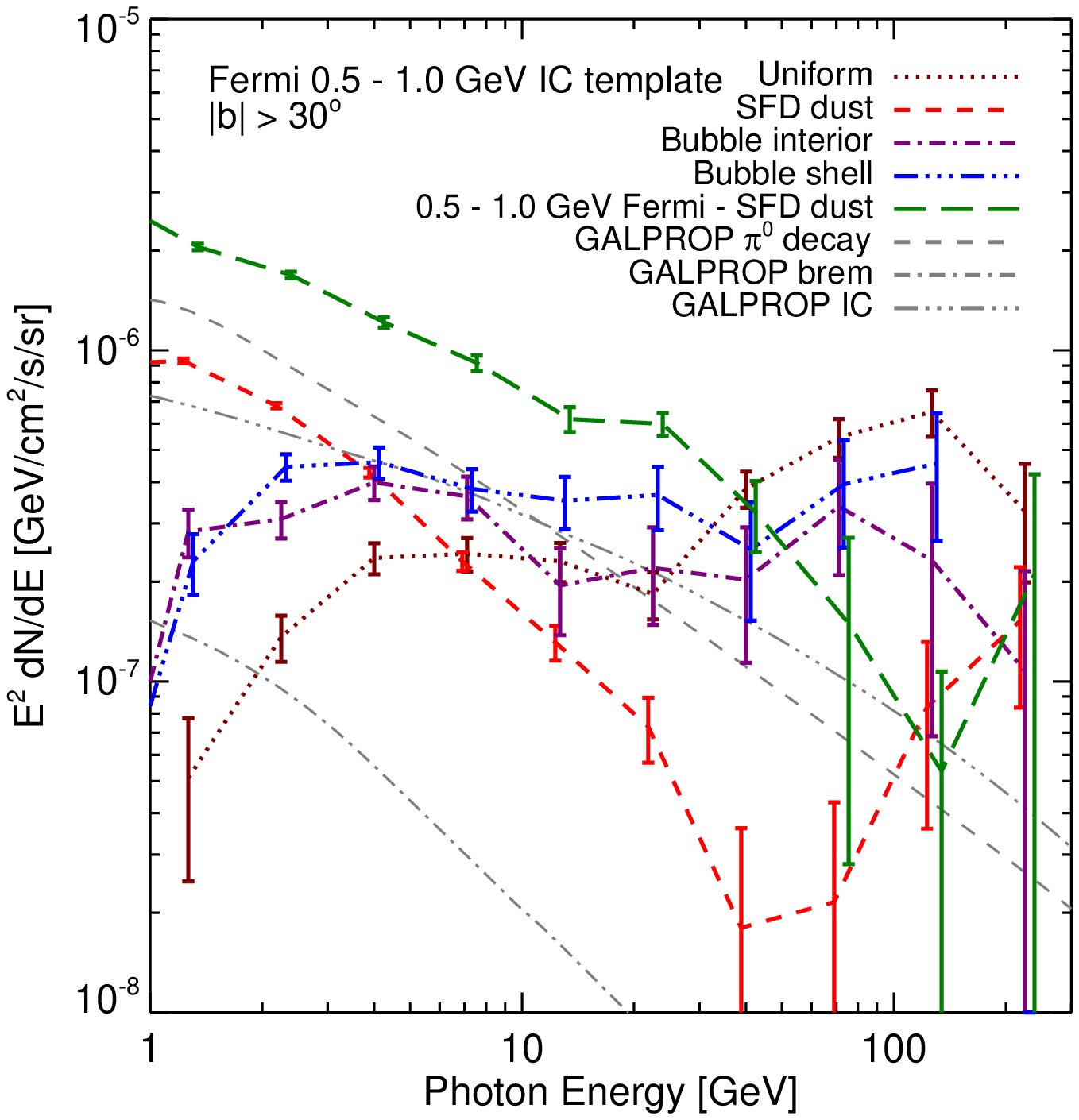}
\includegraphics[width=.48\textwidth]{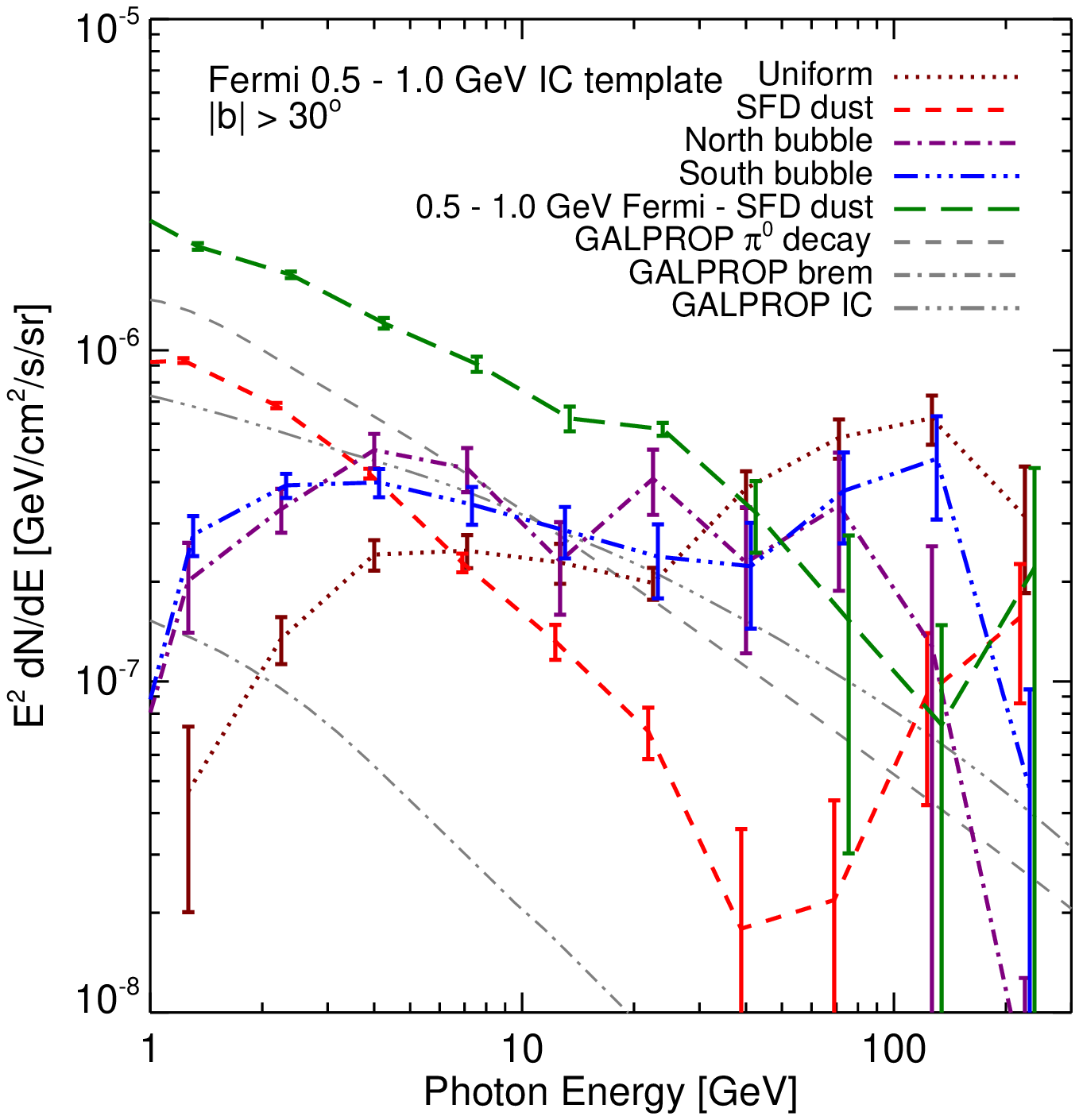}
\end{center}
\caption{ Same as \reffig{bubblespecdisk} and
\reffig{bubblespecfermilowE}, but splitting the \Fermi\ bubble template
into two components for template fitting. The line styles are the same
as \reffig{bubblespecdisk}. \emph{Top row: }Using the simple disk model
as the IC template. In the \emph{left} panel, we split the previous
bubble template into \emph{bubble interior} and \emph{bubble shell}
templates (see \reffig{splittemplate} for the templates). The
correlation coefficients of the 6-template fit involving the two bubble
templates are shown. The purple dash-dotted line and blue
triple-dot-dashed line are for the inner bubble and the outer shell
template respectively. The two templates have a consistent spectrum
which is significantly harder than the other templates, indicating the
\emph{bubble interior} and the \emph{bubble shell} have the same
distinct physical origin. In the \emph{right} panel, we split the bubble
template into \emph{north} and \emph{south} bubbles. As we include the
\emph{Loop I} template (which has a softer spectrum) in the north sky
for regression fitting, the north bubble has a slightly harder spectrum
than the south bubble. Again, both of the templates have harder spectra
than any other components in the fit. \emph{Bottom row: }Employing the
\Fermi\ $0.5 -1$ GeV residual map (after subtracting the SFD dust) as a
template for the \emph{starlight} IC. In the \emph{left} panel, we split
the bubble template into \emph{bubble interior} and \emph{bubble shell}
templates. In the \emph{right} panel, we split the bubble template into
\emph{north bubble} and \emph{south bubble} templates.  As in
\reffig{bubblespecdisk}, the correlation spectra have been normalized to
a reference region; see \refsec{templatefit} for details, and
\reftbl{normfac} for a summary of the normalization factors.  }
\label{fig:bubblespecmore}
\end{figure*}

\begin{figure*}[ht]
\begin{center}
\includegraphics[width=.48\textwidth]{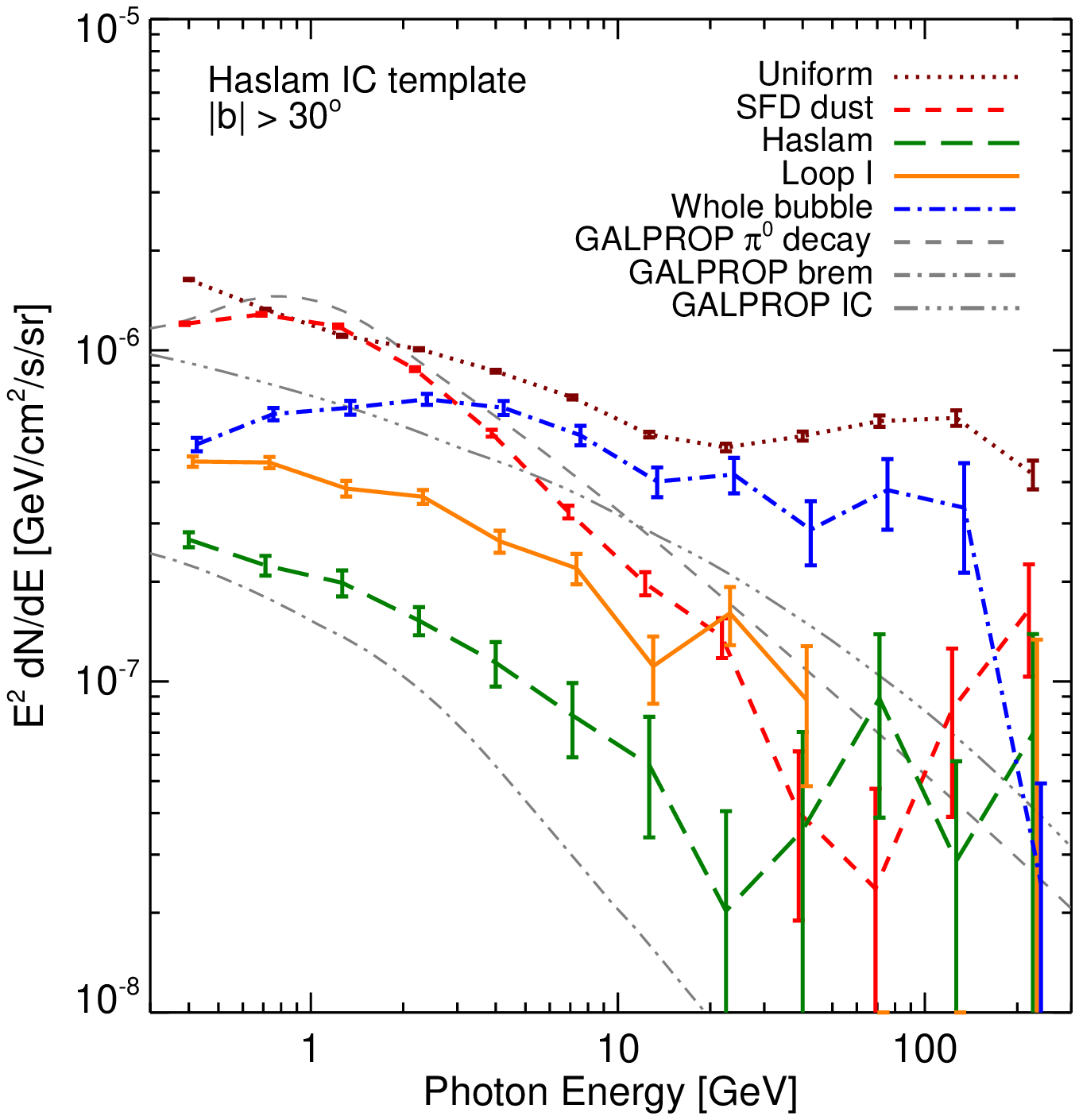}
\includegraphics[width=.48\textwidth]{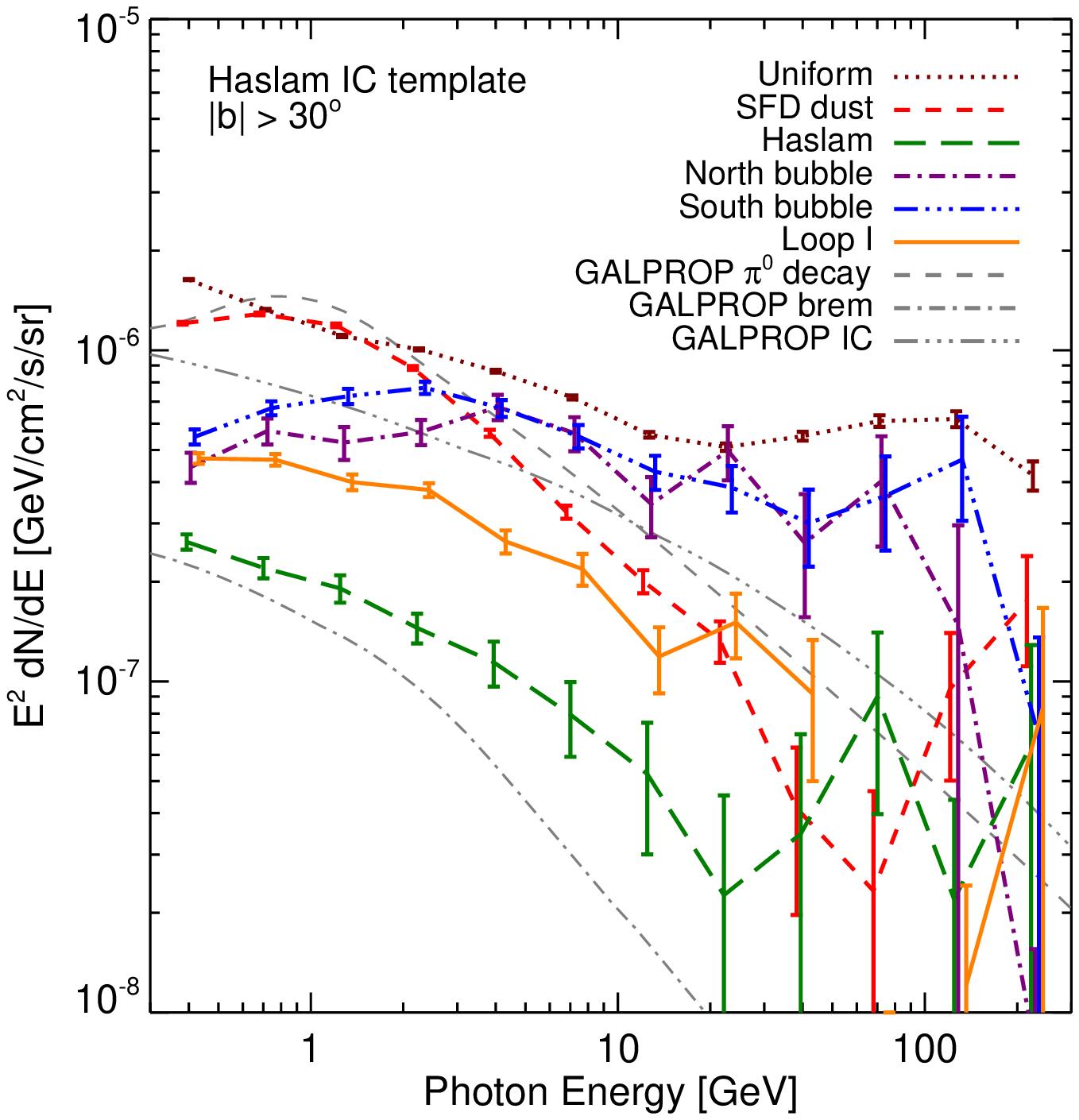}
\end{center}
\caption{ Same as \reffig{bubblespecdisk}, but using the Haslam 408 MHz
map instead of the simple disk model as the IC template. The Haslam map
contains a bright feature associated with \emph{Loop I} (see
\reffig{bubble_compare}) and is dominated by synchrotron emission from
softer electron CRs, of energies around 1 GeV; it is not an ideal tracer
of IC emission which depends on both electron and ISRF distribution. The
resulting best-fit spectrum for the bubble template remains harder than
the other components, but with enhanced lower energy ($\lesssim$ 2 GeV)
correlation coefficients compared to \reffig{bubblespecdisk} and
\reffig{bubblespecfermilowE}. The \emph{right} panel is the same as the
\emph{left} panel, but with the bubble template divided into north and
south bubbles (see \refsec{templatefit} for more discussion).  As in
\reffig{bubblespecdisk}, the correlation spectra have been normalized to
a reference region; see \refsec{templatefit} for details, and
\reftbl{normfac} for a summary of the normalization factors.}
\label{fig:bubblespechaslam}
\end{figure*}

By eye, the \Fermi\ bubbles appear to possess north-south symmetry and are close to spatially uniform in intensity (with a hard cutoff at the bubble edges). To test these hypotheses more quantitatively, we split the \Fermi\ bubble template into the \emph{inner bubble} and \emph{outer shell} templates (\emph{upper row} of \reffig{splittemplate}), or alternatively into the \emph{north} and \emph{south bubble} templates (\emph{lower row} of \reffig{splittemplate}). We then repeat the previous 5-template and 4-template fitting procedure involving either the simple disk IC template or the \Fermi\ $0.5 - 1$ GeV IC template, but splitting the bubble template to either \emph{inner bubble and outer shell} or \emph{north and south bubble} templates. The goal is to identify variations in the intensity and spectral index between the bubble edge and interior, and the northern and southern bubbles. The resulting fitted spectra are shown in \reffig{bubblespecmore}. And the corresponding fitting coefficients and errors of each template are listed in \reftbl{shelldisk}, \reftbl{splitdisk}, \reftbl{shellres}, and \reftbl{splitres} respectively. In \reffig{bubblespechaslam}, we replace the simple disk model with the Haslam 408 MHz map as the IC template, and employ the SFD map, a uniform background, the \emph{Loop I} template, and the double-lobed bubble (\emph{left} panel) or the \emph{north and south bubble} (\emph{right} panel) templates in the fitting. Our conclusion is that the \Fermi\ bubbles appear to be north-south symmetric and spatially and spectrally uniform, with a hard spectrum. This statement is largely independent of our choice of template for the disk IC emission.

\begin{table*}
\begin{center}
\begin{tabular}{@{}rrrrrrrr}
\hline
\hline
North bubble & South bubble  & North arc (outer) & North arc (inner) & Loop I & Donut (outer) & Donut (inner)\\
$(\ell,b)$ [deg] &$(\ell,b)$ [deg] &$(\ell,b)$ [deg] &$(\ell,b)$ [deg] &$(\ell,b)$ [deg]  &$(\ell,b)$ [deg]  &$(\ell,b)$ [deg] \\
\hline
(   0.0,   0.0) & (   5.5,  -5.0) & (  28.5,   5.0) & (  19.5,   5.0) & (  37.5,  25.0) & (  31.9,  -5.0) & (  16.8,  -7.6) \\
(  -9.9,   5.0) & (  10.7, -10.0) & (  31.1,  10.0) & (  23.1,  10.0) & (  43.2,  30.0) & (  34.9,  -9.0) & (  22.0,  -9.4) \\
( -14.2,  10.0) & (  12.9, -15.0) & (  31.9,  15.0) & (  24.9,  15.0) & (  46.3,  35.0) & (  37.0, -14.0) & (  24.6, -14.0) \\
( -14.5,  15.0) & (  15.0, -20.0) & (  34.0,  20.0) & (  26.0,  20.0) & (  47.7,  40.0) & (  36.3, -19.0) & (  23.7, -16.5) \\
( -17.0,  20.0) & (  16.3, -25.0) & (  34.9,  25.0) & (  26.9,  25.0) & (  48.9,  45.0) & (  33.5, -25.0) & (  22.4, -18.7) \\
( -22.3,  25.0) & (  16.0, -30.0) & (  33.7,  30.0) & (  23.7,  30.0) & (  49.8,  50.0) & (  28.7, -29.0) & (  19.2, -21.2) \\
( -22.6,  30.0) & (  15.5, -35.0) & (  32.5,  35.0) & (  23.5,  35.0) & (  46.6,  60.0) & (  11.5, -33.0) & (  13.6, -19.3) \\
( -21.1,  35.0) & (  15.0, -40.0) & (  30.5,  40.0) & (  20.5,  40.0) & (  40.6,  65.0) & (   4.8, -29.0) & (  13.2, -14.3) \\
( -19.9,  40.0) & (  15.0, -45.0) & (  27.5,  45.0) & (  16.5,  45.0) & (  29.4,  70.0) & (  -0.4, -20.0) & (  13.6,  -8.8) \\
( -13.6,  45.0) & (  10.6, -50.0) & (  24.7,  50.0) & (  11.7,  50.0) & (  12.5,  75.0) & (  -1.2, -15.0) & (  16.8,  -7.6) \\
(  -3.0,  50.0) & (   3.7, -52.5) & (  20.0,  52.5) & (   6.0,  52.5) & (  -5.0,  78.0) & (   0.3, -10.0) &                 \\
(   1.5,  50.0) & (  -6.3, -53.5) & (  14.3,  55.0) & (   0.0,  55.0) & ( -19.0,  78.0) & (   5.5,  -5.0) &                 \\
(   8.7,  45.0) & ( -13.8, -50.0) &                 &                 & ( -33.0,  77.0) &                 &                 \\
(  12.3,  40.0) & ( -21.8, -45.0) &                 &                 & ( -49.1,  74.0) &                 &                 \\
(  15.4,  35.0) & ( -25.3, -40.0) &                 &                 & ( -61.5,  72.0) &                 &                 \\
(  17.0,  30.0) & ( -26.7, -35.0) &                 &                 & ( -69.2,  70.0) &                 &                 \\
(  18.3,  25.0) & ( -26.3, -30.0) &                 &                 & ( -75.2,  65.0) &                 &                 \\
(  18.5,  20.0) & ( -25.6, -25.0) &                 &                 & ( -77.6,  60.0) &                 &                 \\
(  18.4,  15.0) & ( -23.0, -20.0) &                 &                 & ( -78.3,  55.0) &                 &                 \\
(  16.0,  10.0) & ( -18.8, -15.0) &                 &                 & ( -77.6,  50.0) &                 &                 \\
(  12.0,   5.0) & ( -13.8, -10.0) &                 &                 & ( -75.5,  45.0) &                 &                 \\
                &                 &                 &                 & ( -73.5,  40.0) &                 &                 \\
                &                 &                 &                 & ( -68.3,  35.0) &                 &                 \\
                &                 &                 &                 & ( -67.0,  25.0) &                 &                 \\
\hline
\end{tabular}
\end{center}
\caption{Coordinates defining the features shown in \reffig{fullsky} and \reffig{energysplit}}
\label{tbl:bubblecoords}
\end{table*}


\begin{table*}
\begin{center}
\begin{tabular}{@{}rrrrrrrrrrr}
\hline
\hline
Uniform & SFD  & bubble & north bubble & south bubble & inner bubble & outer bubble & disk & loop I & $0.5-1.0$ GeV - SFD\\
\hline
1.0 &   0.084 &  1.0 &  1.0 & 1.0 & 1.0 & 1.0 &  0.292 &  1.0 &  1.198 \\
\hline
\end{tabular}
\end{center}
\caption{Normalization factors for different templates.}
\label{tbl:normfac}
\end{table*}


\begin{figure*}[ht]
\begin{center}
\includegraphics[width=.8\textwidth]{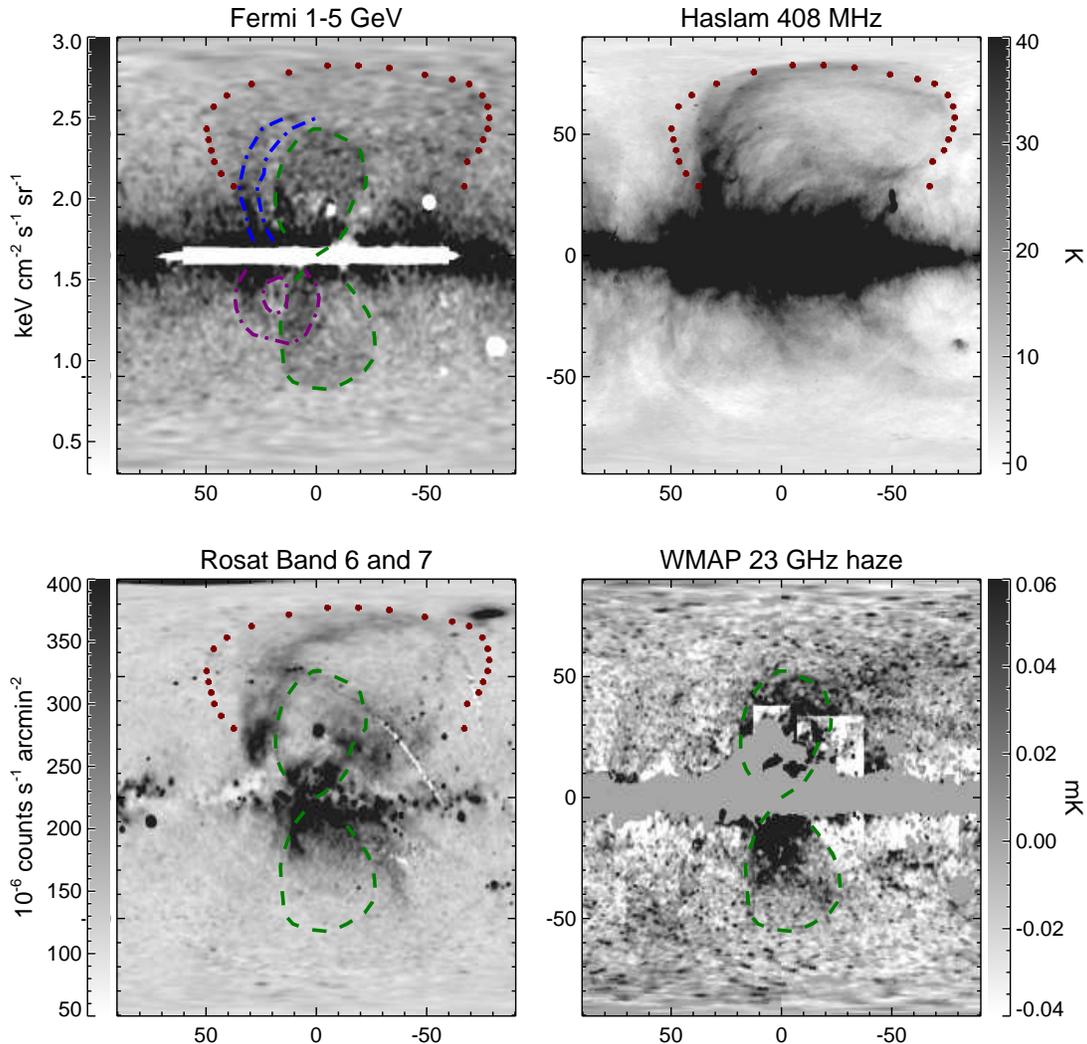}
\end{center}
\caption{Comparison of the \Fermi\ bubbles with features in other maps. \emph{Top left: }Point-source subtracted $1 - 5$ GeV \Fermi-LAT 1.6 yr map, same as the \emph{lower left} panel of \reffig{bubbles} with north and south bubble edges marked with green dashed line, and \emph{north arc} in blue dashed line. The approximate edge of the \emph{Loop I} feature is plotted in red dotted line, and the ``donut'' in purple dot-dashed line. \emph{Top right:} The Haslam 408 MHz map overplotted with the same red dotted line as the \emph{top left} panel. The red dotted line remarkably traces the edge of the bright \emph{Loop I} feature in the Haslam soft synchrotron map. \emph{Bottom left:} the \emph{ROSAT} 1.5 keV X-ray map is shown together with the same color lines marking the prominent \Fermi\ bubble features. \emph{Bottom right:} \WMAP\ haze at K-band 23 GHz overplotted with \Fermi\ bubble edges. The \rosat\ X-ray features and the \WMAP\ haze trace the \Fermi\ bubbles well, suggesting a common origin for these features.
}
\label{fig:bubble_compare}
\end{figure*}


\begin{figure*}[ht]
\begin{center}
\includegraphics[width=.8\textwidth]{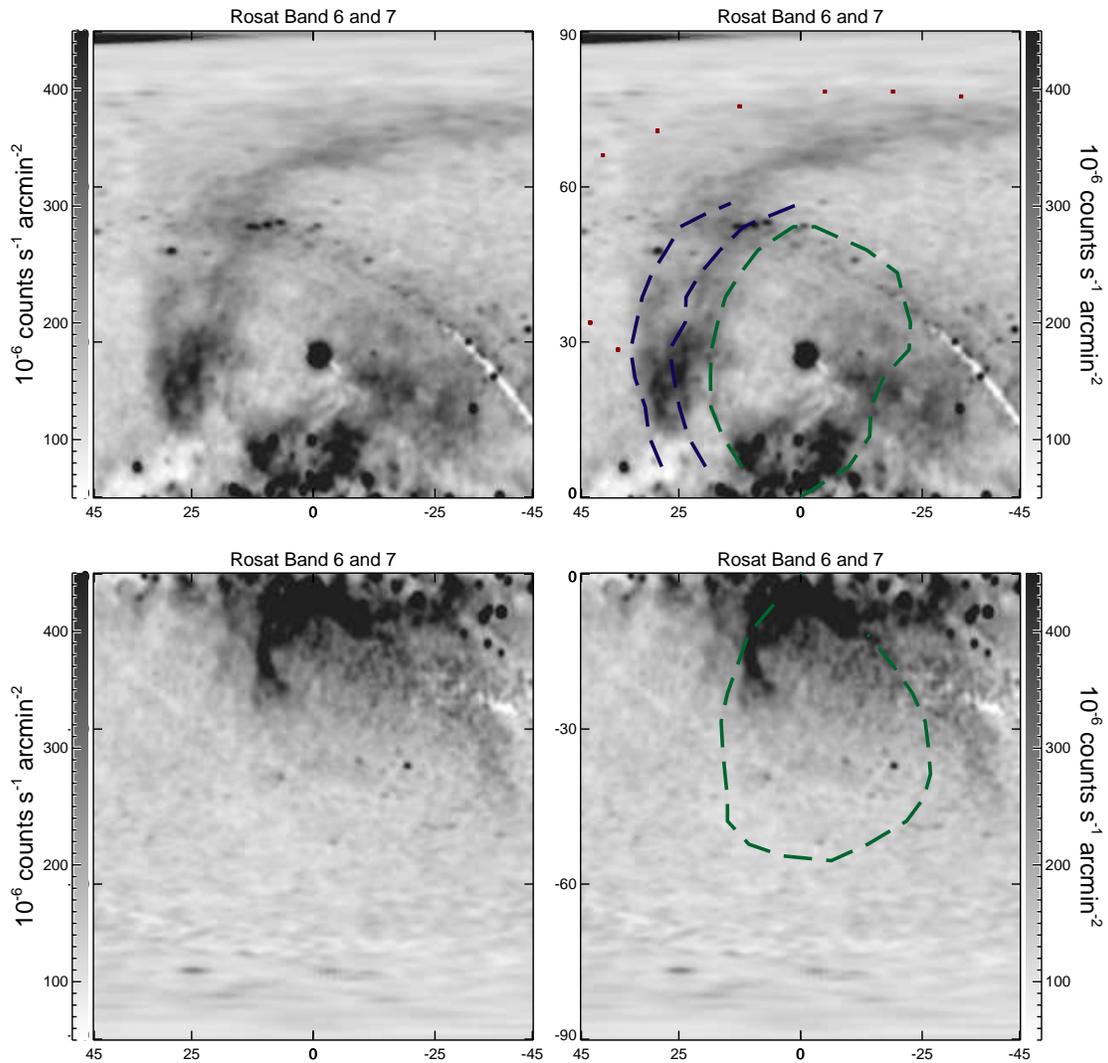}
\end{center}
\caption{
\emph{Top left:}
The \rosat\ X-ray haze features compared with the \Fermi\ bubbles' morphology. \emph{Top row:} The X-ray features of \rosat\ band 6 and 7 in the north sky towards to the GC (\emph{left} panel) compared with \Fermi\ north bubble overplotted with green dashed line, \emph{northern arc} feature in blue dashed line, and \emph{Loop I} feature in red dotted line (\emph{right} panel). \emph{Bottom row:} Same as \emph{top row} but for the south bubble features. }
\label{fig:rosat}
\end{figure*}


\begin{figure*}[ht]
\begin{center}
\includegraphics[width=.8\textwidth]{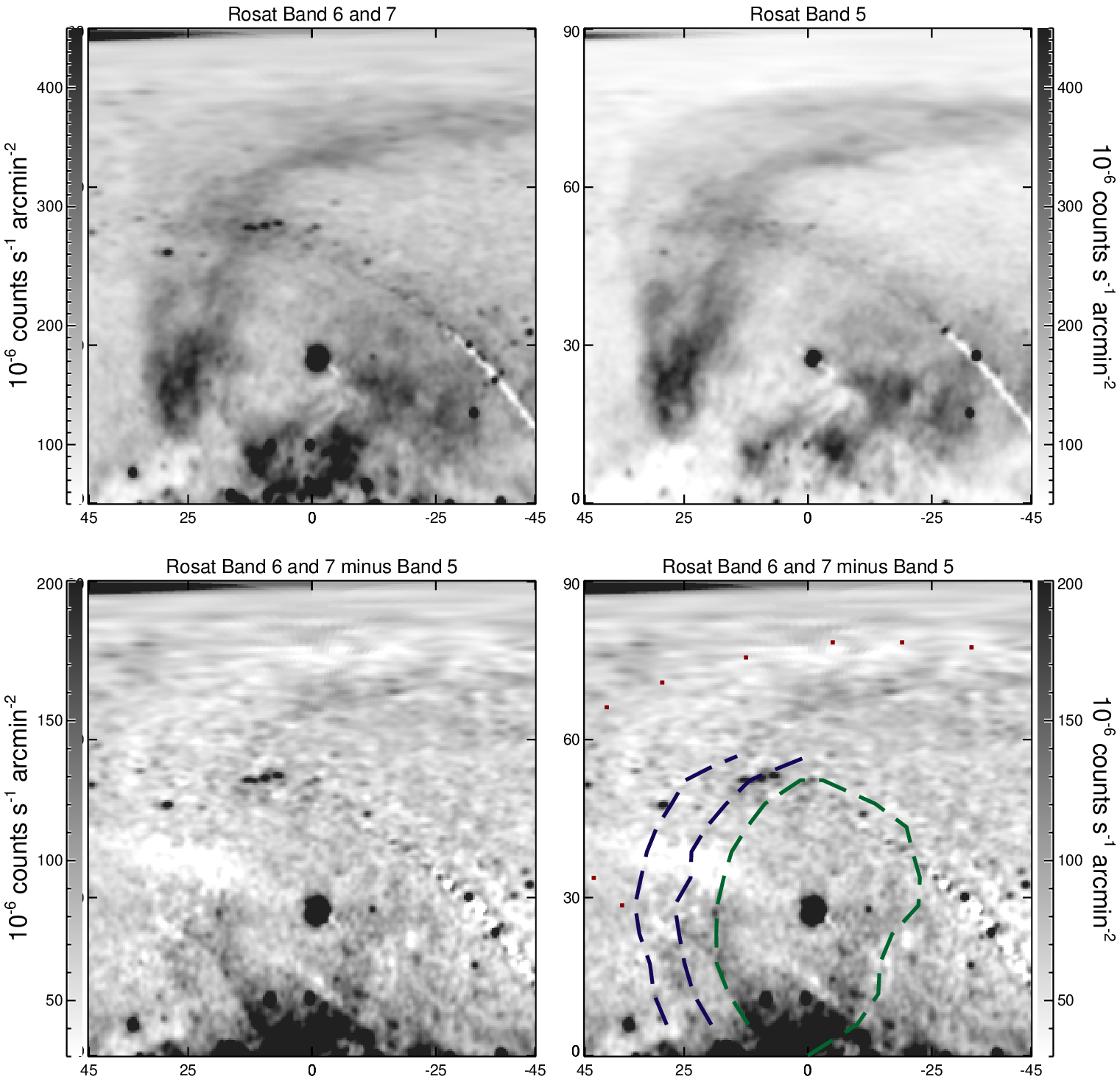}
\end{center}
\caption{
\emph{Top left:} The \rosat\ X-ray features in Band 6 and 7. \emph{Top right:} The same region as the \emph{left} panel, but for the \rosat\ X-ray map in band 5. \emph{Bottom left:} The residual X-ray features after subtracting the \emph{top right} softer Band 5 map from the \emph{top left} harder Band 6 and 7 map. \emph{Bottom right:} The same as the \emph{bottom left} panel, overplotted with the \Fermi\ bubble features, the \emph{northern arc}, and \emph{Loop I} features. The residual X-ray features with harder spectrum than the diffuse \emph{Loop I} feature align well with the \Fermi\ bubble structures.} 
\label{fig:rosatclean}
\end{figure*}

\subsection{The ``\emph{Fermi} Bubbles'' vs the ``\emph{Fermi} Haze''}

\cite{fermihaze} employed essentially the same template regression methods, claimed that the gamma-ray emission not accounted for by the known foregrounds could be well fitted by a bivariate Gaussian with $\sigma_b = 25^\circ$, $\sigma_l = 15^\circ$ . With the improved 1.6 yr data, the edges of the excess at high latitudes are seen to be quite sharp, and are not well-described by a Gaussian fall-off in intensity. However, the question of whether the excess is better modeled as an ``egg'' or a pair of ``bubbles'' is more subtle.

The choice of the bivariate Gaussian template by \cite{fermihaze} was intended to remove as much of the remaining gamma ray signal as possible, once the $\pi^0$ and soft IC emission had been regressed out, minimizing large-scale residuals. The fits were performed with only $|b| < 5$ masked out. In this work, on the other hand, we have masked out all emission with $|b| < 30\degree$; when attempting to subtract the disk-correlated emission, and delineating the ``bubbles'' template, our goal has been to isolate the sharp-edged features from the more slowly spatially varying emission, not to account for all the observed emission.

This difference in approach can be seen in the non-negligible residuals around the inner Galaxy ($|b| \lesssim 20\degree$) in many of our maps (for example, \reffig{bubblefermilowE}). An attempt to fit all the residual emission simultaneously with a simple template may well require a template closer to that used by \cite{fermihaze}, rather than the bubble template. However, the sharp edges now visible in the data, and their alignment with the edges of the \WMAP\ haze and \rosat\ X-ray features (as we will discuss in \refsec{othermaps}), motivate us to consider the bubbles as originating from a distinct physical mechanism. While the bubbles probably do not constitute the entire ``\emph{Fermi} haze'' discussed by \cite{fermihaze}, they are certainly a major component of it, dominating the signal at high latitudes. The question of whether the remaining residual gamma rays represent a separate physical mechanism is an interesting one that we defer to future work.

\begin{table*}
\begin{center}
\begin{tabular}{@{}rrrrrrr}
\hline
\hline
E range (GeV) & Energy & Uniform & SFD dust & simple disk & whole bubble & simple loop I\\
\hline
$   0.3 -   0.5$
 &    0.4 &  1.681 $\pm$ 0.006 &  1.201 $\pm$ 0.011 &  0.689 $\pm$ 0.027 &  0.035 $\pm$ 0.033 &  0.487 $\pm$ 0.015 \\
$   0.5 -   0.9$
 &    0.7 &  1.365 $\pm$ 0.007 &  1.279 $\pm$ 0.012 &  0.608 $\pm$ 0.030 &  0.211 $\pm$ 0.037 &  0.475 $\pm$ 0.016 \\
$   0.9 -   1.7$
 &    1.3 &  1.141 $\pm$ 0.008 &  1.179 $\pm$ 0.014 &  0.503 $\pm$ 0.035 &  0.321 $\pm$ 0.044 &  0.405 $\pm$ 0.019 \\
$   1.7 -   3.0$
 &    2.2 &  1.034 $\pm$ 0.006 &  0.876 $\pm$ 0.011 &  0.393 $\pm$ 0.029 &  0.436 $\pm$ 0.036 &  0.376 $\pm$ 0.016 \\
$   3.0 -   5.3$
 &    4.0 &  0.881 $\pm$ 0.008 &  0.554 $\pm$ 0.013 &  0.420 $\pm$ 0.034 &  0.353 $\pm$ 0.043 &  0.249 $\pm$ 0.018 \\
$   5.3 -   9.5$
 &    7.1 &  0.731 $\pm$ 0.009 &  0.322 $\pm$ 0.014 &  0.282 $\pm$ 0.039 &  0.343 $\pm$ 0.049 &  0.208 $\pm$ 0.021 \\
$   9.5 -  16.9$
 &   12.7 &  0.563 $\pm$ 0.010 &  0.193 $\pm$ 0.015 &  0.251 $\pm$ 0.044 &  0.205 $\pm$ 0.055 &  0.092 $\pm$ 0.023 \\
$  16.9 -  30.0$
 &   22.5 &  0.507 $\pm$ 0.012 &  0.128 $\pm$ 0.018 &  0.191 $\pm$ 0.053 &  0.263 $\pm$ 0.068 &  0.125 $\pm$ 0.029 \\
$  30.0 -  53.3$
 &   40.0 &  0.557 $\pm$ 0.015 &  0.041 $\pm$ 0.021 &  0.096 $\pm$ 0.064 &  0.217 $\pm$ 0.083 &  0.088 $\pm$ 0.036 \\
$  53.3 -  94.9$
 &   71.1 &  0.628 $\pm$ 0.022 &  0.020 $\pm$ 0.030 &  0.183 $\pm$ 0.093 &  0.251 $\pm$ 0.120 & -0.043 $\pm$ 0.048 \\
$  94.9 - 168.7$
 &  126.5 &  0.622 $\pm$ 0.030 &  0.080 $\pm$ 0.043 &  0.012 $\pm$ 0.125 &  0.319 $\pm$ 0.162 & -0.091 $\pm$ 0.063 \\
$ 168.7 - 300.0$
 &  225.0 &  0.436 $\pm$ 0.038 &  0.174 $\pm$ 0.061 & -0.039 $\pm$ 0.158 & -0.015 $\pm$ 0.194 &  0.083 $\pm$ 0.086 \\
\hline
\end{tabular}
\end{center}
\caption{Corresponding template fitting coefficients and errors in \reffig{bubblespecdisk}}.
\label{tbl:bubdisk}
\end{table*}

\begin{table*}
\begin{center}
\begin{tabular}{@{}rrrrrrrrr}
\hline
\hline
E range (GeV) & Energy & Uniform & SFD dust & whole bubble & $0.5-1.0$ GeV - SFD\\
\hline
$   0.3 -   0.5$
 &    0.4 &  1.759 $\pm$ 0.006 &  0.883 $\pm$ 0.012 & -0.026 $\pm$ 0.026 &  1.181 $\pm$ 0.017 \\
$   0.5 -   0.9$
 &    0.7 &  1.446 $\pm$ 0.006 &  0.905 $\pm$ 0.013 &  0.008 $\pm$ 0.029 &  1.275 $\pm$ 0.018 \\
$   0.9 -   1.7$
 &    1.3 &  1.208 $\pm$ 0.008 &  0.929 $\pm$ 0.016 &  0.258 $\pm$ 0.034 &  0.897 $\pm$ 0.020 \\
$   1.7 -   3.0$
 &    2.2 &  1.088 $\pm$ 0.006 &  0.679 $\pm$ 0.012 &  0.375 $\pm$ 0.029 &  0.736 $\pm$ 0.017 \\
$   3.0 -   5.3$
 &    4.0 &  0.921 $\pm$ 0.007 &  0.427 $\pm$ 0.014 &  0.428 $\pm$ 0.034 &  0.530 $\pm$ 0.019 \\
$   5.3 -   9.5$
 &    7.1 &  0.759 $\pm$ 0.008 &  0.231 $\pm$ 0.015 &  0.371 $\pm$ 0.039 &  0.400 $\pm$ 0.021 \\
$   9.5 -  16.9$
 &   12.7 &  0.580 $\pm$ 0.009 &  0.131 $\pm$ 0.016 &  0.269 $\pm$ 0.043 &  0.270 $\pm$ 0.023 \\
$  16.9 -  30.0$
 &   22.5 &  0.522 $\pm$ 0.011 &  0.072 $\pm$ 0.016 &  0.290 $\pm$ 0.053 &  0.261 $\pm$ 0.020 \\
$  30.0 -  53.3$
 &   40.0 &  0.565 $\pm$ 0.014 &  0.015 $\pm$ 0.021 &  0.225 $\pm$ 0.066 &  0.141 $\pm$ 0.034 \\
$  53.3 -  94.9$
 &   71.1 &  0.631 $\pm$ 0.021 &  0.010 $\pm$ 0.033 &  0.364 $\pm$ 0.096 &  0.065 $\pm$ 0.053 \\
$  94.9 - 168.7$
 &  126.5 &  0.615 $\pm$ 0.029 &  0.083 $\pm$ 0.048 &  0.337 $\pm$ 0.127 & -0.032 $\pm$ 0.076 \\
$ 168.7 - 300.0$
 &  225.0 &  0.441 $\pm$ 0.037 &  0.149 $\pm$ 0.068 & -0.104 $\pm$ 0.137 &  0.089 $\pm$ 0.097 \\
\hline
\end{tabular}
\end{center}
\caption{Corresponding template fitting coefficients and errors in \reffig{bubblespecfermilowE}}.
\label{tbl:resdisk}
\end{table*}

\section{The \emph{Fermi} Bubbles Seen in Other Maps}
\label{sec:othermaps}

In this section, we compare the $1 - 5$ GeV \Fermi\ bubble with the \rosat\ 1.5 keV soft X-ray map, the \WMAP\ 23 GHz microwave haze, and the Haslam 408 MHz map. The striking similarities of several morphological features in these maps strongly suggest a common physical origin for the \Fermi\ bubbles, \WMAP\ haze, and X-ray edges towards the GC (\reffig{bubble_compare}).

\subsection{Comparison with \emph{ROSAT} X-ray Features}
\label{sec:rosat}

The \rosat\ all-sky survey provides full-sky images with FWHM 12' at energies from $0.5-2$ keV.\footnote{http://hea-www.harvard.edu/rosat/rsdc.html} We compare the morphology of the X-ray features in \rosat\ 1.5 keV map with the edges of the \Fermi\ bubbles in detail in \reffig{rosat}. The limb brightened X-ray features align with the edges of both the north and south \Fermi\ bubble. Hints of the whole north bubble are also visible in \rosat, as well as two sharp edges in the south that trace the south \Fermi\ bubble close to the disk. We show the \rosat\ 1.5 keV map overplotted with the edges of the \Fermi\ bubbles, the \emph{northern arc}, the \emph{``donut''} and the \emph{Loop I} features in the \emph{right} panels of \reffig{rosat}. The appearance of the X-ray edges in the \rosat\ 1.5 keV map, coincident with the \Fermi\ bubble edges, strongly supports the physical reality of these sharp edges.

In \reffig{rosatclean}, we subtract the \rosat\ 1.0 keV soft X-ray map from the 1.5 keV map to clean up the foreground. We find that the extended \emph{Loop I} feature has a softer spectrum than the X-ray features associated with the bubble edges, and is largely removed in the difference map (\emph{lower left} panel of \reffig{rosatclean}). The residual features strikingly overlap with the edges of the \Fermi\ bubbles (\emph{lower right} panel).  No other noticeable large scale features appear in the residual X-ray map which do not appear in the gamma-rays.

\subsection{Comparison with WMAP Microwave Haze}
\label{sec:wmaphaze}

The \WMAP\ haze is the residual remaining in \WMAP\ microwave data after regressing out contributions from thermal dust, free-free, and ``soft synchrotron'' traced by the Haslam 408 MHz radio survey \citep{1982A&AS...47....1H}. Therefore, it is by construction harder than the Haslam-correlated emission. We will show in this section that the \WMAP\ synchrotron haze appears to be associated with the \Fermi\ bubbles. 

\begin{figure*}[ht]
\begin{center}
\includegraphics[width=.8\textwidth]{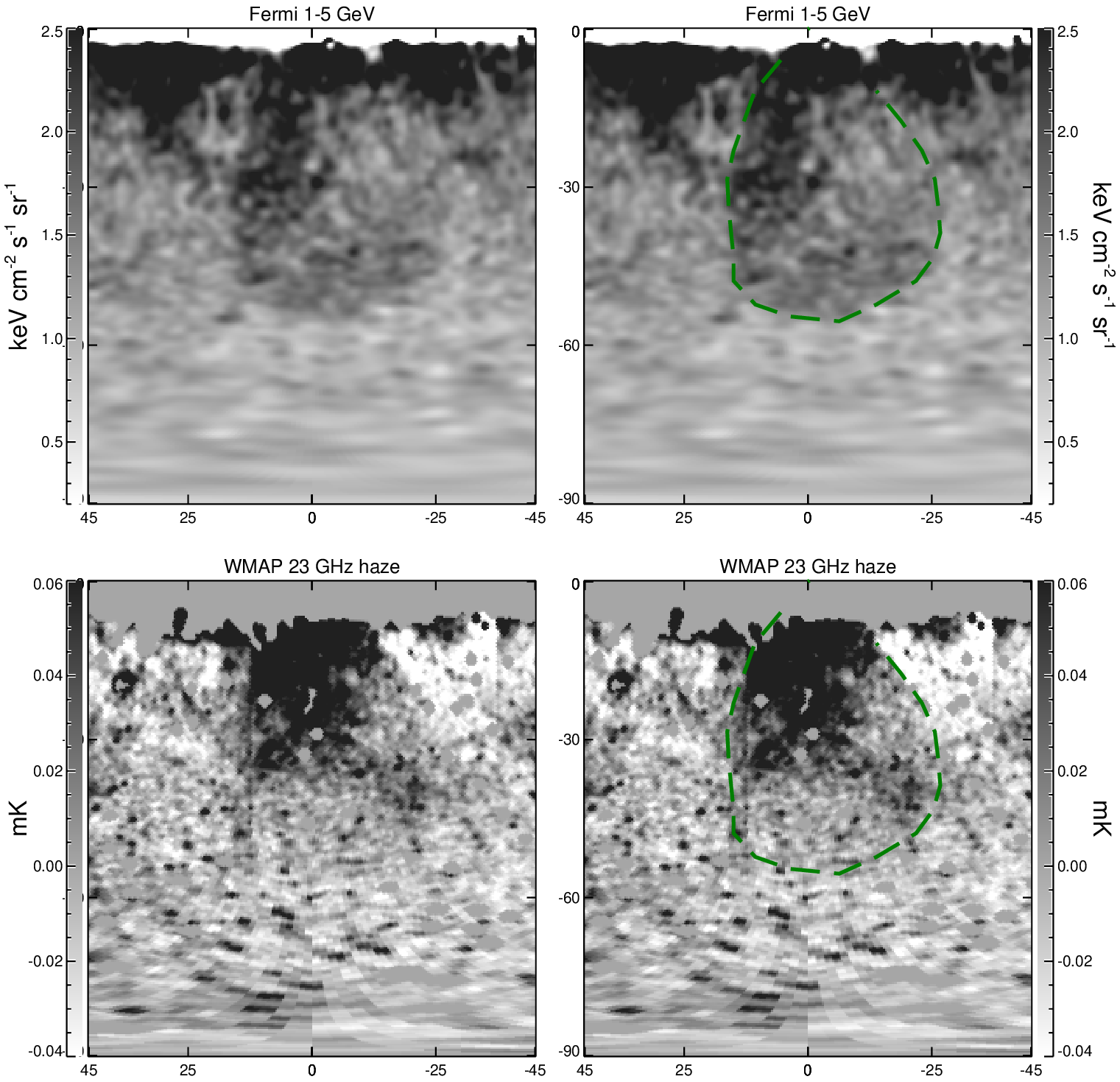}
\end{center}
\caption{
The \Fermi\ bubbles at $1-5$ GeV (the residual map obtained by subtracting the SFD dust map and the disk template) compared with the \WMAP\ K-band (23 GHz) haze \citep{Dobler:2008ww}. \emph{Top row:} The $1-5$ GeV map with $\ell=[-45\degree,45\degree]$ and $b=[-90\degree,0\degree]$ (\emph{left} panel) with the \Fermi\ south bubble edge overplotted in green dashed line (\emph{right} panel); the same as the left column third row panel of \reffig{bubbles}. \emph{Bottom row:} Same sky region as \emph{top row} but displaying the \WMAP\ haze at 23 GHz (\emph{left} panel), with the \Fermi\ south bubble edge overplotted in green dashed line (\emph{right} panel).
}
\label{fig:FermiVsWmap}
\end{figure*}

\reffig{FermiVsWmap} shows a detailed morphological comparison of the south \Fermi\ bubble at $1-5$ GeV with the southern part of the \WMAP\ microwave haze at 23 GHz (K-band). The edge of the \Fermi\ bubbles, marked in green dashed line in the \emph{top right} and \emph{lower right} panels, closely traces the edge of the \WMAP\ haze. The smaller latitudinal extension of the \WMAP\ haze may be due to the decay of the magnetic field strength with latitude. These striking morphological similarities between the \WMAP\ microwave haze and \Fermi\ gamma-ray bubble can be readily explained if the \emph{same} electron CR population is responsible for both excesses, with the electron CRs interacting with the galactic magnetic field to produce synchrotron, and interacting with the ISRF to produce IC emission.

\begin{figure*}[ht]
\begin{center}
\includegraphics[width=.8\textwidth]{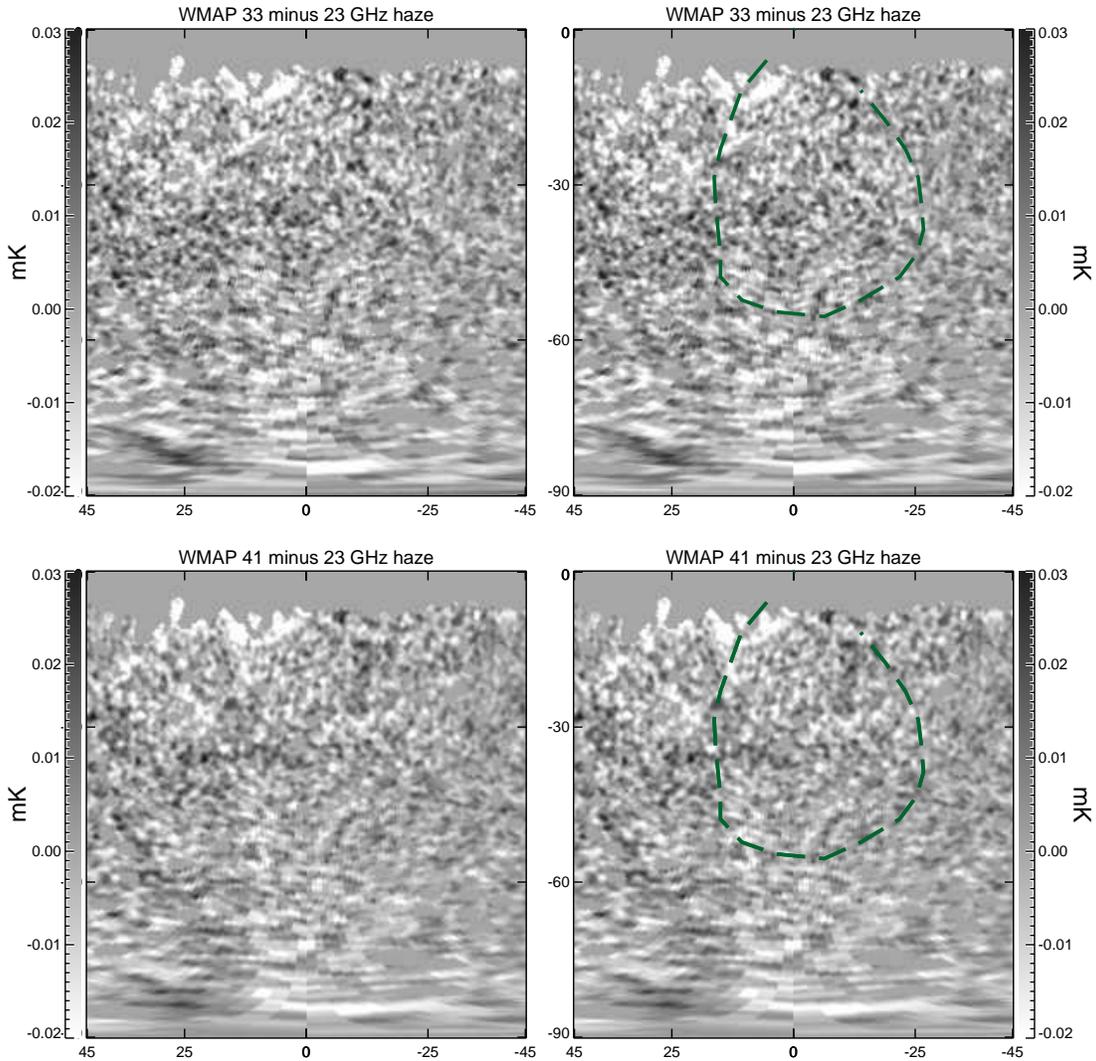}
\end{center}
\caption{
Difference maps of the \WMAP\ haze.  \emph{Top row:} The difference map between the 23 GHz \WMAP\ haze and 33 GHz \WMAP\ haze. The \emph{right} panel is the same as the left but overplotted with the south \Fermi\ bubble edge in green dashed line.
\emph{Bottom row:} The same as the \emph{top row}, but showing the difference between the 41 GHz \WMAP\ haze and 33 GHz \WMAP\ haze. We see no apparent structure in the difference maps, indicating a consistent spectrum across different (spatial) regions of the haze. 
}
\label{fig:Hazecompare}
\end{figure*}

In \reffig{Hazecompare}, we show the difference maps between the 23 GHz \WMAP\ haze and the 33 GHz and 41 GHz haze maps. The difference maps contain no apparent features and indicate a common spectrum of different regions inside the \WMAP\ haze, suggesting a single physical origin for the bulk of the signal. We have reached the same conclusion for the \Fermi\ bubbles in \reffig{energysplit}.

Besides the similarity of the morphology, the relatively hard spectrum of the \Fermi\ bubble also motivates a common physical origin with the \WMAP\ haze. We now provide a simple estimate of the microwave synchrotron and gamma-ray ICS signals from one single population of hard electrons distributed in the inner Galaxy, to demonstrate that the magnitudes and spectral indices of the two signals are consistent for reasonable parameter values.

For a highly relativistic electron scattering on low energy photons, the spectrum of upscattered photons is given by \cite{1970RvMP...42..237B} \citep{Cholis:2009},
\bea \nonumber \frac{dN}{dE_\gamma d\epsilon dt} &=& \frac{3}{4} \sigma_T c \frac{(m_e c^2)^2}{\epsilon E_e^2} \big(2 q \log q + (1 + 2 q) (1 - q) \\ &&+ 0.5 (1 - q) (\Gamma q)^2 / (1 + \Gamma q) \big) n(\epsilon) \label{eq:klein-nishina}, 
\eea
\[ \Gamma =  4 \epsilon E_e / (m_e c^2)^2, \quad q = \frac{E_\gamma}{E_e} \frac{1}{\Gamma (1 - E_\gamma/E_e)},\]
\[\epsilon < E_\gamma < E_e \Gamma / (1 + \Gamma). \]
Here $\epsilon$ is the initial photon energy, $E_e$ is the electron energy, $E_\gamma$ is the energy of the upscattered gamma ray, and $n(\epsilon)$ describes the energy distribution of the soft photons per unit volume. Where $\Gamma \ll 1$, in the Thomson limit, the average energy of the upscattered photons is given by,
\begin{equation} \langle E_\gamma \rangle = (4/3) \gamma^2 \langle \epsilon \rangle, \end{equation}
where $\gamma = E_e / m_e c^2$ is the Lorentz boost of the electron. In the Klein-Nishina (KN) limit, $\Gamma \gg 1$, the spectrum instead peaks at the high energy end, and the upscattered photon carries away almost all the energy of the electron. 

Given a power law steady-state electron spectrum with spectral index $\gamma$, the spectral index of the IC scattered gamma rays is $(\gamma + 1)/2$ in the Thomson limit, and $\gamma + 1$ in the extreme KN limit \citep[e.g.][]{1970RvMP...42..237B}. Photons in the \Fermi\ energy range can be produced by scattering of $\mathcal{O}(10-100)$ GeV electrons on starlight, which is (marginally) in the KN regime, or by Thomson scattering of much higher-energy electrons on IR or CMB photons. Consequently, the spectral index of gamma rays might be expected to vary with latitude even if the electron spectral index is uniform, becoming harder at higher latitudes where scatterings in the Thomson limit dominate. Closer to the disk, where much of the ISRF energy density is in starlight, IC scatterings are in neither the extreme KN nor Thomson limits, and the spectrum needs to be computed carefully. 

\begin{figure*}[ht]
\begin{center}
  \includegraphics[width=0.49\textwidth]{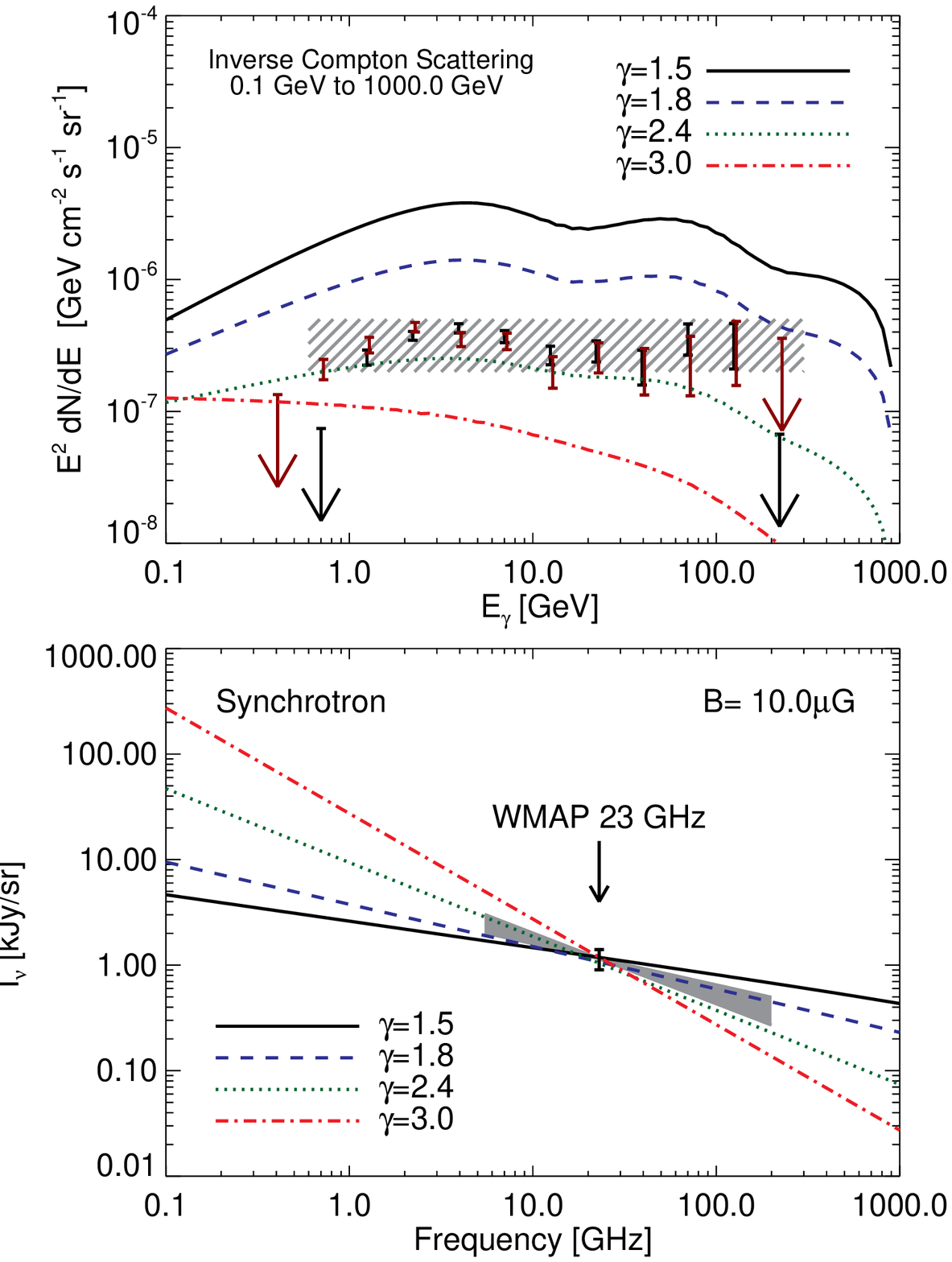}
  \includegraphics[width=0.49\textwidth]{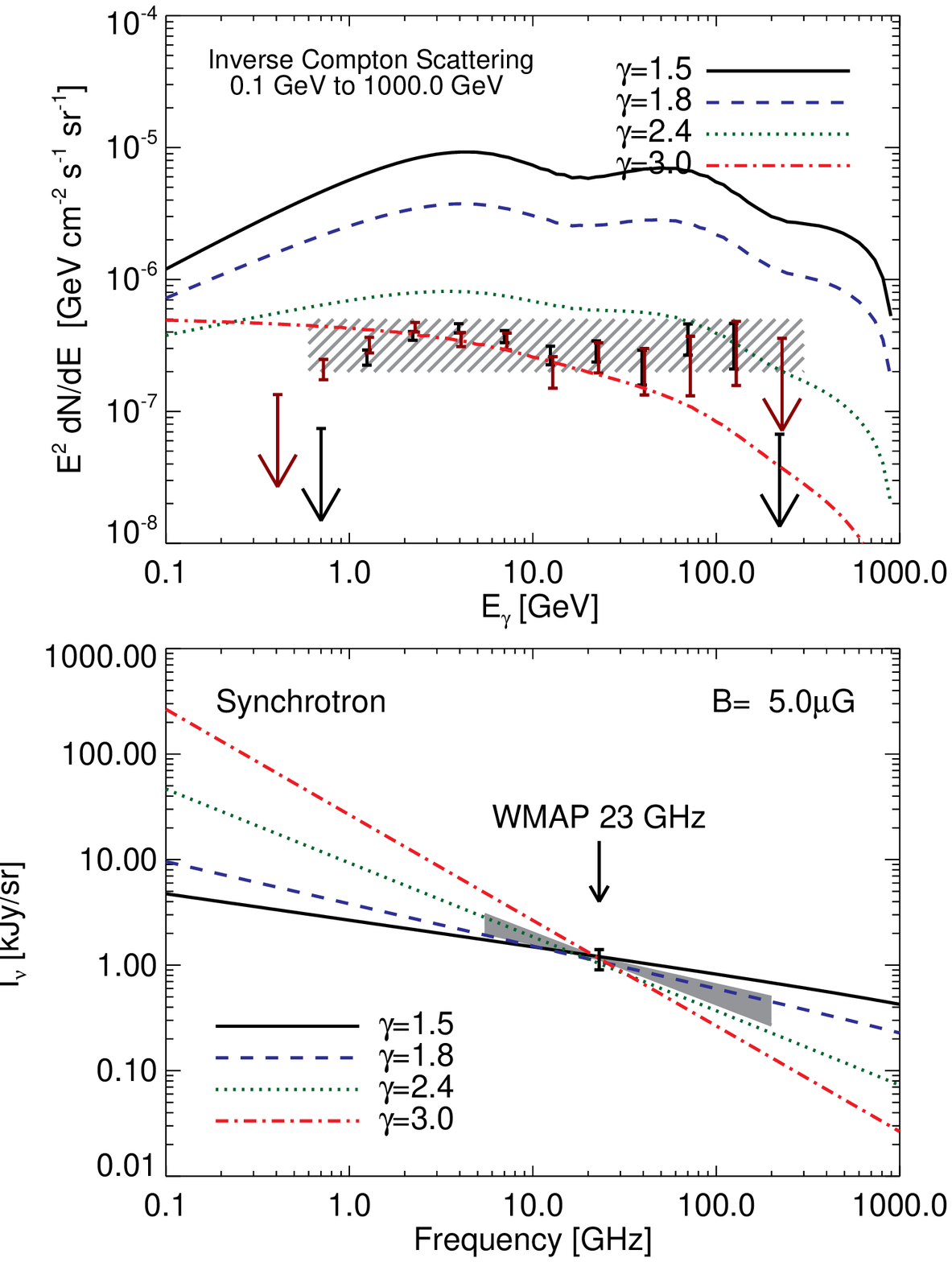}
\end{center}
\caption{The estimated spectrum of IC gamma rays (\emph{upper panel}) and synchrotron radiation (\emph{lower panel}) originating from a hard electron spectrum along a line of sight 4 kpc above the Galactic center (i.e. $b \approx 25\degree$). The steady-state electron spectrum is taken to be a power law, $dN/dE \propto E^{-\gamma}$, with index $\gamma =$ 1.5 (\emph{solid black}), 1.8 (\emph{blue dashed}), 2.4 (\emph{green dotted}), and 3.0 (\emph{red dash-dotted}). In all cases the spectrum has a range of [0.1, 1000] GeV. The interstellar radiation field model is taken from \texttt{GALPROP} version 50p, and the magnetic field is set to be 10 $\mu$G for the \emph{left} panel and 5 $\mu$G for the \emph{right} panel. The data points in the upper panels show the magnitude of the bubble emission obtained from template fitting in \reffig{bubblespecdisk} (\emph{brown}) and \reffig{bubblespecfermilowE} (\emph{black}) including the ``whole bubble'' template, as a function of energy. The lowest and highest bins contain 3$\sigma$ upper limits rather than data points with $1 \sigma$ error bars, due to the large uncertainties in the haze amplitude at those energies. For reference, a rectangular cross hatch region shows a approximate spectrum in the same place in this and subsequent figures. The data point in the lower panel shows the magnitude of the \WMAP\ haze averaged over $b=-20\degree$ to $-30\degree$, for $|\ell| < 10\degree$, in the 23 GHz K-band (the overall normalization is chosen to fit this value), and the gray area indicates the range of synchrotron spectral indices allowed for the \WMAP\ haze by \cite{Dobler:2008ww}. The same population of hard electrons can consistently generate the \WMAP\ synchrotron haze and \Fermi\ ICS bubbles.}
\label{fig:icsvssynchro}
\end{figure*}

We consider a steady-state electron spectrum described by a power law, $dN/dE \propto E^{-\gamma}$, with energy cutoffs at 0.1 GeV and 1000 GeV. The choice of high-energy cutoff is motivated by the local measurement of the cosmic ray electron spectrum by \Fermi\ \citep{Abdo:2009zk}. We consider a region $\sim 4$ kpc above the Galactic center, as an example (and since both the \WMAP\ haze and \Fermi\ bubbles are reasonably well measured there), and employ the model for the ISRF used in \texttt{GALPROP} version 50p \citep{Porter:2005qx} at 4 kpc above the GC. We normalize the synchrotron to the approximate value measured by \WMAP\ in the 23 GHz K-band \citep{Hooper:2007kb}, $\sim 25^\circ$ below the Galactic plane, and compute the corresponding synchrotron and IC spectra. The \WMAP\ haze was estimated to have a spectrum $I_\nu \propto \nu^{-\beta}$, $\beta = 0.39-0.67$ \citep{Dobler:2008ww}, corresponding approximately to an electron spectral index of $\gamma \approx 1.8-2.4$;  \reffig{icsvssynchro} shows our results for a magnetic field of 10 $\mu$G and 5 $\mu$G at 4 kpc above the GC, and electron spectral indices $\gamma = 1.8-3$. We find good agreement in the case of $\alpha \approx 2-2.5$, consistent with the spectrum of the \WMAP\ haze.

In the default \texttt{GALPROP} exponential model for the Galactic magnetic field, $|B| = |B_0| e^{-z/z_s}$ with scale height $z_s \approx 2$ kpc, this field strength would correspond to $B_0 \approx 30-40\ \mu$G or even higher. This value is considerably larger than commonly used \citep[e.g.][]{Page:2006hz}. However, models with a non-exponential halo magnetic field, as discussed by e.g. \cite{Alvarez-Mu:2002,Sun:2008}, can have $\sim$ 10 $\mu$G fields well off the plane.

We note also that the extrapolated value of $B_0$ required to obtain good agreement between the IC and synchrotron amplitudes, in the exponential model, is somewhat higher than found by \cite{fermihaze}, who performed the comparison at 2 kpc. This apparent discrepancy originates from the fact that in the haze latitudinal profile given by \cite{Hooper:2007kb}, the emission falls off rapidly with latitude for $0 > b > -15^\circ$, but then plateaus at $b \sim - 15-35^\circ$, contrary to expectations based on a B-field profile exponentially falling away from $z=0$. This suggests either that the magnetic field inside the bubble does not fall exponentially with $|z|$ inside the bubbles, or that the \WMAP\ haze contains a significant free-free component at high latitude.

\begin{figure*}[ht]
\begin{center}
  \includegraphics[width=0.49\textwidth]{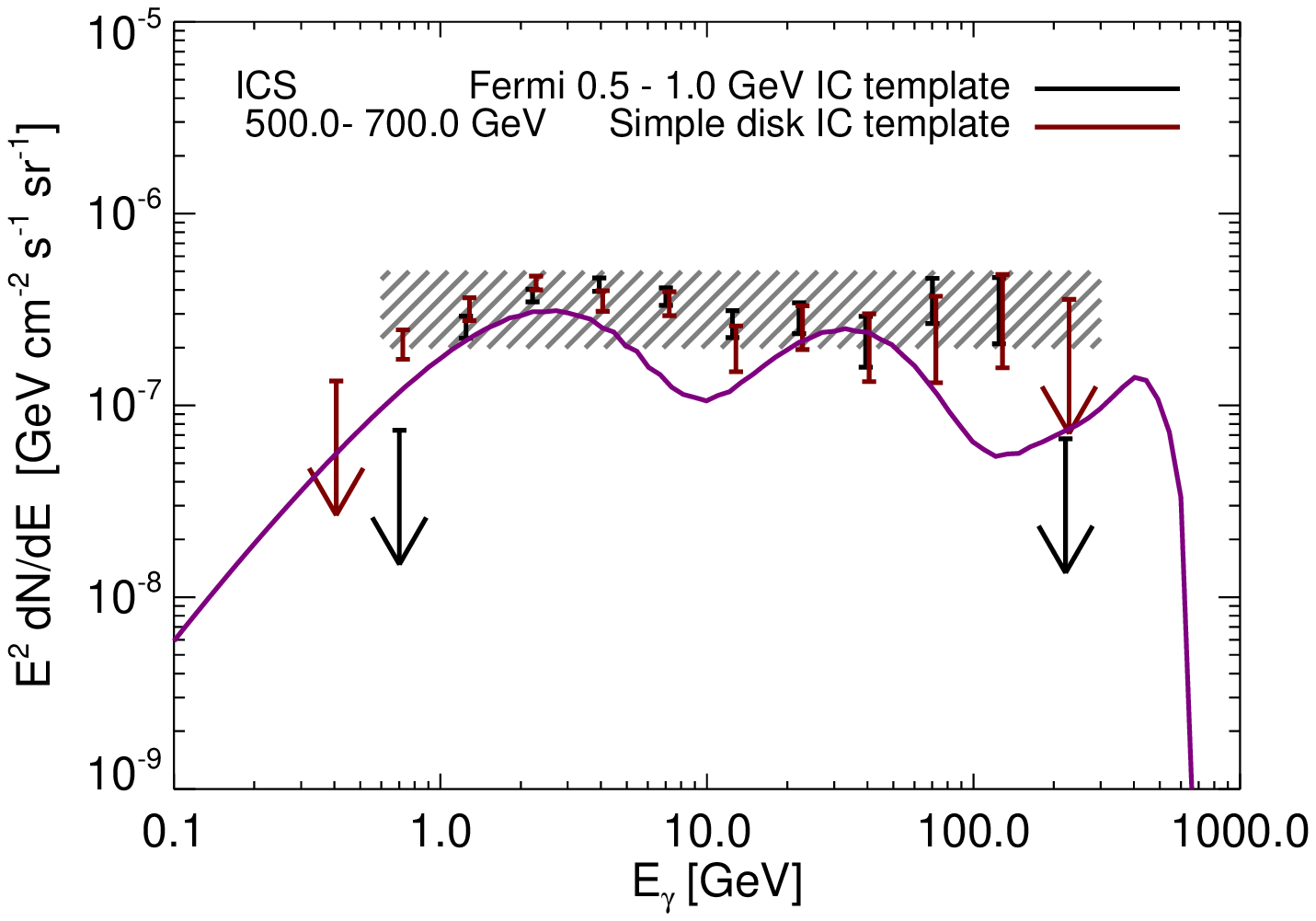}
  \includegraphics[width=0.49\textwidth]{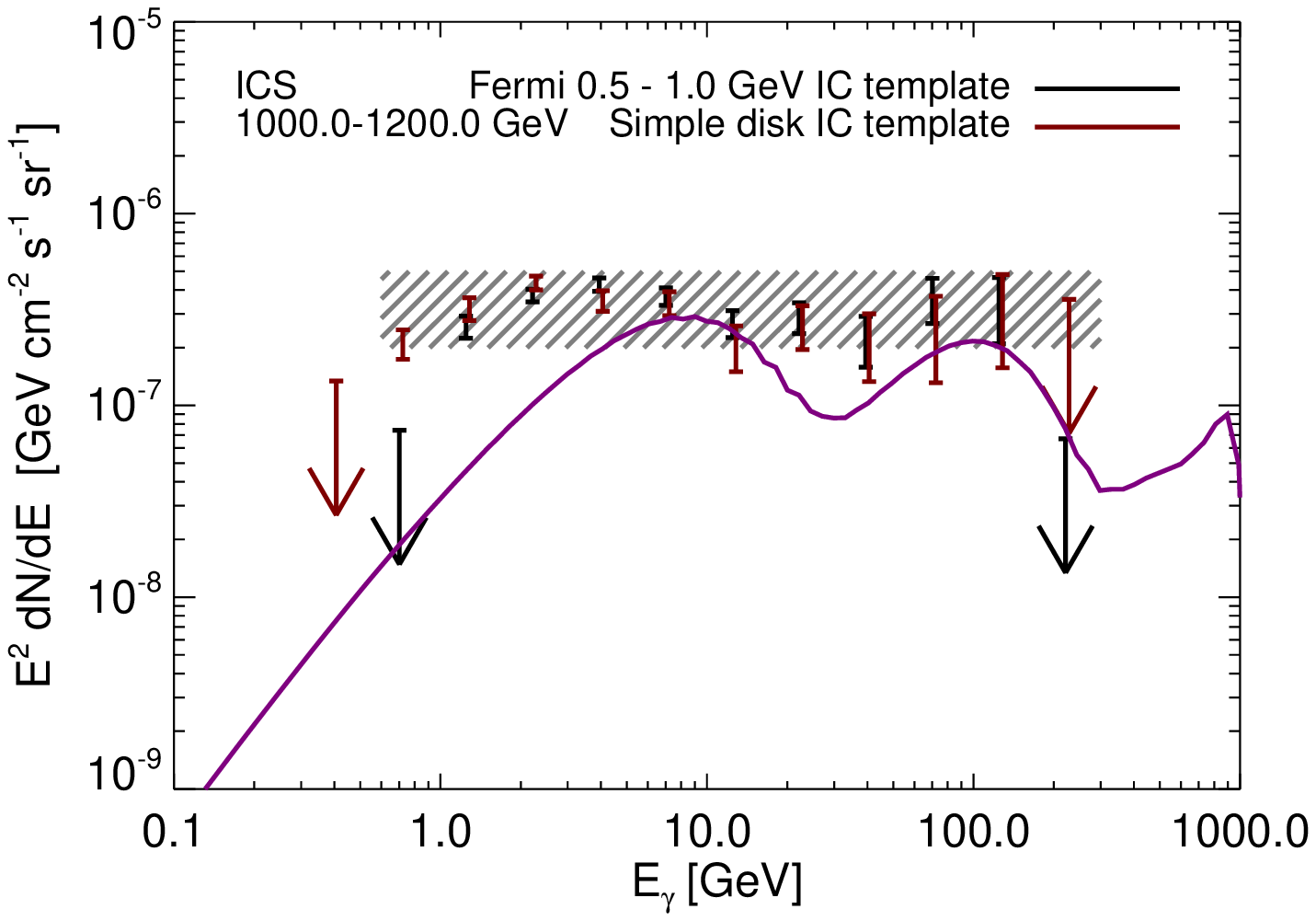}
  \includegraphics[width=0.49\textwidth]{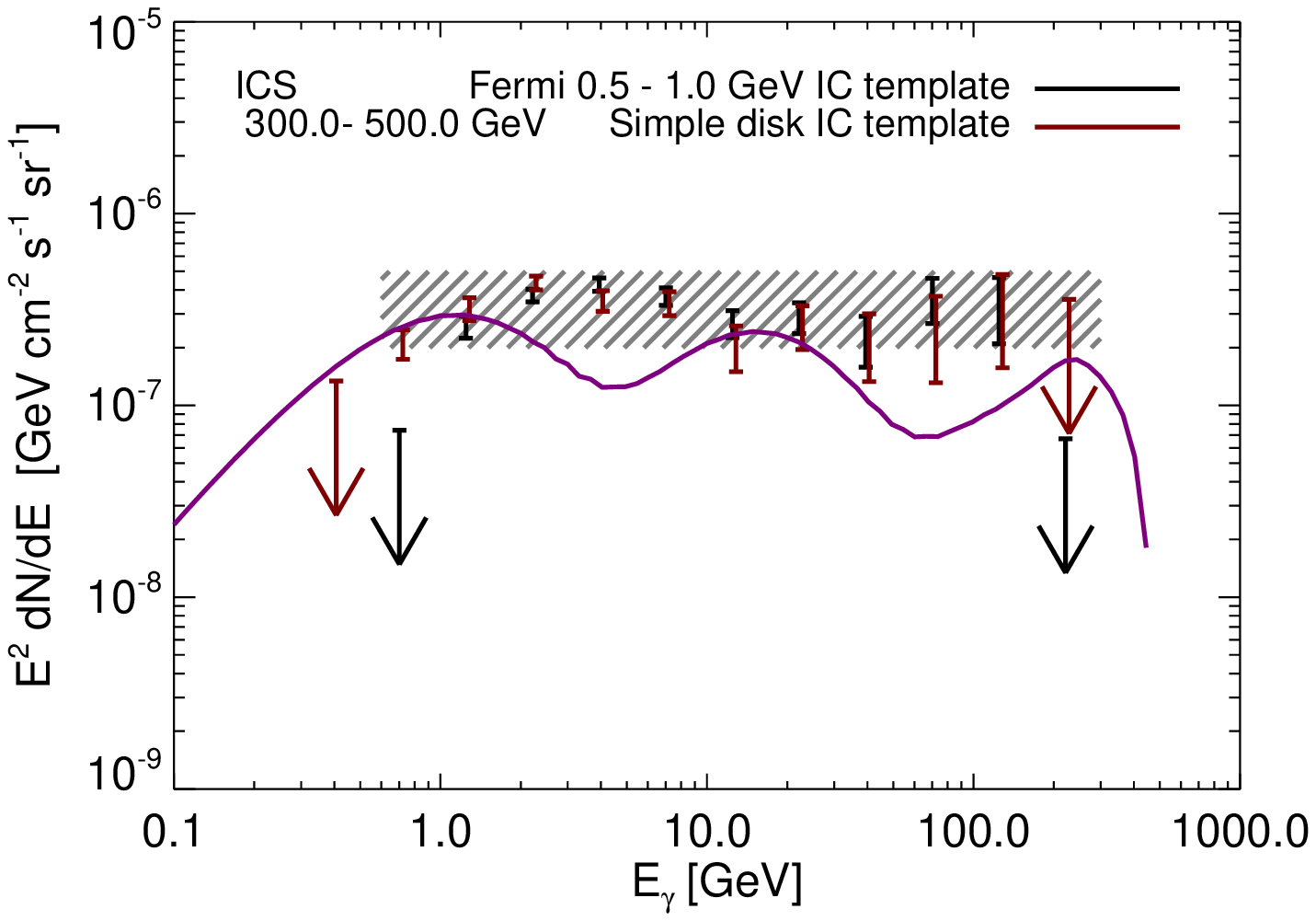}
  \includegraphics[width=0.49\textwidth]{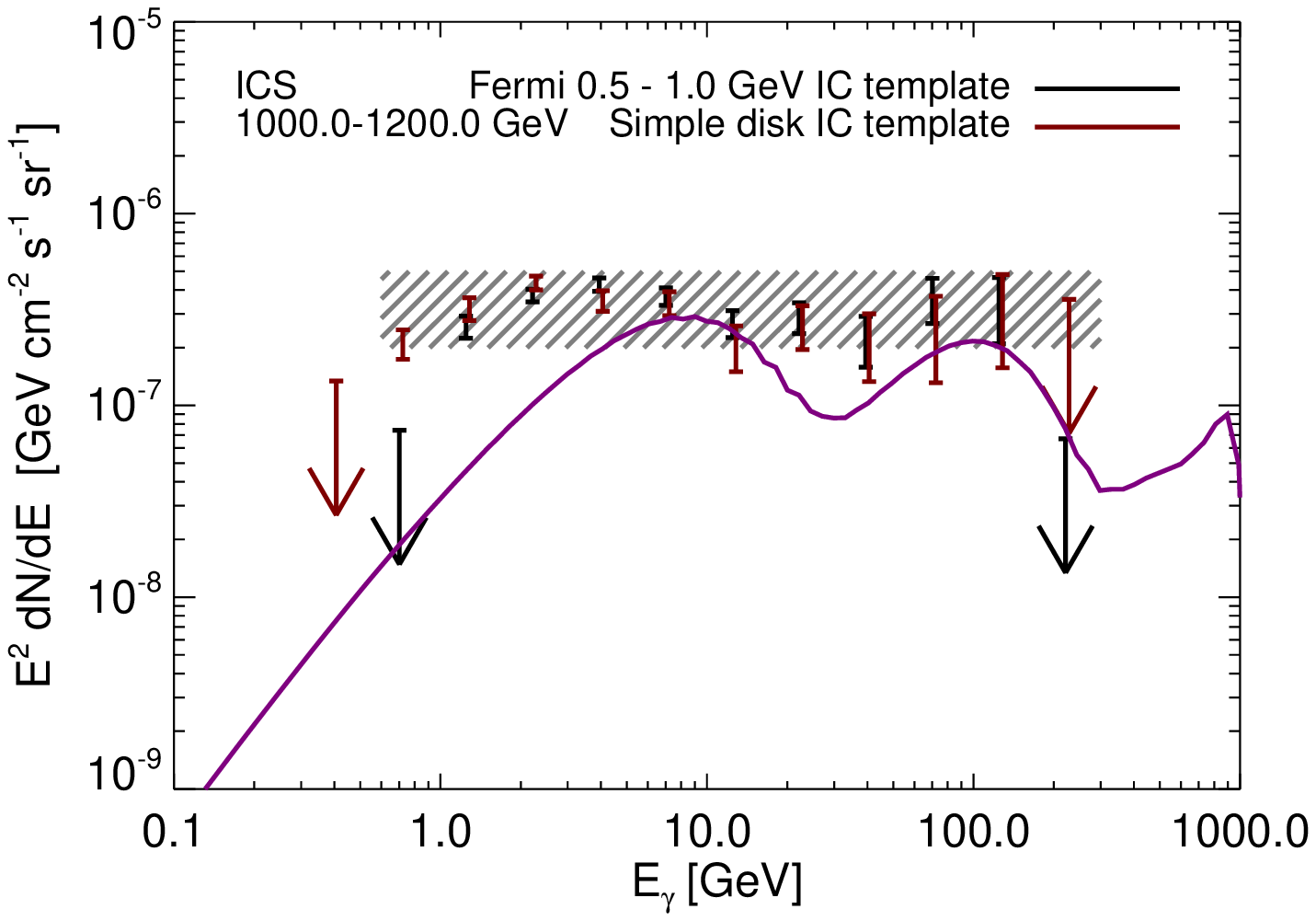}
\end{center}
\caption{The estimated spectrum of IC gamma rays originating from a hard electron spectrum ($dN/dE \propto E^{-2}$) with a limited energy range, as in the \emph{top row} of \reffig{icsvssynchro}, but with different minimum and maximum energies. The normalization of the ICS signal is fitted to the data. The three peaks are from ICS of the CMB (left peak), FIR (middle peak), and optical/UV interstellar radiation field. The ISRF model is taken from \texttt{GALPROP} version 50p. A hard electron CR population with a low energy cutoff at about 500 GeV can fit the data better than a power law electron spectrum extending from 0.1 GeV to 1000 GeV (see \reffig{icsvssynchro}). 
}
\label{fig:icslowEcut}
\end{figure*}

\begin{figure*}[ht]
\begin{center}
  \includegraphics[width=0.49\textwidth]{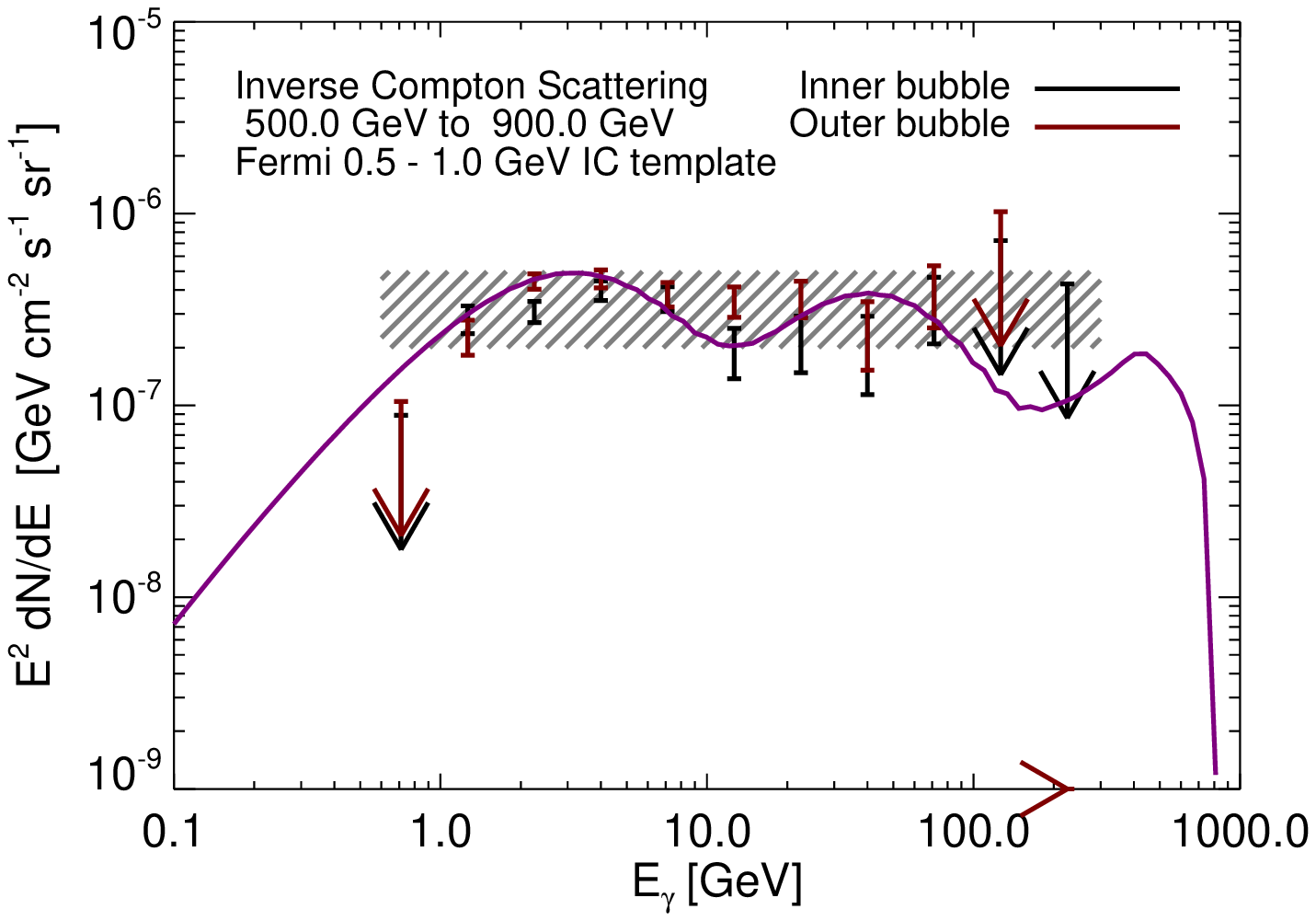}
  \includegraphics[width=0.49\textwidth]{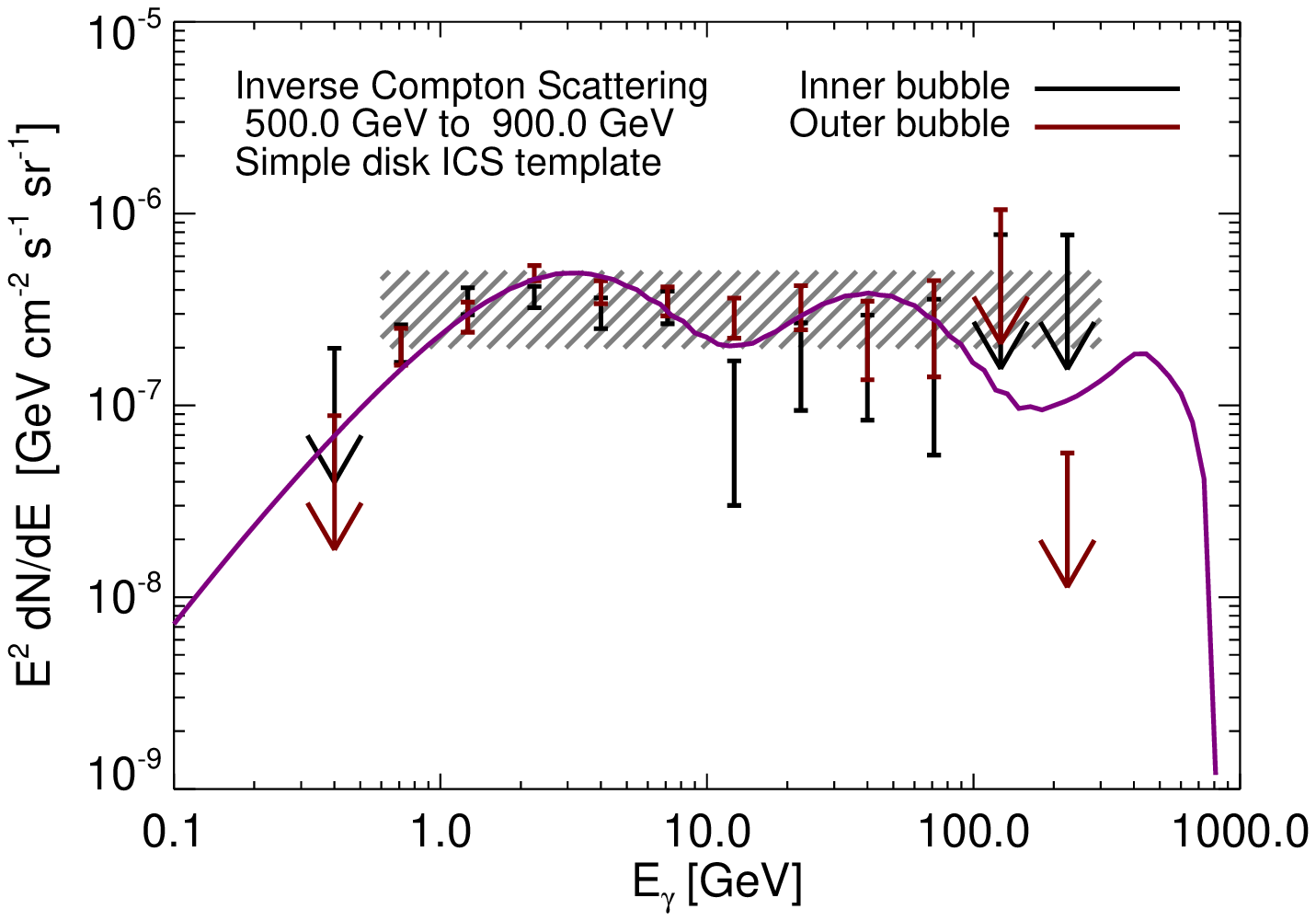}
  \includegraphics[width=0.49\textwidth]{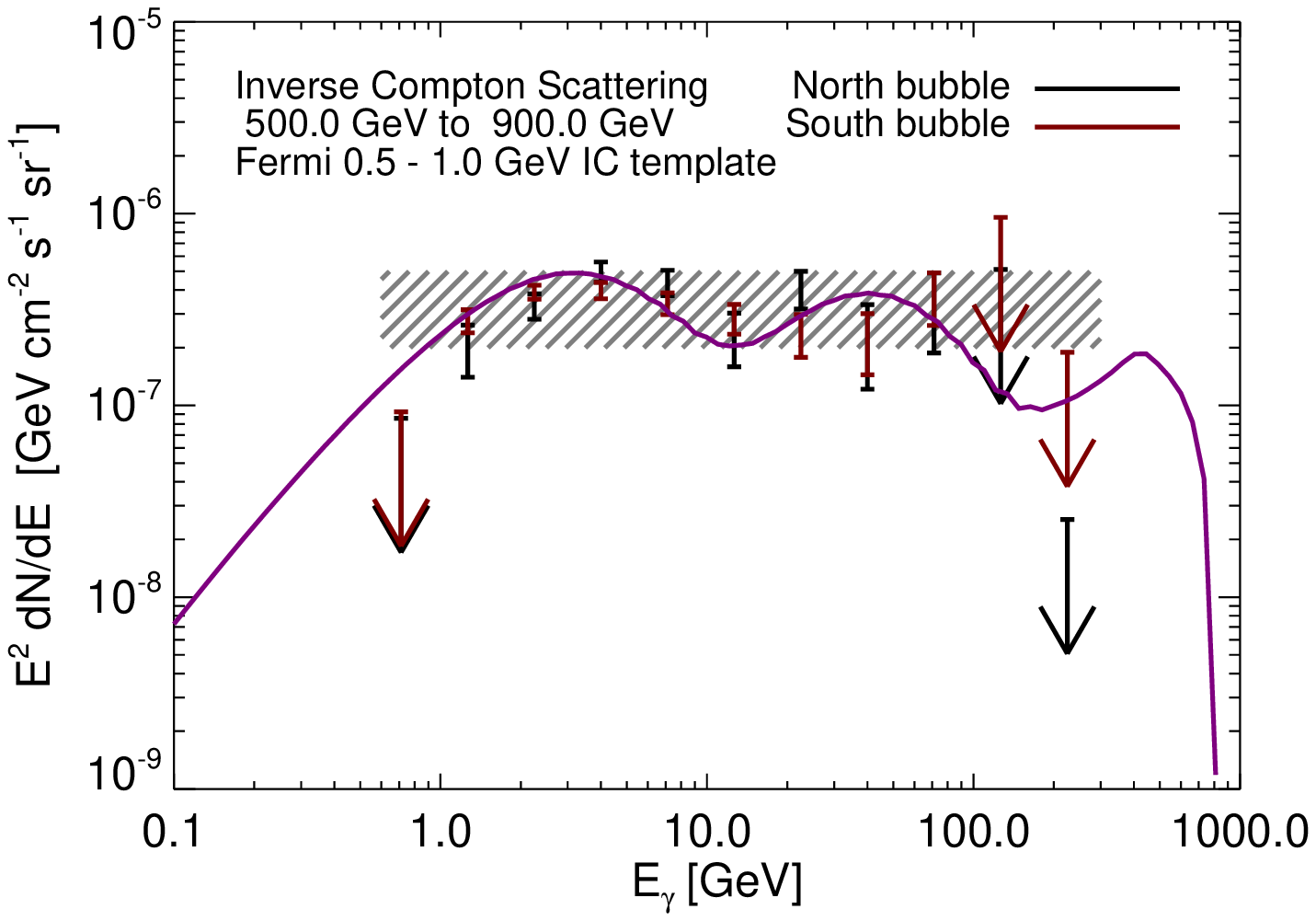}
  \includegraphics[width=0.49\textwidth]{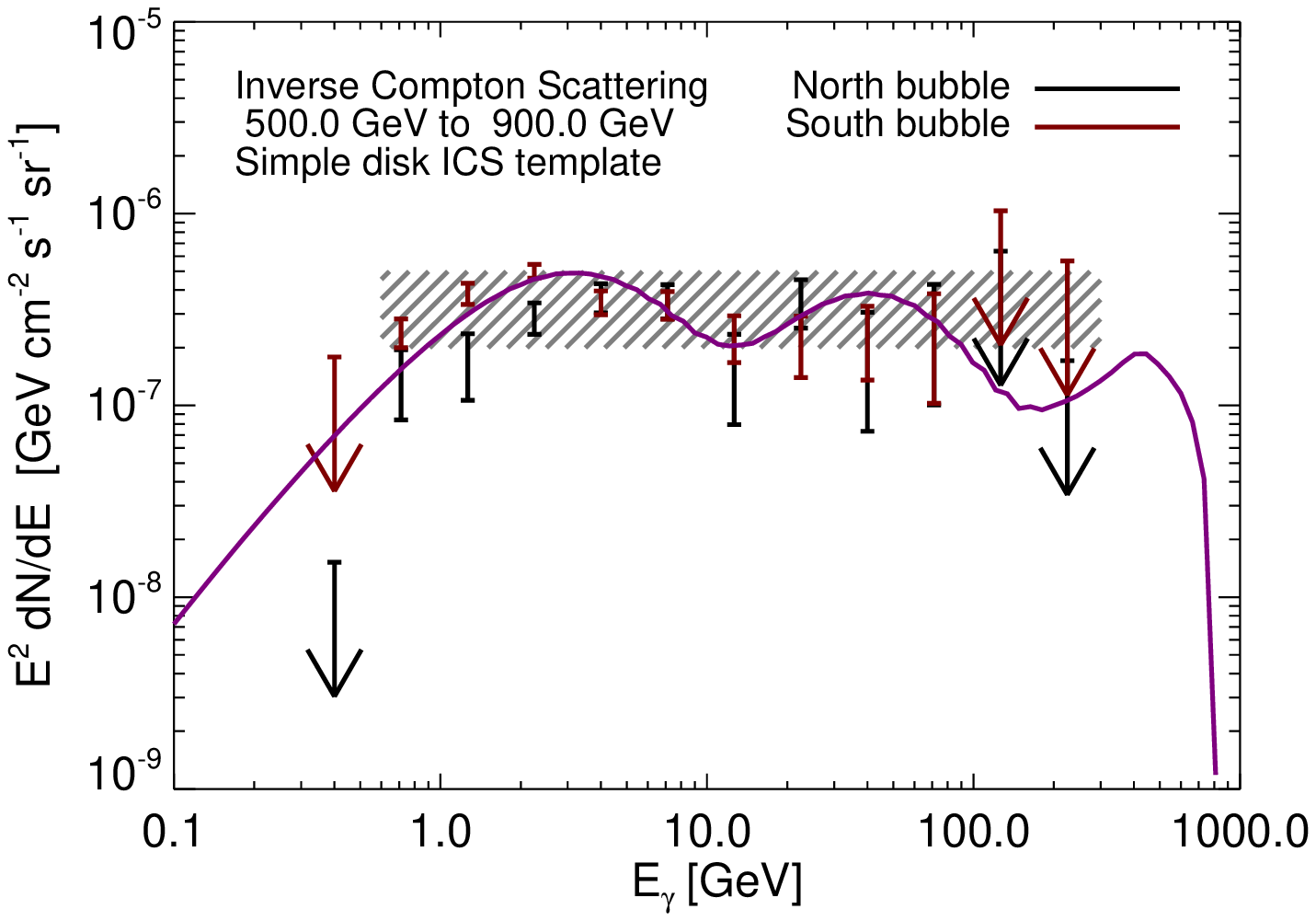}
\end{center}
\caption{The estimated spectrum of IC gamma rays originating from a hard electron spectrum, the same as \reffig{icslowEcut}, but for the $500-900$ GeV energy range. \emph{Top row:} data points show the separately fitted spectra for the bubble interior and outer shell templates (as defined in \reffig{splittemplate}), with the template for the disk IC emission given by (\emph{left} panel) the $0.5-1$ GeV \Fermi\ map with dust-correlated emission subtracted, and (\emph{right} panel) the simple geometric disk model defined in \reffig{bubbles}. \emph{Bottom row:} as top row, but showing the separately fitted spectra for the north and south \Fermi\ bubbles (as defined in \reffig{splittemplate}).}
\label{fig:icssplit}
\end{figure*}

\begin{table*}
\begin{center}
\begin{tabular}{@{}rrrrrrrrr}
\hline
\hline
E range (GeV) & Energy & Uniform & SFD dust & simple disk & inner bubble & outer bubble & simple loop I\\
\hline
$   0.3 -   0.5$
 &    0.4 &  1.681 $\pm$ 0.006 &  1.201 $\pm$ 0.011 &  0.683 $\pm$ 0.027 &  0.071 $\pm$ 0.042 &  0.004 $\pm$ 0.040 &  0.490 $\pm$ 0.015 \\
$   0.5 -   0.9$
 &    0.7 &  1.365 $\pm$ 0.007 &  1.279 $\pm$ 0.012 &  0.607 $\pm$ 0.030 &  0.215 $\pm$ 0.048 &  0.207 $\pm$ 0.045 &  0.475 $\pm$ 0.016 \\
$   0.9 -   1.7$
 &    1.3 &  1.142 $\pm$ 0.008 &  1.179 $\pm$ 0.014 &  0.498 $\pm$ 0.036 &  0.354 $\pm$ 0.057 &  0.293 $\pm$ 0.053 &  0.407 $\pm$ 0.019 \\
$   1.7 -   3.0$
 &    2.2 &  1.034 $\pm$ 0.006 &  0.876 $\pm$ 0.011 &  0.403 $\pm$ 0.029 &  0.370 $\pm$ 0.047 &  0.491 $\pm$ 0.045 &  0.373 $\pm$ 0.016 \\
$   3.0 -   5.3$
 &    4.0 &  0.880 $\pm$ 0.008 &  0.554 $\pm$ 0.013 &  0.426 $\pm$ 0.035 &  0.307 $\pm$ 0.056 &  0.393 $\pm$ 0.054 &  0.247 $\pm$ 0.018 \\
$   5.3 -   9.5$
 &    7.1 &  0.731 $\pm$ 0.009 &  0.322 $\pm$ 0.014 &  0.284 $\pm$ 0.039 &  0.330 $\pm$ 0.064 &  0.354 $\pm$ 0.061 &  0.207 $\pm$ 0.021 \\
$   9.5 -  16.9$
 &   12.7 &  0.562 $\pm$ 0.010 &  0.193 $\pm$ 0.015 &  0.265 $\pm$ 0.044 &  0.100 $\pm$ 0.070 &  0.293 $\pm$ 0.069 &  0.087 $\pm$ 0.023 \\
$  16.9 -  30.0$
 &   22.5 &  0.506 $\pm$ 0.012 &  0.128 $\pm$ 0.018 &  0.201 $\pm$ 0.054 &  0.182 $\pm$ 0.088 &  0.335 $\pm$ 0.087 &  0.121 $\pm$ 0.029 \\
$  30.0 -  53.3$
 &   40.0 &  0.556 $\pm$ 0.015 &  0.041 $\pm$ 0.021 &  0.098 $\pm$ 0.065 &  0.190 $\pm$ 0.106 &  0.243 $\pm$ 0.107 &  0.087 $\pm$ 0.036 \\
$  53.3 -  94.9$
 &   71.1 &  0.627 $\pm$ 0.022 &  0.020 $\pm$ 0.030 &  0.187 $\pm$ 0.093 &  0.206 $\pm$ 0.151 &  0.294 $\pm$ 0.153 & -0.045 $\pm$ 0.048 \\
$  94.9 - 168.7$
 &  126.5 &  0.620 $\pm$ 0.030 &  0.081 $\pm$ 0.043 &  0.037 $\pm$ 0.129 &  0.180 $\pm$ 0.209 &  0.431 $\pm$ 0.207 & -0.098 $\pm$ 0.064 \\
$ 168.7 - 300.0$
 &  225.0 &  0.435 $\pm$ 0.038 &  0.179 $\pm$ 0.062 & -0.065 $\pm$ 0.155 &  0.145 $\pm$ 0.242 & -0.178 $\pm$ 0.206 &  0.097 $\pm$ 0.086 \\
\hline
\end{tabular}
\end{center}
\caption{Corresponding template fitting coefficients and errors in \emph{upper left panel} of \reffig{bubblespecmore}.}
\label{tbl:shelldisk}
\end{table*}

\begin{table*}
\begin{center}
\begin{tabular}{@{}rrrrrrrrr}
\hline
\hline
E range (GeV) & Energy & Uniform & SFD dust & simple disk & north bubble & south bubble & simple loop I\\
\hline
$   0.3 -   0.5$
 &    0.4 &  1.679 $\pm$ 0.006 &  1.205 $\pm$ 0.011 &  0.683 $\pm$ 0.027 & -0.042 $\pm$ 0.050 &  0.069 $\pm$ 0.037 &  0.496 $\pm$ 0.015 \\
$   0.5 -   0.9$
 &    0.7 &  1.363 $\pm$ 0.007 &  1.282 $\pm$ 0.012 &  0.603 $\pm$ 0.030 &  0.140 $\pm$ 0.056 &  0.241 $\pm$ 0.041 &  0.483 $\pm$ 0.017 \\
$   0.9 -   1.7$
 &    1.3 &  1.138 $\pm$ 0.008 &  1.187 $\pm$ 0.015 &  0.493 $\pm$ 0.035 &  0.171 $\pm$ 0.065 &  0.385 $\pm$ 0.049 &  0.421 $\pm$ 0.020 \\
$   1.7 -   3.0$
 &    2.2 &  1.030 $\pm$ 0.007 &  0.883 $\pm$ 0.012 &  0.384 $\pm$ 0.029 &  0.288 $\pm$ 0.054 &  0.503 $\pm$ 0.041 &  0.391 $\pm$ 0.016 \\
$   3.0 -   5.3$
 &    4.0 &  0.881 $\pm$ 0.008 &  0.553 $\pm$ 0.013 &  0.421 $\pm$ 0.035 &  0.368 $\pm$ 0.064 &  0.346 $\pm$ 0.049 &  0.247 $\pm$ 0.019 \\
$   5.3 -   9.5$
 &    7.1 &  0.731 $\pm$ 0.009 &  0.321 $\pm$ 0.014 &  0.283 $\pm$ 0.039 &  0.354 $\pm$ 0.072 &  0.337 $\pm$ 0.055 &  0.207 $\pm$ 0.022 \\
$   9.5 -  16.9$
 &   12.7 &  0.562 $\pm$ 0.010 &  0.195 $\pm$ 0.016 &  0.248 $\pm$ 0.044 &  0.157 $\pm$ 0.078 &  0.230 $\pm$ 0.063 &  0.097 $\pm$ 0.024 \\
$  16.9 -  30.0$
 &   22.5 &  0.509 $\pm$ 0.012 &  0.124 $\pm$ 0.018 &  0.196 $\pm$ 0.053 &  0.353 $\pm$ 0.100 &  0.216 $\pm$ 0.076 &  0.116 $\pm$ 0.030 \\
$  30.0 -  53.3$
 &   40.0 &  0.556 $\pm$ 0.015 &  0.042 $\pm$ 0.021 &  0.094 $\pm$ 0.064 &  0.190 $\pm$ 0.117 &  0.232 $\pm$ 0.097 &  0.092 $\pm$ 0.038 \\
$  53.3 -  94.9$
 &   71.1 &  0.628 $\pm$ 0.022 &  0.020 $\pm$ 0.030 &  0.184 $\pm$ 0.093 &  0.264 $\pm$ 0.164 &  0.243 $\pm$ 0.140 & -0.045 $\pm$ 0.049 \\
$  94.9 - 168.7$
 &  126.5 &  0.614 $\pm$ 0.030 &  0.091 $\pm$ 0.044 &  0.004 $\pm$ 0.126 &  0.110 $\pm$ 0.208 &  0.450 $\pm$ 0.195 & -0.068 $\pm$ 0.067 \\
$ 168.7 - 300.0$
 &  225.0 &  0.429 $\pm$ 0.039 &  0.186 $\pm$ 0.064 & -0.058 $\pm$ 0.160 & -0.169 $\pm$ 0.254 &  0.061 $\pm$ 0.222 &  0.107 $\pm$ 0.092 \\
\hline
\end{tabular}
\end{center}
\caption{Corresponding template fitting coefficients and errors in \emph{upper right panel} of \reffig{bubblespecmore}.}
\label{tbl:splitdisk}
\end{table*}

\begin{table*}
\begin{center}
\begin{tabular}{@{}rrrrrrr}
\hline
\hline
E range (GeV) & Energy & Uniform & SFD dust & inner bubble & outer bubble & $0.5-1.0$ GeV - SFD\\
\hline
$   0.3 -   0.5$
 &    0.4 &  1.759 $\pm$ 0.006 &  0.883 $\pm$ 0.012 &  0.011 $\pm$ 0.035 & -0.069 $\pm$ 0.037 &  1.180 $\pm$ 0.017 \\
$   0.5 -   0.9$
 &    0.7 &  1.446 $\pm$ 0.006 &  0.905 $\pm$ 0.013 &  0.005 $\pm$ 0.039 &  0.012 $\pm$ 0.041 &  1.275 $\pm$ 0.018 \\
$   0.9 -   1.7$
 &    1.3 &  1.208 $\pm$ 0.008 &  0.929 $\pm$ 0.016 &  0.284 $\pm$ 0.047 &  0.230 $\pm$ 0.048 &  0.896 $\pm$ 0.020 \\
$   1.7 -   3.0$
 &    2.2 &  1.088 $\pm$ 0.006 &  0.680 $\pm$ 0.012 &  0.309 $\pm$ 0.039 &  0.444 $\pm$ 0.041 &  0.737 $\pm$ 0.017 \\
$   3.0 -   5.3$
 &    4.0 &  0.921 $\pm$ 0.007 &  0.427 $\pm$ 0.014 &  0.399 $\pm$ 0.047 &  0.459 $\pm$ 0.049 &  0.530 $\pm$ 0.019 \\
$   5.3 -   9.5$
 &    7.1 &  0.759 $\pm$ 0.008 &  0.231 $\pm$ 0.015 &  0.362 $\pm$ 0.053 &  0.382 $\pm$ 0.056 &  0.400 $\pm$ 0.021 \\
$   9.5 -  16.9$
 &   12.7 &  0.580 $\pm$ 0.009 &  0.131 $\pm$ 0.016 &  0.195 $\pm$ 0.057 &  0.351 $\pm$ 0.063 &  0.271 $\pm$ 0.023 \\
$  16.9 -  30.0$
 &   22.5 &  0.523 $\pm$ 0.011 &  0.069 $\pm$ 0.015 &  0.218 $\pm$ 0.071 &  0.368 $\pm$ 0.079 &  0.261 $\pm$ 0.018 \\
$  30.0 -  53.3$
 &   40.0 &  0.565 $\pm$ 0.012 &  0.014 $\pm$ 0.015 &  0.202 $\pm$ 0.088 &  0.249 $\pm$ 0.096 &  0.144 $\pm$ 0.025 \\
$  53.3 -  94.9$
 &   71.1 &  0.631 $\pm$ 0.021 &  0.010 $\pm$ 0.033 &  0.337 $\pm$ 0.128 &  0.394 $\pm$ 0.140 &  0.065 $\pm$ 0.053 \\
$  94.9 - 168.7$
 &  126.5 &  0.614 $\pm$ 0.029 &  0.084 $\pm$ 0.048 &  0.232 $\pm$ 0.164 &  0.455 $\pm$ 0.190 & -0.029 $\pm$ 0.076 \\
$ 168.7 - 300.0$
 &  225.0 &  0.440 $\pm$ 0.037 &  0.152 $\pm$ 0.069 &  0.021 $\pm$ 0.195 & -0.262 $\pm$ 0.175 &  0.088 $\pm$ 0.096 \\
\hline
\end{tabular}
\end{center}
\caption{Corresponding template fitting coefficients and errors in \emph{lower left panel} of \reffig{bubblespecmore}.}
\label{tbl:shellres}
\end{table*}

\begin{table*}
\begin{center}
\begin{tabular}{@{}rrrrrrrrr}
\hline
\hline
E range (GeV) & Energy & Uniform & SFD dust & north bubble & south bubble & $0.5-1.0$ GeV - SFD\\
\hline
$   0.3 -   0.5$
 &    0.4 &  1.759 $\pm$ 0.006 &  0.882 $\pm$ 0.012 & -0.011 $\pm$ 0.047 & -0.031 $\pm$ 0.029 &  1.180 $\pm$ 0.017 \\
$   0.5 -   0.9$
 &    0.7 &  1.446 $\pm$ 0.006 &  0.905 $\pm$ 0.013 & -0.009 $\pm$ 0.052 &  0.014 $\pm$ 0.032 &  1.276 $\pm$ 0.018 \\
$   0.9 -   1.7$
 &    1.3 &  1.207 $\pm$ 0.008 &  0.931 $\pm$ 0.016 &  0.201 $\pm$ 0.061 &  0.277 $\pm$ 0.038 &  0.899 $\pm$ 0.020 \\
$   1.7 -   3.0$
 &    2.2 &  1.087 $\pm$ 0.006 &  0.681 $\pm$ 0.013 &  0.331 $\pm$ 0.050 &  0.391 $\pm$ 0.033 &  0.738 $\pm$ 0.017 \\
$   3.0 -   5.3$
 &    4.0 &  0.922 $\pm$ 0.007 &  0.424 $\pm$ 0.014 &  0.500 $\pm$ 0.060 &  0.399 $\pm$ 0.039 &  0.527 $\pm$ 0.019 \\
$   5.3 -   9.5$
 &    7.1 &  0.760 $\pm$ 0.008 &  0.228 $\pm$ 0.015 &  0.440 $\pm$ 0.067 &  0.342 $\pm$ 0.045 &  0.397 $\pm$ 0.021 \\
$   9.5 -  16.9$
 &   12.7 &  0.580 $\pm$ 0.009 &  0.132 $\pm$ 0.016 &  0.231 $\pm$ 0.072 &  0.286 $\pm$ 0.051 &  0.272 $\pm$ 0.023 \\
$  16.9 -  30.0$
 &   22.5 &  0.524 $\pm$ 0.011 &  0.068 $\pm$ 0.015 &  0.406 $\pm$ 0.091 &  0.236 $\pm$ 0.060 &  0.257 $\pm$ 0.008 \\
$  30.0 -  53.3$
 &   40.0 &  0.565 $\pm$ 0.014 &  0.014 $\pm$ 0.015 &  0.227 $\pm$ 0.105 &  0.222 $\pm$ 0.077 &  0.143 $\pm$ 0.015 \\
$  53.3 -  94.9$
 &   71.1 &  0.631 $\pm$ 0.021 &  0.011 $\pm$ 0.033 &  0.339 $\pm$ 0.151 &  0.376 $\pm$ 0.115 &  0.067 $\pm$ 0.054 \\
$  94.9 - 168.7$
 &  126.5 &  0.608 $\pm$ 0.029 &  0.091 $\pm$ 0.049 &  0.067 $\pm$ 0.189 &  0.470 $\pm$ 0.162 & -0.013 $\pm$ 0.077 \\
$ 168.7 - 300.0$
 &  225.0 &  0.437 $\pm$ 0.038 &  0.156 $\pm$ 0.070 & -0.216 $\pm$ 0.229 & -0.066 $\pm$ 0.160 &  0.095 $\pm$ 0.098 \\
\hline
\end{tabular}
\end{center}
\caption{Corresponding template fitting coefficients and errors in \emph{lower right panel} of \reffig{bubblespecmore}.}
\label{tbl:splitres}
\end{table*}

\subsection{Evidence of a $\sim 700$ GeV Electron Excess?}
\label{sec:atic}

In \reffig{icslowEcut}, we calculate the gamma-ray spectrum from IC scattering, using the standard ISRF model taken from \texttt{GALPROP} -- as in \reffig{icsvssynchro}, but with a different energy range for the electron CRs. An electron CR population with a hard low-energy cutoff at about 500 GeV can fit the \Fermi\ bubble spectrum better than a single power law extending from $0.1-1000$ GeV, due to the downturn in the spectrum in the lowest energy bin. Even a rather hard ($dN/dE \sim E^{-2}$) power law component at $300-500$ GeV produces a long tail at low energies. Interestingly, this preferred $500-700$ GeV energy range is rather close to the peak in local $e^+ + e^-$ cosmic rays observed by \emph{ATIC} \citep{aticlatest}\footnote{However, note that the large peak observed by \emph{ATIC}-2 and \emph{ATIC}-4 appears to be in conflict with the \Fermi\ measurement of the spectrum, which contains no such feature.}. We note that although the estimated error bars of energy lower than 1 GeV mildly depend on the templates we use in the fitting procedure, the fall-off of the bubble intensity in the lowest energy bins is robust. \reffig{icsvssynchro} shows the same analysis for 500-900 GeV electrons with two different templates for the disk IC emission, for the bubble interior and shell separately, and for the north and south bubbles. In all cases the same cut-off in the spectrum below 1 GeV is observed.

In this case, while the lowest energy bin in the gamma-rays is better fitted, the lack of low-energy CRs means that synchrotron can contribute to the \WMAP\ haze only at the sub-percent level, unless the magnetic field in the inner galaxy is extreme (and even then the spectrum does not reproduce the observations). In this case an alternate explanation for the \WMAP\ haze would need to be considered. As suggested initially in \cite{Finkbeiner:2003im}, the \WMAP\ haze could originate from free-free emission (thermal bremsstrahlung), and this would explain the lack of a clear haze signal in \WMAP\ polarization maps. However, the spectrum of the haze is somewhat softer than generally expected from free-free emission; the gas temperature required is also thermally unstable, requiring a significant energy injection ($\sim 10^{54-55}$ ergs) to maintain its temperature \citep{Hooper:2007kb, McQuinn:2010}. 


\subsection{Gamma-ray power and $e^-$ cosmic ray density}

In order to estimate the total gamma-ray power emitted by the bubbles,
we must estimate the surface brightness integrated over energy, the
solid angle subtended, and the distance (suitably averaged).  From \reffig{bubblespecdisk} we take the intensity to be $E^2 dN/dE = 3\times 10^{-7} \gevflux$ from $1-100$ GeV, integrating to 1.4$\times 10^{-6}\gevflux$.  The bubble template used in our analysis (\reffig{bubbles}) subtends 0.808 sr, yielding a total bubble flux of 1.13$\times 10^{-6}$ GeV\,cm$^{-2}$s$^{-1}$.  To obtain an average distance for the emission, we
approximate the bubbles as 2 spheres centered at $b=\pm 28\degree$, and directly above and below the Galactic center.
For a Sun-GC distance of 8.5 kpc, this implies a distance of 9.6 kpc,
and a total power (both bubbles) in the $1-100$ GeV band of 2.5$\times
10^{40}$ GeV/s or $4.0\times 10^{37}$ erg/s, which is $\sim 5\%$ of the total Galactic gamma-ray luminosity between $0.1-100$ GeV \citep{strong:2010}.

The electron cosmic-ray density in the bubbles required to generate the observed gamma rays, at any given energy, depends strongly on the assumed electron spectrum. However, typically the required values are comparable to the locally measured electron CR density. For example, for the model in the first panel of \reffig{icslowEcut} ($dN/dE \propto E^{-2}$ for 500 GeV $\le E \le$ 700 GeV), the inferred bubble electron density is $\sim 10 \times$ greater than the local electron density (as measured by \Fermi) at an energy of 500 GeV. For a representative model from the first panel of \reffig{icsvssynchro} ($dN/dE \propto E^{-2.3}$ for 0.1 GeV $\le E \le$ 1000 GeV, with a 10 $\mu G$ magnetic field generating the WMAP Haze via synchrotron), at 500 GeV the bubble electron density is a factor $\sim 2 \times$ greater than the local density.

\section{Interpretation}
\label{sec:Interp}

As discussed in \cite{fermihaze}, the \Fermi\ bubbles seem most likely to originate from IC scattering, since the required electron CR population can also naturally generate the \WMAP\ haze as a synchrotron signal. The \rosat\ X-ray measurements suggest that the bubbles are hot and hence \emph{underdense} regions, and thus argue against the gamma rays originating from bremsstrahlung or $\pi^0$ decay. 

Even though the material in the bubbles is likely high pressure, it is also probably very hot ($\sim10^7$ K) and has lower gas density than the ambient ISM. This would explain why the \rosat\ 1.5 keV map shows a ``cavity'' of soft X-rays toward the center of the \Fermi\ bubble structure, like the X-ray cavity in galaxy clusters \citep{McNamara:2007}, especially in the north \Fermi\ bubble. Furthermore, allowing the \Fermi\ bubbles to have lower density than the ambient medium means they would experience a buoyant force moving the bubble material away from the GC (see \refsec{Buoyant} for further discussion), which may help generate the observed morphology. Because $\pi^0$ and bremsstrahlung gamma-ray emission both scale as the CR density $\times$ the gas density, an underdense region cannot be brighter unless the cosmic ray densities are greatly increased to compensate; for protons, in particular, the propagation lengths are great enough that a proton overdensity cannot reasonably explain the sharp bubble edges observed in the data, if the bubbles are in a steady state. 

If the bubbles are expanding rapidly and highly accelerated protons responsible for the gamma-ray emission are trapped behind shock fronts, then sharp edges for the \Fermi\ bubbles could occur naturally. However, in the presence of such a shock, electrons would also be accelerated, and would generally produce more gamma-rays than the protons via ICS (since the cooling time for electron CRs is much shorter than the cooling time for proton CRs of comparable energy).

It might be thought that the presence of a bright X-ray edge could lead to a sharp edge in the gamma-ray signal, via IC scattering of electron CRs on the X-ray photons. In the Thomson limit, the energy of IC scattered photons is of order $(\Gamma_e/2) E_e$, with $\Gamma_e = 4 E_e E_\gamma / m_e^2$ (where $E_\gamma$ and $E_e$ are the initial photon and electron energies, respectively, and $m_e$ is the electron mass), and the scattering cross section is independent of the initial electron and photon energies. Thus a higher-energy photon population, leading to a larger value of $\Gamma_e$, allows IC gamma-rays at a given energy to originate from lower-energy electrons, which are much more abundant for typical electron spectra with $dN/dE \sim E^{-\gamma}$, $\gamma \gtrsim 2$. However, in the Klein-Nishina regime where $\Gamma_e \gtrsim 1$ this picture changes: the energy of scattered photons is determined mostly by the energy of the initial electron, and the cross section scales as $1/\Gamma_e$. Scatterings of $\sim 50$ GeV electrons already produce $\sim 10$ GeV gamma rays; when compared to the scattering of 10 GeV electrons on X-rays in the extreme KN limit, the KN suppression of the cross section in the latter case more than counteracts the greater abundance of $\sim 10$ GeV electrons, unless the electron spectrum is very soft (which is inconsistent with the observed signal).

Thus hard UV or X-rays in the bubbles, which might be naturally expected in a high-temperature region, would \emph{not} make the IC spectrum harder, and IC scattering on these photons is subdominant to IC scattering on the usual ISRF for the electron energies in question, unless the X-ray photon number density is much \emph{greater} than the starlight photon number density. Furthermore, there is no reason to think the bubbles contains more $\lesssim1$ eV photons, at least not in a region with a well defined spatial edge. Thus the sharp edges of the \Fermi\ bubbles, and the non-uniformity in the emissivity, most likely arise from the electron CR density rather than the photon density. The presence of similar sharp edges in the \WMAP\ haze (at $|b| \le 30\degree$) supports this hypothesis, if the \WMAP\ haze is attributed to synchrotron radiation from the electron CRs. 

Similarly, the elongated shape of the \Fermi\ bubble structures perpendicular to the Galactic plane suggests that the electron CR distribution itself is extended perpendicular to the plane. The \Fermi\ bubble morphology is a strong argument against the possibility that the \WMAP\ haze originates from a disk-like electron distribution with significant longitudinal variation of the magnetic field, as suggested by \cite{Kaplinghat:2009ix}.

\begin{figure*}[ht]
\begin{center}
\includegraphics[width=0.8\textwidth]{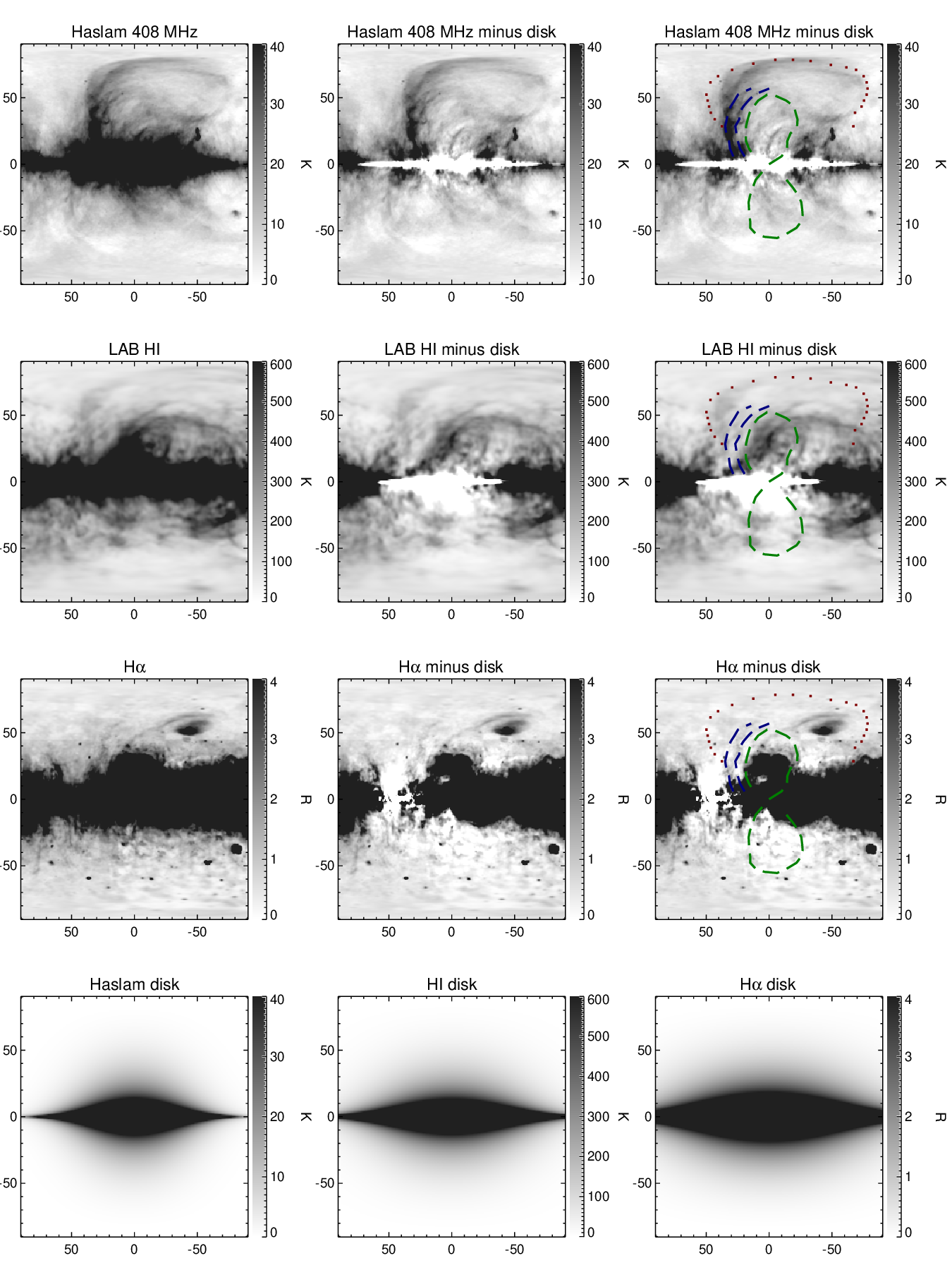}
\end{center}
\caption{
\Fermi\ bubble features in other maps. \emph{Top row:} The \emph{left} panel shows the half sky Haslam 408 MHz map \citep{1982A&AS...47....1H} with $-90\degree < \ell < 90\degree$, the \emph{middle} panel subtracts a simple geometric disk template (shown in the \emph{bottom left} panel) to better reveal the structures deeper into the Galactic plane. The \emph{right} panel is the same as the \emph{middle} panel but overplotted with the \Fermi\ bubbles, the \emph{northern arc}, and the \emph{Loop I} features identified from the $1-5$ GeV \Fermi\ gamma-ray map (see \reffig{fullsky}). The \emph{Loop I} feature (\emph{red dotted line}) align with the extended diffuse features in the Haslam 408 MHz synchrotron map (known as \emph{North Polar Spur}). The inner and outer edges of the \emph{northern arc} (\emph{dashed blue lines}) overlap with two arcs in the Haslam synchrotron map. However, the \Fermi\ bubbles have no apparent counterparts in this map. \emph{Second row:} The same as the \emph{top row}, but for the Leiden/Argentine/Bonn (LAB) Survey of Galactic \HI\ \citep{Kalberla:2005}. The \emph{middle} panel  subtracts a simple disk template shown in the \emph{bottom middle} panel to better reveal structures towards the GC. No apparent features have been identified that correlate with the \Fermi\ bubbles and other features in $1-5$ GeV \Fermi\ map (there may be some faint filaments morphologically tracing the gamma-ray features). \emph{Third row:} The same as the \emph{top row} but for the H$\alpha$ map \citep{Finkbeiner:2003}. The \emph{middle} panel subtracts a simple disk template shown in the \emph{bottom right} panel to reveal more structures in the inner Galaxy. No corresponding features have been identified morphologically similar to the structures in the $1-5$ GeV \Fermi\ gamma-ray maps (color line in the \emph{right} panel).
}
\label{fig:HasLabHalpha}
\end{figure*}

The limb brightening of the X-rays in the \rosat\ data (as shown in
\reffig{rosat}), and the flat intensity profile of the \Fermi\ bubbles,
suggest the presence of a shell or shock, with increased electron CR
density, coinciding with a hot thermal plasma.  If the ambient medium
several kpc above and below the GC were neutral, then bubbles of ionized
gas could produce a void in the \HI\ map (\reffig{HasLabHalpha}).
We see no evidence for features aligned with the bubbles in
these maps, suggesting that the \HI\ map in this part of the sky is
dominated by disk emission, and has nothing to do with the bubbles. If
the bubbles are in a static state, the bubble edges should have lower
temperature than the bubble interior and thus higher gas density,
although shocks or MHD turbulence might lead to higher temperatures at
the bubble wall. The X-rays in \rosat\ may be thermal bremsstrahlung
emission, so the emissivity is proportional to the thermal electron
density $\times$ ion density. They could also arise from charge exchange
reactions occurring when the high-speed gas in the bubbles collides with
the denser gas at the bubble edge \citep[see][and reference
therein]{Snowden:2009}; this mechanism could explain the pronounced limb
brightening of the X-rays. As an alternative explanation, the \rosat\
X-ray feature might be synchrotron emission from \emph{very high} energy
electron CRs. Typically, though, one needs $\sim50$ TeV ($\sim5$ TeV)
electrons with $\sim10$ $\mu$G (1 mG) magnetic field to produce $\sim1$
keV synchrotron photons.

The \Fermi\ bubble features do \emph{not} appear to be associated with \emph{Loop I}, a giant radio loop spanning over 100 degrees \citep{Large:1962}, which is thought to be generated from the local \emph{Sco-Cen} OB association. Detections of \emph{Loop I} in high-energy gamma-rays have been claimed by \cite{Bhat:1985} and also recently by \Fermi\ \citep{Casandjian:2009}; we have also discussed its presence in this work (see \refsec{templatefit}). 

The \emph{Loop I} gamma-rays may be the IC counterpart of the synchrotron emission seen in the Haslam 408 MHz map, although some of the emission might be $\pi^0$ gammas associated with \HI\ (\reffig{HasLabHalpha}). We compare structures identified from the \Fermi\ $1-5$ GeV maps with \emph{Loop I} features in the Haslam 408 MHz map in the \emph{top row} of \reffig{bubble_compare}, and see that the \Fermi\ bubbles are spatially distinct from the arcs associated with \emph{Loop I}; as we have shown in \reffig{bubblespecmore}, the \emph{Loop I} correlated emission also has a softer spectrum than the \Fermi\ bubble emission. The \emph{Loop I} feature in the \rosat\ map similarly has a softer spectrum than the limb-brightened X-ray bubble edges: as shown in \reffig{rosatclean}, when a low-energy map is subtracted from a higher-energy map in such a way that \emph{Loop I} vanishes, the bubble edges remain bright. We also see additional shell structures which follow the \Fermi\ bubble edges and the \emph{northern arc} in the Haslam 408 MHz map (\emph{top row} of \reffig{HasLabHalpha}). 

The \Fermi\ bubbles are morphologically and spectrally distinct from both the $\pi^0$ emission and the IC and bremsstrahlung emission from the disk electrons. As we have shown in \reffig{bubblespecdisk} to \reffig{bubblespechaslam}, the \Fermi\ bubbles have a distinctly hard spectrum, $dN_\gamma/dE \sim E^{-2}$, with no evidence of spatial variation across the bubbles. As shown in \reffig{icsvssynchro}, an electron population with $dN_e/dE \sim E^{-2-2.5}$ is required to produce these gamma rays by IC scattering: this is comparable to the spectrum of electrons accelerated by supernova shocks or polar cap acceleration \citep{Biermann:2010}. However, diffusive propagation and cooling would be expected to soften the spectrum, making it difficult to explain the \Fermi\ bubbles by IC scattering from a steady-state population of these electrons (a single brief injection of electrons with $dN/dE \sim E^{-2}$ could generate a sufficiently hard spectrum for the bubbles \emph{if} there was a mechanism to transport them throughout the bubble without significant cooling). The facts strongly suggest that a distinct electron component with a harder spectrum than steady-state SNR generated electrons is responsible for the \Fermi\ bubbles and associated signals in the \WMAP\ and \rosat\ data.

It has been suggested that a large population of faint millisecond pulsars (MSPs) in the Milky Way halo could contribute to the \WMAP\ and \Fermi\ haze signals, via both pulsed gamma-ray emission and $e^+ e^-$ production \citep{Malyshev:2010xc}. With a halo population of $3 \times 10^4$ MSPs, roughly half of the spin-down power going into $e^+ e^-$ pairs and $\sim 10\%$ going to pulsed gamma-rays, consistency with both the \WMAP\ data and the first-year \Fermi\ photon data is possible; however, this model does not immediately explain either the rather sharp edge in the distribution of gamma-ray emission, or the features in the \rosat\ X-ray data (the same can of course be said for models which generate the hazes via dark matter annihilation or decay). Other attempts to explain the \WMAP\ haze with pulsars have generally employed a disk population of pulsars, either peaking in the GC or peaking at small Galactocentric radius but going to zero in the GC \citep{Kaplinghat:2009ix, Harding:2009ye}; such models have difficulty explaining the spherical morphology of the haze. Furthermore, as pointed out by \cite{McQuinn:2010}, the regression of the 408 MHz Haslam map should remove much of the contribution from young pulsars in the \WMAP\ data, since young pulsars have a similar spatial distribution to supernovae.

\begin{figure*}[ht]
\begin{center}
\includegraphics[width=.8\textwidth]{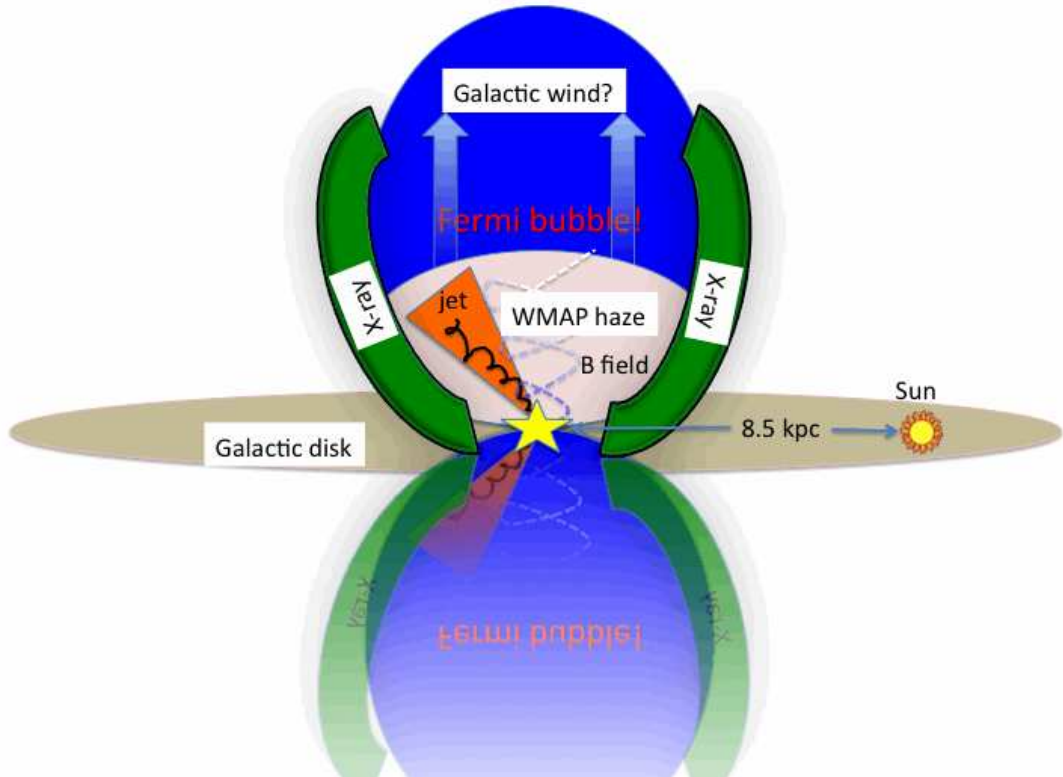}
\end{center}
\caption{
A cartoon picture to summarize the observations of the \Fermi\ bubble structures. Two blue bubbles symmetric to the Galactic disk indicate the geometry of the gamma-ray bubbles observed by the \Fermi-LAT. Morphologically, we see corresponding features in \rosat\ soft X-ray maps, shown as green arcs embracing the bubbles. The \WMAP\ haze shares the same edges as the \Fermi\ bubbles (the pink egg inside the blue bubbles) with smaller extension in latitude. These related structures may have the same physical origin: past AGN activities or a nuclear starburst in the GC (the yellow star).
}
\label{fig:stack}
\end{figure*}

How could the electron CRs possess the same hard spectrum everywhere within the bubble and extend up to 10 kpc, while experiencing a steep fall-off at the bubble edge? A large population of CRs might be entrained in large scale Galactic outflows from the GC and enrich the bubbles (see more discussion in \refsec{starburst}). CRs could be produced along with jets, or shock accelerated CRs from magnetic reconnection inside the bubble or near its surface (see more discussion in \refsec{CR}). However, it is challenging to produce a flat intensity profile for the bubble interior with a sharp edge. The ambient gas should be compressed to a higher density on the shell by shocks (probably also enhancing the magnetic field), and brighter synchrotron emission on the shell would then be expected, but the haze emission observed in \WMAP\ is not limb-brightened and shows no evidence for a shell of finite thickness (although we do see shell structure in the X-rays). A cartoon picture summarizing the morphology of the \Fermi\ bubbles and associated signals at other wavelengths is shown in \reffig{stack}.


\section{The origin of the bubble structure}
\label{sec:explanation}

As we have shown in \refsec{bubbles}, the bilobular \Fermi\ bubble structures are apparently well centered on $\ell=0\degree$, and symmetric with respect to reflection across the Galactic plane (\emph{bottom row} in \reffig{bubbles}). The bubbles extend down to the plane, where they appear even closer to $\ell=0\degree$.  This alignment and north-south symmetry are unlikely unless the bubbles originate from the GC, and motivate explanations involving Massive Black Hole (MBH) activity (\refsec{BH}) or recent starbursts towards the GC (\refsec{starburst}). We note that Sco X-1 is approximately centered on the north bubble at the present, but this appears to be a coincidence. Given the similarity between the northern and southern bubbles and the absence of any similar feature in the south, we believe the \Fermi\ bubbles have a GC origin.

While the origin of the \Fermi\ bubbles is unknown, a rough estimate can be made for their age and the total energy required (although the latter quantity depends linearly on the gas density in the bubbles, which is poorly constrained). From the \rosat\ data, we envisage the bubbles as hot low density ($n\sim 10^{-2} \cm^{-3}$) cavities filled with $\sim 2$ keV gas, with (from the \Fermi\ data) height $\sim 10$ kpc, expanding at velocity $v \lesssim 10^3$ km/s: thus we estimate the energy of the \Fermi\ bubbles to be about $10^{54-55}$ erg, with an age of $\sim10^{7} (v/1000\,\kms)^{-1}$ yr.

Energetic galactic outflows are common phenomena which have been found in both nearby and high redshift galaxies \citep{Veilleux:ARAA}. Galactic winds are believed to have significant impact on galaxy formation, morphology, and their environments. The main sources of the energy are stellar winds, supernovae, and/or active galactic nuclei (AGN); CRs and magnetic field pressure can also help drive galactic outflows \citep{Everett:2010,Socrates:2008}. 

In this section we present some ideas for past activity in the GC that could help to generate the shape of the \Fermi\ gamma-ray bubbles and the associated signals at other wavelengths. \Refsec{GC} presents and discusses evidence of past GC activity. \Refsec{BH} discusses the hypothesis that outflows from a black hole accretion event, including possible AGN jets, could form the \Fermi\ bubbles. \Refsec{starburst} focuses on the possibility of a previous starburst phase of the Milky Way, with the \Fermi\ bubbles being inflated by the subsequent energetic Galactic wind. 

\subsection{Observational Evidence of Previous GC Activity}
\label{sec:GC}

Observations across the electromagnetic spectrum have provided
constraints on dynamics and evolution of the central gas and stellar
populations, and current and prior accretion activity of the central
MBH. We highlight some of the evidence of previous activities
towards the GC, which may relate to the production of the \Fermi\
bubbles.

\paragraph{X-ray Reflection Nebulae in the GC}
There are indications of previous GC activity from X-ray echoes and time variability of reflected X-ray lines from cold iron atoms in molecular clouds around \sgras\ including Sgr B1 and B2, Sgr C, and M0.11-0.11 \citep{Sunyaev:1993,Sunyaev:1998}. The changes in the intensity, spectrum, and morphology of the fluorescent iron nebulae near the GC, observed by \asca\ and \emph{INTEGRAL} are likely due to reflected X-rays from previous activity of \sgras\ with high luminosity $\sim$300 yr ago. The luminosity is $\sim$1.5$\times$10$^{39}$ erg s$^{-1}$ in $2-200$ keV with a power law spectrum $dN/dE \propto E^{-\gamma}$ with $\gamma$=1.8$\pm$0.2 \citep{Ponti:2010,Revnivtsev:2004,Nobukawa:2008}. The changes in the intensities and morphologies of hard X-ray nebulosities on parsec scales have been discovered \citep{Muno:2007}.



\paragraph{Outflows: Galactic Center Lobe and Expanding Molecular Ring}

\cite{2000ApJ...540..224S} and \cite{2003ApJ...582..246B} have previously noted the presence of the extended bipolar
structure in the \emph{ROSAT} data, and have attributed it to a
large-scale bipolar wind, powered by a starburst in the
GC. In this picture, overpressured bubbles rise and
expand adiabatically away from the injection region in the Galactic
plane, driving shocks into the surrounding gas by ram
pressure. Reflection of the shocks increases the density and
temperature of the post-shocked gas, which has been suggested to have $T \approx 5 \times 10^6$ K and $n \approx 0.1$ cm$^{-3}$ at 2 kpc above the plane \citep{2003ApJ...582..246B}. Free-free emission in the resulting high-temperature plasma produces the observed X-ray signals. The Galactic Center Lobe (GCL) on \emph{degree} scale has estimated total kinetic energy $\sim10^{55}$ erg and a dynamical time scale of $\sim10^6$ yr. The size, energy, and time scales are similar to those of the expanding molecular ring (EMR) around the GC \citep{Kaifu:1972,Scoville:1972,Totani:2006}. \cite{2000ApJ...540..224S} interpreted the North Polar Spur (NPS) with tens of degrees scale to be an outflow from the GC with energy scale of $\sim10^{55-56}$ erg and time scale of $\sim10^7$ yr. \cite{Totani:2006} suggested that all these outflows can be attributed to the past high activity of \sgras\ of a duration of $\sim10^7$ yr, comparable to the reasonable estimation of the lifetimes of AGNs.

\paragraph{Diffuse X-ray Emission}

\cite{Muno:2004} studied the diffuse X-ray emission within $\sim $ 20 pc of the GC in detail using \emph{Chandra} observations. The hard component plasma with $kT \sim 8$ keV is spatially uniform and correlated with a softer component with $kT \sim 0.8$ keV. Neither supernova remnants nor W-R/O stars are observed to produce thermal plasma hotter than $kT \sim 3$ keV. A $kT \sim 8$ keV plasma would be too hot to be confined to the GC and would form a plasma wind. A large amount of energy input $\sim 10^{40}$ erg/s is required to sustain such hot plasma. If the hot plasma is truly diffused, the required power is too large to be explained by supernova explosions and the origin of this hot plasma, and might be explained as a result of shock heating by the wind or AGN activities. Similar diffuse hard X-ray emission has been detected from the starburst galaxy M82 \citep{Strickland:2007}. However, \cite{Revn:2009} have resolved $\sim80$ percent of the hard diffuse gas into faint point sources.

\paragraph{Bipolar Hard X-ray in the GC}

\emph{Chandra} observations show the morphology of hot gas with a few keV seems to be bipolar with each lobe extending to $\sim$10 pc \citep[see][and reference therein]{Markoff:2010}. Three explanations have been suggested for the origin of the lobes: thermal wind from the central cluster of massive young stars, steady outflows from \sgras, or repeated episodic outbursts from \sgras \citep{Markoff:2010}. For the collective stellar wind interpretation, it is unclear why the lobes only extend up to 10 pc which has a estimated flow time $\sim3\times10^4$ yr, whereas the star cluster is $\sim6$ millon years old. Moreover, discrete blobs have been found within the lobes, and quasi-continuous winds are hard pressed to explain the origin of the blobs. The possibility of a transient jet-like feature is intriguing. The jets can be produced by accreting the debris of tidal disruption stars (see \refsec{Jets} for more discussion).

\paragraph{OB Stellar Disk}

There are two young star disks that have been identified in the central parsec of the GC \citep{Paumard:2006}. \emph{In situ} star formation from dense gas accretion disks is favored over the inspiraling star cluster scenario \citep[see e.g.][]{Bartko:2009}. The gas disks could be formed as a consequence of a large interstellar cloud captured by the central MBH, which then cooled and fragmented to form stars. Interestingly, the two star formation events happened near-simultaneously about $6\pm2$ Myr ago and the two disks are coeval to within $\sim1$ Myr \citep{Paumard:2006} and little activity has occurred since. The two young massive star clusters \emph{Arches} and \emph{Quintuplet} in the central 50 pc, with similar stellar mass, content and mass functions, were formed $\sim10^7$ yr ago. It has been suggested in \cite{Paumard:2006} that a global event may cause an increase in the rate of star formation, such as a passing satellite galaxy that enhanced the clouds' collision frequency and may also provide the gas for active star formation.

The unique characteristics of stellar clusters towards the GC have been used to explain the origin of the magnetized nonthermal radio filaments, threads, and streaks \citep{LaRosa:2004,Yusef-Zadeh:2004}. Collective winds of massive W-R and OB stars within such a dense stellar environment can produce terminal shocks that accelerate relativistic particles. The abundance and characteristics of these nonthermal radio filaments within the inner 2 degrees of the GC region can be evidence of an earlier starburst \citep{Rosner:1996}. \cite{Yusef-Zadeh:2004a} propose a jet model in which the characteristics common to both protostellar and extragalactic jets are used to explain the origin of nonthermal filaments in the Galactic center region.




\subsection{Outflow from Black Hole Accretion Events}
\label{sec:BH}

The central MBH in our Milky Way, with an estimated mass of M$_{BH}\sim$4~$\times10^6$ M$_{\odot}$ \citep[e.g.][]{Ghez:2008, Gillessen:2009}, is currently quiescent, radiating at about 8 orders of magnitude lower than the Eddington luminosity ($L_{Edd} \sim10^{44}$ erg/s). The X-rays of \sgras\ is weak and thermal, in contrast to the hard nonthermal power law typically observed in most low-luminosity active galactic nuclei \citep[LLAGN; see e.g.][]{Ho:2008}. Fast X-ray flaring of \sgras\ via nonthermal processes has been discovered by \emph{Chandra} ACIS observation \citep{Baganoff:2003}. The observed submillimeter/IR bump is seen to flare simultaneously with the X-rays. Due to the short timescale of the flares and the lack of evidence of any standard thin accretion disk emission \citep{Shakura:1973}, a magnetic origin of the flares has been suggested, either from synchrotron or synchrotron-self Compton emission \citep{Eckart:2004,yusef-Zadeh:2006,Dodds-Eden:2009}. Both radiative inefficient accretion flow (RIAF) models and outflow-dominated models have been invented \citep{Falcke:2000,Blandford:1999,Yuan:2002}.

Clearly the MBH has not always been so underluminous: it may have experienced a long active state in the past few million or tens of million years through one or more accretion events, driving jets out of the disk, shocking the ambient material, producing both gamma-rays and CRs, and appearing more similar to normal low-luminosity AGN. If the MBH were radiating at the Eddington luminosity, it would take only $\sim10^{3-4}$ years to reach the estimated energy of the \Fermi\ bubbles; for a percent level accretion rate ($\sim10^{42}$ erg/s), it would take $\sim10^{5-6}$ yr, comparable to the estimated cooling time of the electron CRs.

\subsubsection{Scenarios for MBH Jet Formation in the GC}
\label{sec:Jets}

Currently, the X-ray flares are the only unambiguous AGN-like activity that \sgras\ displays \citep{Markoff:2010}. However, the synchrotron radio emission is comparable to typical LLAGN with compact jets in terms of spectral characteristics. 
It is well known that jets associated with accretion disks surrounding black holes can efficiently generate high energy particles. BL Lacs can form relativistic jets to produce TeV gamma rays, as can microquasars. However, we know through multiwavelength observations that the central MBH in our Milky Way has extraordinarily low bolometric luminosity of $\sim10^{36}$ ergs/s, and so is currently in its quiescent dim state, and no jets toward the GC have been physically resolved. Detailed examinations of the GCL have shown that the gas shell is deep into the disk, and do not support a jet origin for that structure \citep{Law:2010}. However, weak jets consistent with the spectrum of \sgras\ can be easily hidden by the blurring of their photosphere \citep{Markoff:2007}. 

The jets related to the energetic accretion in the GC can go in any direction; they need not be aligned with the minor axis of the Galaxy. This mechanism does not obviously provide a natural explanation for the north-south symmetry of the \Fermi\ bubbles, and the relatively flat gamma-ray intensity inside the bubbles. Also, there is as yet no conclusive evidence for the presence of jets toward the GC \citep{Muno:2008}. However, CR rich wide jets might have existed in the past, and the \Fermi\ bubbles might have been inflated by the jets in a relatively short time scale without significant cooling. Although the past AGN phase could be the primary heating source of the \Fermi\ bubbles, the low density and momentum associated with the jets do not readily distribute the thermal energy isotropically, making it harder to explain the morphology and flat intensity of the \Fermi\ bubbles. If a starburst phase in the GC coincided with the energetic jets, SNe in the starburst might provide a large injection of momentum and turbulence, which could help isotropize the energy distribution.  

\paragraph{Accretion of Stars}
One way to form jets in the GC is for the MBH to tidally disrupt and swallow stars in the nuclear star cluster \citep{Hills:1975,Rees:1988}. The typical picture is that when a star trajectory is sufficiently close to the MBH, the star is captured and disrupted by tidal forces, and after about an orbital timescale, a transient accretion disk forms from the debris of the disrupted star. The capture rate of the GC MBH has been estimated at $\sim4.8\times10^{-5} \yr^{-1}$ for main-sequence stars and $\sim8.5\times10^{-6} \yr^{-1}$ for red giant stars. \sgras\ could temporarily behave like an AGN and produce a powerful jet by accreting the debris of such stars, which may be ejected from surrounding molecular disks. If a 50 solar mass star is captured by the MBH in the GC, it gives an energy in relativistic protons as high as $\sim10^{54-55}$ ergs on a very short timescale ($\sim10^{3-4}$ yr), at a rate of about $\sim10^{43}$ ergs/s.



\paragraph{Accretion of ISM}

Quasi-periodic starbursts in the GC have been recently suggested as a result of the interactions between the stellar bar and interstellar gas \citep{Stark:2004}. In this scenario, gas is driven by bar dynamics toward the inner Lindblad resonance, and accumulates in a dense gas ring at 150 pc until the critical density threshold is reached. Then giant clouds can be formed on a short timescale, and move toward the center by dynamical friction. The timescale for this cycle to repeat is highly uncertain but is estimated to be of order 20 Myr \citep{Stark:2004}.
 
\paragraph{Accretion of IMBH}

A single $10^4$ solar mass BH spiraling in to the GC may also trigger starbursts and change the spin of the \sgras \cite{Gualandris:2009}, thereby producing precessing jets. It has been argued that one such event happens approximately every $10^7$ years in order to create a core of old stars in the GC, of radius 0.1 pc. Chandra has detected more than 2000 hard X-ray ($2-10$ keV) point sources within 23 pc of the GC \citep{Muno:2003}. Some of them may harbor intermediate mass black holes. However, to our knowledge, no evidence of a collimated outflow has been found from the GC. There may be aligned, VLBI-scale ($\sim$1 pc in the GC) radio knots that have not been discovered.

\paragraph{Models of Past Enhanced \sgras Accretion}

 \cite{Totani:2006} suggested a RIAF model of \sgras\ to explain both the past high X-ray luminosity and the kinetic luminosity of the outflows inferred from observations. The required boost factor of the accretion rate $\sim10^{3-4}$ with a time scale of $\sim10^7$ yr is naturally expected in the model. The induced outflow is energetic enough to support the hot ($\sim$8 keV) plasma halo towards the GC. The sudden destruction of the accretion flow of \sgras\ is caused by the remnant Sgr A East passing through the MBH in the past $\sim10^2$ yr. Such a model is claimed by \cite{Totani:2006} to better explain the positron production which is required to generate the 511 keV line emission towards the Galactic bulge observed by \emph{INTEGRAL/SPI} \citep{Weidenspointner:2008}.

\paragraph{Galactic Jets in Starburst Galaxies}

In other galaxies, powerful AGN radio jets interacting strongly with the hot gas have been observed \citep{McNamara:2007}. In relatively radio-quiet galaxies such as NGC 4636, NGC 708, and NGC 4472, \citep[see e.g.][for a summary]{Wang:2005}, faint X-ray ``ghost'' cavities appear without corresponding radio lobes. It has been suggested that capture of red giant stars and accretion onto the central MBH can power jets/outflows with typical energy $\sim10^{56}$ ergs, which is the required energy to form the observed cavities. 

Interferometric monitoring of the Seyfert galaxy NGC 3079 has found evidence of radio jet components undergoing compression by collision with the clumpy ISM, within a few pc of the central engine. This result supports the idea that the kpc-scale superbubble originates from the spread of the momentum of jets impeded from propagating freely. The generalization of this scenario provides an explanation for why jets in Seyfert galaxies are not able to propagate to scales of kpc as do jets in radio-loud AGN \citep{Middelberg:2007}.


\paragraph{Precessing Jets}

Several processes may lead to jet precession, including magnetic
torques, warped discs, and gravitational torques in a binary system
\citep[see e.g.][]{Falceta:2010}. If the momenta of the MBH and the accretion disk are not perfectly aligned, the Bardeen-Petterson effect could be a likely mechanism for precession \citep{Bardeen:1975}, which will force the alignment of the disk and the MBH angular momentum.

\cite{Irwin:1988} suggested that the ``bubble'' structures that have been seen in nearby starburst galaxies could have been blown by a \emph{precessing} VLBI-scale jet, a flat-spectrum radio core was recognized in NGC 3079 \citep{Baan:1995,Irwin:1992}. Typically, narrow, relativistic and non-precessing jets can carve out a hot gas bubble by interaction with ISM and release most of their energy far from the GC. Wide jets with large opening angles are capable of transferring momentum into a larger area resulting in the inflation of fat bubbles (also for precessing AGN jets see \citealt{Sternberg:2008_1}). 


\subsubsection{Shocked Shells Driven by AGN Jet}
\label{sec:shock}

Relativistic jets dissipate their kinetic energy via interactions with the ISM. Jets in radio-loud AGN can inflate a bubble composed of decelerated jet matter which is often referred to as a cocoon. Initially, the cocoon is highly overpressured against the ambient ISM and a strong shock is driven into the ambient matter. Then a thin shell is formed around the cocoon by the compressed medium. The shells are expected to be a promising site for particle acceleration.

As a simple estimate of the dynamics of the expanding cocoon and shell, we assume the bubbles and shells are spherical, and also assume that  the ambient mass density profile has a form of a power-law given by $\rho_a(r) = \rho_{0} (r / {\rm 1kpc})^{-1.5}$, where ${\rho_0}$ is the mass density at $r=1~{\rm kpc}$. We further assume that the kinetic power of the jet, $L_{\rm j}$, is constant in time. Under these assumptions, the dynamics can be approximately described based on the model of stellar wind bubbles.
 The radius of the shock is given by 
 $R(t) \sim  6  {\rho}_{0.01}^{-2/7} L_{42}^{2/7} t_{7}^{6/7}~{\rm kpc} $, where $\rho_{0.01} = \rho_0 / 0.01 m_p ~{\rm cm}^{-3}$, $L_{42} = L_{\rm j}/10^{42}~{\rm ergs~s^{-1}}$ and $t_{7} = t / 10^7~{\rm yr}$. Taking the expected numbers for the \Fermi\ bubbles, we get a approximate estimate of the bubble size. The total internal energy stored in the shell can be expressed as $E_{\rm s} \sim 0.1\, L_{\rm j} t$, implying that roughly $10\%$ of the total energy released by the jet is deposited in the shell.



\subsubsection{Buoyant Bubbles}
\label{sec:Buoyant}

X-ray images have revealed shock fronts and giant cavities, some with bipolar structure, in the inner regions of clusters, surrounded with X-ray emitting gas. It is believed that the power in radio jets is comparable to the energy required to suppress cooling in giant elliptical galaxies or even rich clusters \citep{McNamara:2007}.

The depressions in the X-ray surface brightness of the \emph{ROSAT} map may themselves indicate the presence of empty cavities or bubbles embedded in hot gas, and could be interpreted as a signature of previous AGN feedback with hot outflows. For adiabatic or supersonic bubbles first inflated by AGN jets, once the bubble reaches pressure equilibrium with the surrounding gas, it becomes buoyant and rises, because the mass density is lower in the bubble than in its surroundings. As the bubble moves away from the GC, toward regions with even lower density and pressure, it expands. The velocity at which the bubbles rise depends on the buoyancy and the drag forces.


The ISM is in turn pushed by the rising bubble, which causes a upward displacement behind the bubble called ``drift''. This trailing fluid can give rise to filaments of cool gas. There are indeed filamentary structures in the inner Galaxy in both \HI\ and H$\alpha$ maps. Their identification could support the buoyant bubble scenario. 


\subsection{Nuclear Starburst}
\label{sec:starburst}

Another possible source of dramatic energy injection is a powerful starburst in the nucleus. Starburst induced Galactic winds are driven by the energy released by supernova explosions and stellar winds following an intense episode of star formation, which create an over-pressured cavity of hot gas. The galactic wind fluid is expected to have an initial temperature within the starburst region in the range of $10^{7-8}$ K even if it has been lightly mass loaded with cold ambient gas \citep{Chevalier:1985}. The ISM can be swept up by the mechanical energy of multiple SN explosions and stellar winds. Large-scale galactic outflows have been observed in starburst galaxies both in the local Universe and at high redshifts \citep{Veilleux:ARAA,Bland-Hawthorn:2007}. Starburst episodes near the GC have been discussed in \citep[e.g.][]{Hartmann:1995}.

\subsubsection{Morphology of Outflows in Starburst Galaxies}
\label{sec:starburstmorphology}

Starburst-driven galactic winds have been studied extensively in both multi-waveband observations and hydrodynamical simulations \citep{Strickland:2000,Veilleux:ARAA}.

AGN or starburst galaxies show bipolar outflows \citep{Gallimore:2006,Sharp:2010}. The total energy of the superwind has been estimated as $\sim 10^{55-56}$ ergs, comparable to the estimated energy of the \Fermi\ bubbles. The \emph{Spitzer} Space Telescope has found a shell-like, bipolar structure in \emph{Centaurus A}, 500 pc to the north and south of the nucleus, in the mid-infrared \citep{Quillen:2006}. The shell has been estimated to be a few million years old and its mechanical energy of $10^{53-55}$ erg depends on the expanding velocity. A small, few-thousand solar mass nuclear burst of star formation, or an AGN origin has been proposed to explain the formation of the shell. 

Recently, \cite{Westmoquette:2009} showed that ionized gas in the starburst core of M82 is dynamically complex with many overlapping expanding structures located at different radii, with compressed, cool, photo-ionized gas at the roots of the superwind outflow. Extra-planar warm $H_2$ knots and filaments extending more than $\sim$3 kpc above and below the galactic plane have also been found \citep[e.g.][]{Veilleux:2009}

NGC 253 is one of the most famous nearby starburst galaxies and is similar to our Milky Way in its overall star formation rate, except for a starburst region toward the center of the galaxy with spatial extent of a few hundred pc. A galactic wind in NGC 253 was found in H$\alpha$ \citep{McCarthy:1987}. \cite{Strickland:2000} discussed the spatial structure in detail in X-ray. The wind reaches out to $\sim$9 kpc perpendicular to the disk. Filamentary structures as part of projected conical outflow are found in H$\alpha$ and near-infrared $H_{2}$ emission. The relatively warm gas ($\sim10^4$ K) exists close to the hot gas ($\sim10^6$ K). The X-ray filaments tend to be located in inner regions compared to H$\alpha$, and are brighter where the H$\alpha$ emissions are locally weak. The separation between the H$\alpha$ and X-ray filaments is $\sim$70 pc. The spatial distributions of H$\alpha$ and X-ray indicate that the inner Galactic wind has higher temperature than the outer part. The UV emission seems to form a shell around the X-ray emission \citep{Bauer:2008}. VLBI and VLA observations of the nuclear region of NGC 253 at 22 GHz shows no detection of any compact continuum source on milliarcsecond scales, indicating no low-luminosity AGN in the central region of NGC 253. It seems that the starburst region is the most plausible explanation for the source powering the wind \citep{Brunthaler:2009}.




\subsubsection{The Mechanism of Galactic Winds}
\label{sec:wind}

The energy and momentum transfer of the Galactic wind could be dominated by high thermal and/or ram pressure. Materials have been swept-up and entrained as part of the wind, and the wind fluid comprises merged SN ejecta and massive star stellar wind material with ambient gas from the starburst region. Two popular extrinsic feedback mechanisms have been suggested: thermally driven winds powered by core-collapse supernovae and momentum-driven winds powered by starburst radiation \citep{Cox:2005}. 

On the other hand, the idea that CRs and magnetic fields can help to drive galactic winds has been known for decades \citep{Ipavich:1975,Breitschwerdt:1991}. For every core-collapse SN, about 10 percent of the energy release is converted into CRs ($\sim 10^{50}$ erg). These CRs interact with the magnetized ISM extensively and exchange momentum through Alfvenic disturbances: the characteristic mean free path for a CR proton in the starburst phase of the GC is $\sim$1 pc. The effective cross section for CR protons and nuclei interacting with the ambient gas is much higher than the Thomson cross section for electrons.  The luminosity at which CR collisions with gas balance gravity is about $10^{-6}$ of the usual Eddington luminosity \citep{Socrates:2008}. 

Momentum wind outflowing galactic supershells can also be driven by Ly$\alpha$ radiation pressure around star-forming regions \citep{Dijkstra:2009}. The supershell velocity can be accelerated to $10^{2-3}$ km/s, and it may even be able to escape from the host galaxy. The radii are predicted to be $r_{sh}=0.1-10$ kpc, with ages $t_{sh}=1-100$ Myr and energies $E_{sh}=10^{53-55}$ erg.

However, the morphology of the galactic wind in nearby starburst galaxies inferred from synchrotron, H$\alpha$, and \HI\ maps is asymmetric about both the galactic minor axis and galactic plane, which may suggest the inhomogeneous nature of the ISM. On the other hand, as we have shown, the north and south \Fermi\ bubbles are approximately symmetric with respect to the galactic plane and the minor axis of the disk. The symmetric structure of the \Fermi\ bubbles might indicate that they are not generated by subsequent interactions with ambient gas throughout the wind. 

Furthermore, the typical speed of galactic winds is about $200-300$ km/s. It takes about $5\times10^7$ yr for CR electrons to reach 10 kpc, but we have not seen any evidence of cooling in the gamma ray intensity and spectrum. We probably need a faster transport mechanism of CRs if they were generated from the GC. However, the higher the velocity of the wind which entrains the CRs, the greater the kinetic energy the wind contains. The estimated energy of the \Fermi\ bubbles only includes thermal energy; if they are actually kinetic energy dominated, then the energy requirement to form the \Fermi\ bubbles is even larger \citep{Gebauer:2009,Jokipii:1987,Lerche:1982}.

The estimated supernova rate in the NGC 253 starburst region is about 0.1/yr \citep{Engelbracht:1998}. Assuming each supernova explosion releases $10^{51}$ erg, and 10 percent of the energy is transferred to heating the \Fermi\ bubbles, this gives a rate of energy injection comparable to $\sim 10^{42}$ erg/s. The star formation activity in NGC 253 has been underway for about $20-30$ Myr, and is considered to be in a steady state for the CR transport, presumably with a smaller time scale \citep{Engelbracht:1998}.  

It is possible that although the center of our Milky Way is currently in its quiet phase, it was recently (in the past $10^7$ yr) in a starburst phase similar to NGC 253 \cite{Heesen:2009}. To our knowledge, however, there is no evidence of massive supernova explosions ($\sim10^{4-5}$) in the past $\sim10^{7}$ yr towards the GC, and no apparent Galactic wind features have been found in H$\alpha$, indicating no strong recent ($\sim10^{4}$ yr) star formation activity.

\cite{Everett:2008,Everett:2010} compared the synchrotron and soft X-ray emission from large-scale galactic wind models to \rosat\ and Haslam 408 MHz maps. They show that a CR and thermally-driven wind could consistently fit the observations and constrain the launching conditions of the wind, including the launching region and magnetic field strength. The comparison of the gamma-ray prediction of the wind model with \Fermi-LAT diffuse emission, especially to the \Fermi\ bubbles might have important implications for the CR driven wind of our Milky Way.

\subsubsection{Cosmic Rays from a Starburst}
\label{sec:starburstCR}

A central starburst might also generate the increased population of electron CRs in the GeV-TeV energy range required to produce the gamma-ray bubble signals. Starburst galaxies host a greatly increased rate of supernovae in the central region. The shocks from supernovae can merge and produce energetic galactic scale winds, and the enhanced population of supernova remnants is believed to accelerate CRs, resulting in orders of magnitude higher CR density than currently expected in the GC. It is likely that the GC exhibited comparable CR density in the past, with a starburst phase turned on by boosted formation of massive stars. As previously discussed, during this starburst phase the CR protons produced in the inner Galaxy scatter on the ISM with very short path lengths, producing gamma rays (via $\pi^0$ production and decay), and electron (and positron) CRs. Although the immediately-produced gamma-rays and high density gas and ISRF associated with the starburst phase would not be observable today, the secondary electrons might be leftover from the past active starburst phase, and could have been transported to $\sim$10 kpc by the magnetic field entrained in the Galactic winds. GeV and TeV gamma-rays have recently been detected in nearby starburst galaxies NGC 253 \citep{Acero:2009}, M 82, and Large Magellanic Cloud (LMC), by \Fermi-LAT \citep{Fermi:M82}, High Energy Stereoscopic System (H.E.S.S.) \citep{Itoh:2007} and VERITAS \citep{Karlsson:2009}, support the starburst galaxies as a rich source of high energy gamma-rays. If a transient starburst did occur in the GC of our Milky Way, and produced a large population of CRs responsible for the observed \Fermi\ bubbles, and \WMAP\ haze, what triggered and terminated the starburst phase is unclear.

If the CRs are driven by winds, the halo magnetic field can carry CRs along the field lines from the inner disk/bulge into the halo, which could help the CR electrons to reach 10kpc without significant diffusive softening of their spectrum. The vertical CR bulk velocity is typically hundreds of km/s, which is remarkably constant over the entire extent of the disk and for galaxies with different mass. In the standard picture, CRs can not stream faster than the Alfv\'{e}n speed with respect to the static frame of the magnetic field, due to the well-known streaming instability \citep{Kulsrud:1969}. Considering nearby starburst galaxies, the vertical CR bulk speed has been measured to be $v_{\rm CR}=300\pm30\,\rm km\,s^{-1}$ for NGC 253. For a typical magnetic field strength of $B\approx 15\,\rm \mu G$ and a density of the warm gas of $n\approx0.05\,\rm cm^{-3}$ leads to an Alfv\'{e}n speed of $v_{\rm  A}=B/\sqrt{4\pi\rho}\approx\, 150\,\rm km\,s^{-1}$ \citep{Heesen:2009}. The super-Alfv\'{e}nic CR bulk speed requires that the CRs and the magnetic field which is frozen into the thermal gas of the wind are advectively transported \emph{together}. The measured CR bulk speed is the superposition $\upsilon_{\rm  CR}=\upsilon_{\rm wind} + \upsilon_{\rm A}$.

In this picture, the spectrum of electrons injected in the starburst region might be harder than elsewhere in the Galaxy if the transient starburst phase led to a top-heavy mass function of stars. Magnetic turbulence in the GC could also be much stronger than in the Solar neighborhood: the stronger the turbulence, the faster CRs are transported, and the higher the transition energy at which synchrotron/ICS losses overtake diffusive losses. It has been shown that the high star formation rate per area in the GC leads to short transport times \citep{Becker:2009}. 

\subsubsection{Constraints on Recent Starburst Activity}
If the \Fermi\ bubbles were generated by previous starburst activity in the GC, we would expect to see many more supernova remnants towards the GC than have been discovered. Moreover, radioactive $^{26}$Al (half-life $\sim$7.2$\times10^5$ yr) is believed to be mainly produced by massive stars, supernovae, and novae in the Galaxy \citep{Prantzos:1996}. $^{26}$Al decays into $^{26}$Mg, emitting a gamma-ray photon of 1808.65 keV. Observations of the spectrometer (SPI) on the \emph{INTEGRAL} gamma-ray observatory seem to disfavor the starburst scenario: the Galaxy-wide core-collapse supernova rate has been estimated at $1.9 \pm 1.1$ per century from the flux of $^{26}$Al gamma-rays \citep{Diehl:2006}. Within its half-life, $^{26}$Al can only travel $\sim$0.1 kpc with the typical Galactic outflow velocity of $\sim100$ km/s. One expects strong $^{26}$Al gamma-ray emission concentrated towards the GC, with a flux comparable with the total gamma-ray flux from the disk, if the outflow was produced by a starburst \citep{Totani:2006}. However, such a strong concentration at the Galactic center is not found \citep{Prantzos:1996}, indicating that the accretion activity of \sgras\ is more plausible as the origin of the mass outflow. Future observations of the ratio of the Galactic $^{60}$Fe to $^{26}$Al may provide better constraints on the starburst scenario.


\subsection{Other Ways to Generate the Bubbles}
\label{sec:otherway}

The molecular loops in the GC could possibly be explained in terms of a buoyant rise of 
magnetic loops due to the Parker instability. For a differentially rotating, 
magnetically turbulent disk, such magnetic loops can easily be formed and rise out of the disk. The typical scale of such a loop is 1kpc \citep{machida:2009}.

\section{The Origin of the Cosmic Rays}
\label{sec:CR}
    
It is not necessary that the physical mechanism that creates the bubbles also injects the electron CRs responsible for the \Fermi\ bubbles. It is possible that the bubble structures were formed earlier and the electron CRs were injected by an alternative mechanism that then lights up the bubble structure with gamma-ray emission. In this section, we would like to separate the CR production from the bubble formation, and address the spatial origin of the CRs. 

In any case, the production mechanism should generate electron CRs inside the \Fermi\ bubbles, and also prevent them from efficiently leaving the bubbles, in order to produce the observed ``sharp edge''.  However, electrons with $\sim10^2$ GeV diffuse on the order of $1$ kpc before losing half their energy \citep{McQuinn:2010}. Higher energy electrons lose energy more rapidly and so have shorter path lengths; if the gamma-ray emission from the bubbles is dominated by IC scattering from TeV-scale electrons injected inside the bubble, then the sharp edge of the bubbles may be natural. This would in turn imply that the gamma rays observed by \Fermi\ are largely upscattered \emph{CMB} photons (starlight and far infrared photons are upscattered to much higher energies), which is advantageous for generating such a latitudinally extended IC signal. If instead $\sim 100$ GeV electrons scattering on starlight are primarily responsible, the electron CR density must increase markedly at high Galactic latitudes to compensate for the falling of the starlight density to higher latitude. However, generating a very hard CR electron spectrum extending up to $\mathcal{O}$(TeV) energy may be challenging for conventional CR production and acceleration mechanisms. These difficulties may be ameliorated if the hard gamma-ray signal from the \Fermi\ bubbles is a transient rather than a steady-state solution.



\subsection{CRs from the Galactic Center}
\label{sec:CRI}

As discussed in \refsec{explanation}, the electron CRs could be produced in the inner Galaxy by mechanisms such as OB associations \citep{Higdon:2005}, accretion events, and supernova explosions, then entrained in subsequent jets or outflows and rapidly carried up to large scales, avoiding diffusive softening of the spectrum. \cite{Breitschwerdt:1994} suggested that the all-sky soft X-ray emission can be explained by delayed recombination in a large-scale CR and thermal-gas pressure driven wind.  Such a wind model has been applied to the Milky Way which could explain the observed Galactic diffuse soft X-ray emission and synchrotron. The model indicates that the Milky Way may possess a kpc-scale wind. 

The cooling time (denoted $\tau$) of TeV-scale electron CRs can impose stringent constraints on such models. In the Thomson regime, the cooling time scales as (electron energy)$^{-1}$, leading to estimates of $\tau \sim 10^5$ years for TeV-scale electrons. However, where scattering on starlight photons is important, electron energies of $100-1000$ GeV are no longer in the Thomson regime, so the scattering cross section is suppressed and the cooling time can be longer than naively expected.

\begin{figure*}[ht]
\begin{center}
\includegraphics[width=.6\textwidth]{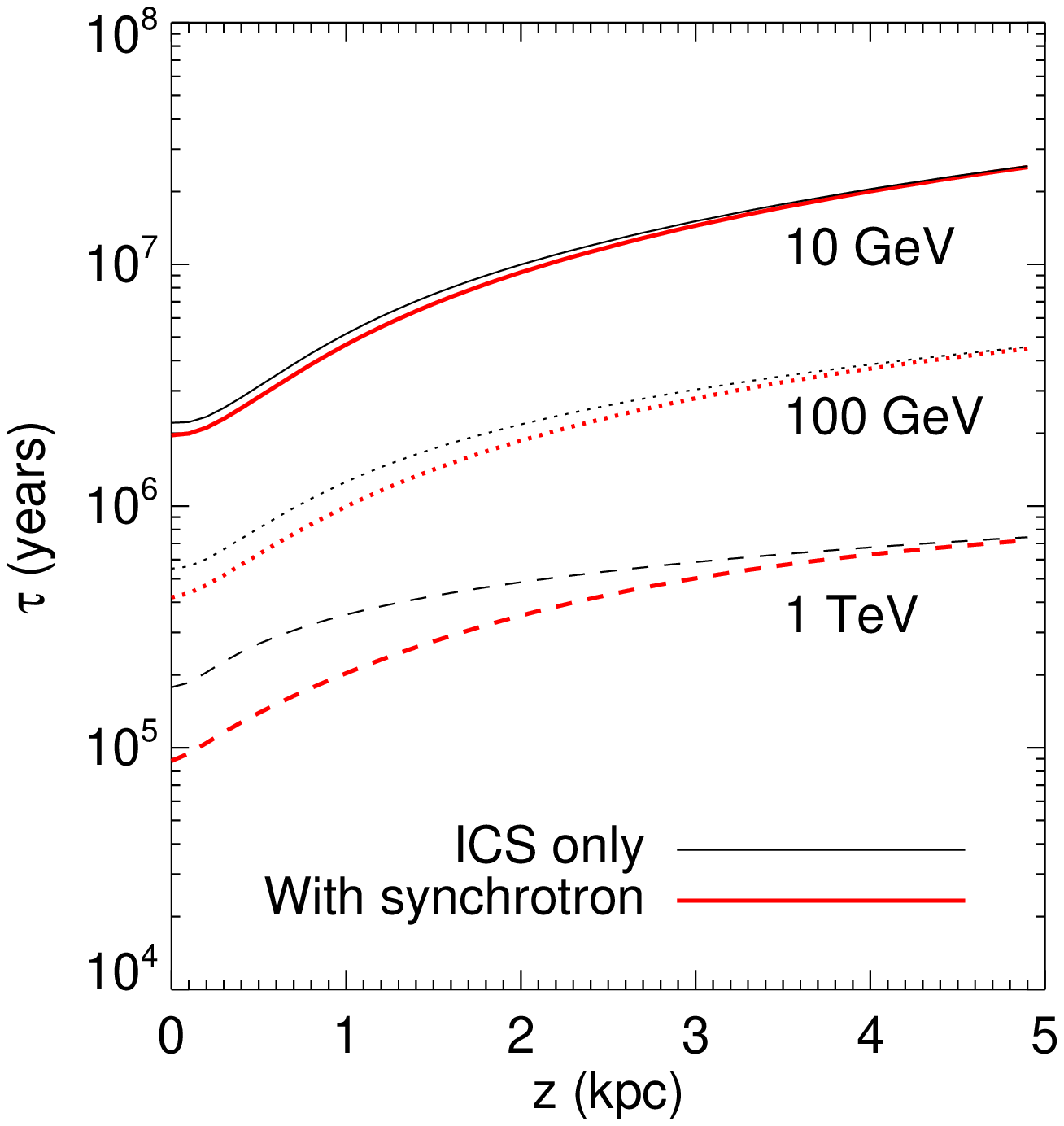}
\end{center}
\caption{The cooling time for electron CRs as a function of energy and height above the Galactic plane, for $r=0$. \emph{Thin black lines:} Synchrotron losses are neglected; equivalently, the B-field is assumed to be negligible everywhere. \emph{Thick red lines:} The cooling time calculation includes synchrotron losses in a magnetic field given by $|B| = 30 e^{-z / 2 \mathrm{kpc}} \mu\mathrm{G}$. We use the standard radiation field model from \texttt{GALPROP} version 50p, and define the cooling time $\tau = 1/(d \ln E / dt)$.}
\label{fig:electroncooling}
\end{figure*}

\reffig{electroncooling} shows the variation in cooling time (defined as $1/(d \ln E / dt)$) as a function of electron energy and height above (or below) the Galactic plane, for the standard ISRF model, both with and without the inclusion of synchrotron losses (for a simple exponential magnetic field profile)\footnote{However, note that we have treated the target photon distribution as isotropic.}. For example, at $z = 2$ kpc, the cooling time for TeV electrons is $\sim 3-4 \times 10^5$ years, rising to $\sim 7 \times 10^5$ years at $z=5$ kpc. Scatterings purely on the CMB give an upper bound on the lifetime of $\sim 5 \times 10^6\, (1\,\mathrm{TeV} / E)$ years, but even several kpc from the GC, scatterings on the infrared photons dominate at TeV-scale energies. The effect of the KN cross-section suppression at higher electron energies can be seen in the small-$z$ limit where synchrotron losses are neglected, so IC scattering of the electrons on starlight is an important contribution to the total energy loss rate.

These relatively short lifetimes, especially close to the Galactic plane, may lead to severe difficulties in propagating CR electrons from the GC out to fill the bubbles. Propagation over such large distances may also lead to significant diffusive softening of the electron spectrum, which must be reconciled with the apparent spatial uniformity of the bubbles' (gamma-ray) spectral index. With electron injection primarily at the GC there is also no obvious natural explanation for the flat projected intensity profile, which seems to require sharp increases in the CR density at the bubble walls.

\subsection{CRs from the Bubble Edge}
\label{sec:CRIII}


If the majority of the electron CRs are produced from shock acceleration within the edge of the \Fermi\ bubbles, the electron CRs in the bubble interior might be leftover CRs which undergo cooling after the shock passes through. The CRs continue to diffuse inward from the shock front while also diffusing outward; if the shock is moving faster than the electrons diffuse out, a sharp edge in the resulting \Fermi\ bubble gamma-rays is still expected. It is also possible that the CRs may be secondary electrons, produced by enhanced proton-ISM interaction in shocks (within the bubble shell), where protons could be ejected from the GC and entrained in the shocks with high gas density due to shock compression. 

We can estimate the diffusion path length of $100-1000$ GeV electrons, given the lifetimes calculated in \refsec{CRI}. We use the estimated diffusion constant from \texttt{GALPROP}, $K = 5.3 \times 10^{28}$ cm$^2$/s at a reference rigidity of $4$ GV, and take the diffusion coefficient index to be $\delta = 0.43$ following the results of \cite{Ahn:2008my}. Then the path length is given by,
\begin{equation} \sqrt{K \tau} \approx 1.4 \left(\frac{E}{1 \mathrm{TeV}} \right)^{0.43/2} \left(\frac{\tau}{10^6 \mathrm{yr}} \right)^{0.5} \mathrm{kpc}. \end{equation}
Thus we expect the diffusion scale to be small relative to the bubble size, although not negligible.

Consequently, the electrons in the interior of the \Fermi\ bubbles are unlikely to maintain a hard spectrum due to diffusion inward from the bubble walls. In this scenario, one needs to tune the electron CR distribution to get near-flat projected intensity. Although the \Fermi\ bubble gamma-rays along any line of sight include contributions from both the bubble interior and bubble shell (due to the integration along the line of sight), the flat intensity and consistent spectrum of the inner bubble and outer shell templates (\reffig{bubblespecmore}) implies that the CR spectrum in the 3D bubble interior cannot be very different from the spectrum at the 3D bubble shell with electron CRs generated \emph{in situ}. 

The edge of the \Fermi\ bubbles might contain MHD turbulent fluid with compressed gas and magnetic field. Fast magnetic reconnection rather than shocks might drive the CR acceleration. Significant magnetic reversals within the bubble shell are naturally expected, just like the heliosheath region of the solar system \citep{Drake:2006,Innes:1997}: the crossing of the termination shock by Voyager 1 and 2 may indicate the acceleration within regions of fast magnetic reconnection \citep{Lazarian:2009}. It is well known that magnetic field reversal can cause magnetic reconnection. When two magnetic flux tubes of different directions become too close, rearrangement of the magnetic field lines takes place, converting the energy of the magnetic field into energy of plasma motion and heating. Such phenomena have been investigated extensively in e.g. the solar sphere and gamma-ray bursts. In the reconnection region, the energetic particles bounce between two magnetic tubes, undergoing first-order Fermi acceleration as the dominant acceleration process \citep{Beck:1996,Biskamp:1986,Elmegreen:2004}.

On the other hand, maintaining the shape of the bubble, and preventing it from breaking out, is a non-trivial process. As the bubble rises through the ISM, it tends to fatten simply because the fluid moves faster on its sides than its front. The classical Rayleigh-Taylor instability appears at the leading edge of the bubble, as the inside density is lower than that of the ISM; Kelvin-Helmholtz instability occurs at the sides of the bubble due to discontinuities in the velocity and density. Although hydrodynamical processes may be capable of stabilizing the bubble structures against these instabilities, it has been suggested that a large-scale coherent magnetic field could help to prevent disruption. We will discuss in \refsec{CRIV} the possibility that the magnetic field can be coherent on the \Fermi\ bubble scale.












\subsection{CR from Diffuse Production in the Bubble}
\label{sec:CRIV}

The hard, spatially uniform spectrum of the \Fermi\ bubbles motivates the possibility of some diffuse injection of hard CR electrons throughout the bubble volume. Such a mechanism, if present, would solve the issues of short electron cooling times, relative to the propagation time from the GC, and the hardness of the required spectrum. A uniform diffuse injection would give rise to a centrally brightened signal, so in models of this type there would most likely need to be some other mechanism increasing the electron intensity at the bubble wall, perhaps associated with a shock there. 

This requirement for multiple mechanisms may seem unwieldy, and perhaps also unnecessary, if diffusion or cooling of electrons produced at the shock can explain the flat projected intensity profile and hard spectrum across the entire bubble. Given the large size of the bubble, however, it is unclear whether production or acceleration of electron CRs solely at the bubble walls can give rise to a sufficiently centrally bright and hard signal.

Decay or annihilation of dark matter has previously been proposed as a mechanism for diffuse injection of very hard electrons and positrons; in particular, dark matter annihilation has provided good fits to the data (at least in the inner $\sim 15-20^\circ$) in previous studies of the \WMAP\ haze \citep{Hooper:2007kb, Cholis:0811, Cholis:0802, Harding:2009ye, McQuinn:2010, Lin:2010fb, Linden:2010eu}. The IC signal from electrons produced in dark matter annihilation would naively be expected to have approximate radial symmetry about the Galactic center and fall off sharply at increasing $r$ and $z$, roughly tracing the distribution of dark matter density squared, but smoothed and broadened by diffusion of the electrons. This expectation is in conflict with the bubble morphology, with a gamma-ray distribution elongated \emph{perpendicular} to the Galactic disk, and extending to 10 kpc without much change in intensity. However, this naive picture is based on isotropic diffusion with a spatially uniform diffusion constant, in a tangled magnetic field; including the effects of an ordered magnetic field and anisotropic diffusion can lead to a much more bubble-like morphology \citep{gregprivate}.

Another way to produce CRs \emph{in situ} inside the bubble is through magnetic reconnection. Electrons could be accelerated directly, or produced as secondaries from accelerated protons. The magnetic fields in the underdense bubbles, which may be inflated by AGN outflow, may relax into an equilibrium filling the entire volume of the bubbles. The timescale depends on the magnetization and helicity of the outflow and also the properties of the ISM. The magnetic field could undergo reconnection on a short timescale, converting magnetic energy into heat. This mechanism can explain how the bubbles could move a large distance through the ISM without breaking up. The reconnection in the bubbles can also accelerate energetic particles, circumventing the problem of synchrotron emitters having a shorter lifetime than the age of the bubble they inhabit \citep{Braithwaite:2010}. The \Fermi\ bubbles might be initially highly turbulent, with a disordered magnetic field far from equilibrium (gas pressure and Lorentz force are not balanced); the kinetic energy would then be dissipated by viscosity, especially in the low density bubble.  

The time scale of the relaxation to equilibrium can be estimated \citep{Braithwaite:2010} by,
\begin{eqnarray}\label{eq:recon} 
\tau_{\rm relax} & \approx & \frac{r}{\alpha v_{\rm A}}  = \frac{r\sqrt{4\pi \rho}}{\alpha B} = \frac{r}{\alpha}\left(\frac{4\pi r^3 \rho}{6E}\right)^\frac{1}{2}\\
& \approx & 4.4\times 10^6 \left(\frac{\alpha}{0.1}\right)^{-1} \left(\frac{r}{4 {\rm kpc}}\right) \left(\frac{\rho}{{10^{-5}m_{\rm p} \rm cm}^{-3}}\right)^\frac{1}{2} \nonumber\\ && \left(\frac{B}{5 \mu {\rm G}}\right)^{-1} {\rm yr},
\end{eqnarray}
where $v_{\rm A}$, $B$ and $E$ are the Alfv\'en speed, magnetic field and energy {\it at equilibrium}. We use the estimated characteristic values for the \Fermi\ bubbles. The resulting time scale is on the order of $10^{6-7} yr$.  Furthermore, the reconnection time scale is orders of magnitude shorter than the relaxation time scale: we expect to see ongoing reconnection if equilibrium has not already been reached.



\section{Potential connections to other open questions}
\label{sec:discussion}








\subsection{The Cosmic Gamma-ray Background}

Measurement of the intensity and spectrum of cosmic gamma-ray background \citep{Fermibg:2010,Fields:2010} has suggested that instead of rare but intrinsically bright active galaxies, it is numerous but individually faint normal galaxies that comprise the bulk of the \Fermi\ gamma-ray background. This result infers a tighter correlation between cosmic star formation history and the cosmic gamma-ray background \citep{Thompson:2007, Lacki:2010}. Galactic outflows have been identified from near infrared observations at redshift $z\sim2$ \citep{Alexander:2010} indicating that such outflows are common features of ultraluminous infrared galaxies. Such outflows can entrain energetic cosmic rays to escape the galaxies, thus providing a way to contribute to the cosmic gamma-ray background at energies in the \Fermi\ range. The \Fermi\ bubbles may provide local evidence for such a scenario. 

\subsection{The Origin of Hypervelocity Stars} 

Recent surveys of hypervelocity stars (HVSs) have found 16 HVSs which are mostly B-type stars with ages $\gsim(1-2)\times 10^8$ yr \citep[see e.g.][]{Brown:2009}. These HVSs distributed in the Galactic halo are believed to originate from the GC involving interactions of stars with the MBH \citep{Hills:1988,Yu:2003}. Thus energetic energy release from the GC which generate the \Fermi\ bubbles could be dynamically related to the ejection of HVSs. Recently \cite{2010ApJ...709.1356L} have shown that the spatial distribution of the discovered HVSs is consistent with being located on two thin disk planes. The orientation of the planes are consistent with the inner/outer clockwise-rotating young stellar (CWS) disk. One possibility could be that the HVSs originate from some unknown and previously existing disk-like stellar structures. HVSs might have been ejected periodically, and related accretion events produce jets which generate the \Fermi\ bubbles.

\subsection{The Future of the \Fermi\ Bubbles}

What is the future of the \Fermi\ bubbles, are they in a ``breakout'' stage? An interesting possibility is that the \emph{northern arc} and even part of the \emph{Loop I} feature are parts of the relics of previous bubbles, and the bubble production is a periodic process. The bubbles might be fast expanding shocks which might finally expand freely into the galactic halo, thus contaminating the ISM with entrained hot gas and CRs. An intriguing possibility is that CRs and gas released from previous such bubbles to the Galactic halo may contribute to the observed diffuse X-ray and gamma-ray background. In any case, the study of \Fermi\ bubbles would have potential implications for the understanding of the \emph{feedback} mechanism from the Galaxy. 





\subsection{Missing Baryons and High-Velocity Clouds}

N-body/gas dynamical galaxy formation simulations have shown that for Milky Way like galaxies, about $70\%$ extra baryonic mass should reside around the galaxy in form of hot gas \citep{Sommer:2006}. 
Warm clouds are confined by the pressure of the ambient hot halo gas, which contains mass at least two orders of magnitude more than these warm clouds. 
The study of the \Fermi\ bubbles may provide hints of hot gas feedback from the Galaxy, the search for the missing baryons \citep{Bregman:2007} and the puzzle of high-velocity clouds. 

\subsection{Metallicity}

Although the Galactic outflow can not inject a large fraction of the ISM, a significant amount of the freshly produced metals could be channeled along the galactic wind. In ordinary photo-ionization it is difficult to make the [\ion{N}{2}]/H$\alpha$ ratio exceed about 1.0; shock ionization/excitation is plausible once [\ion{N}{2}]/H$\alpha$ is detected. [\ion{N}{2}]/H$\alpha$ has been estimated $\sim$1.5 in NGC 253, less than 1.0 in M82 where the ratio trends to increase far from the disk. 

The presence of metals in the IGM has been interpreted as the consequence of energetic metal-rich outflows from galaxies. Active star formation in the inner Galaxy may contaminates the surrounding ISM, with (periodic) Galactic winds entraining the metal-rich gas to tens of kpc. The \Fermi\ bubbles may give some hints to understanding the feedback and metallicity of the IGM.  Jets from GC in general do not imply a high metallicity, and detections of metal rich outflows may essentially constrain the energetic injection from jets or Galactic outflows from previous starburst towards the GC.





\section{Conclusions}
\label{sec:conclusion}

We have identified two large gamma-ray bubbles at $1 \la E \la 50\gev$ in \Fermi\ maps containing 1.6 years of data.  They have approximately uniform surface brightness with sharp edges, neither limb brightened nor centrally brightened, and are nearly symmetric about the Galactic plane.  The bubbles extend to $50\degree$ above and below the Galactic center, with a maximum width of about $40\degree$ in longitude.  At $|b| \la 30\degree$, these ``\Fermi\ bubbles'' have a spatial morphology similar to the \WMAP\ microwave haze \citep{Finkbeiner:2003im, Dobler:2008ww}. The \rosat\ soft X-ray 1.5 keV map also reveals hard-spectrum features that align well with the edges of the \Fermi\ bubbles. The similarities of the morphology and hard spectrum strongly suggest that the \WMAP\ haze and the \Fermi\ bubbles share a common origin.

In contrast, the \Fermi\ bubble features are not aligned with \emph{Loop I} or any other feature in the Haslam 408 MHz map; while \emph{Loop I} and other shell structures appear in the gamma rays, their spectra are softer than the bubble spectrum. Furthermore, there are no convincing features spatially correlated with the bubbles in the LAB \HI\ or H$\alpha$ maps.

To better reveal the bubble structures, we use spatial templates to regress out known emission mechanisms. To remove $\pi^0$ and bremsstrahlung gammas we use the SFD dust map, assuming that the interstellar dust traces gammas produced by cosmic ray protons ($\pi^0$ decay) and electrons (bremsstrahlung) colliding with the ISM. To trace the gamma rays from IC scattering of disk electrons on the ISRF, we consider 3 different templates: a simple geometric disk model, the 408 MHz Haslam map, and the $0.5-1$ GeV \Fermi\ map after removal of the (dust-correlated) $\pi^0$ gammas.  We verify that our results are insensitive to this choice.  As an additional cross check, we subtract the \Fermi\ diffuse Galactic model from the data, finding that this also reveals the bubble structures. Although our template fitting technique is subject to significant uncertainties due to uncertain line of sight gas and CR distributions, these uncertainties mainly affect the intensity profile at low latitudes.  For $|b| > 30\degree$ and $1 < E < 50\gev$ the morphology and spectrum are completely consistent for different template choices.  Indeed, in the $1-5\gev$ maps, a significant part of the southern bubble is easily visible before \emph{any} template subtractions. 

The \Fermi\ bubbles have an energy spectrum of $dN/dE \sim E^{-2}$, significantly harder than other gamma-ray components. There is no apparent spatial variation in the spectrum between the bubble edge and interior, and the north and south bubbles have consistent spatial and spectral profiles. Both the morphology and spectrum are consistent with the two bubbles having the same origin and being the IC counterpart to the electrons which generate the microwave haze seen in \WMAP\ \citep{Finkbeiner:2003im, Dobler:2008ww}. The spectrum of the CR electrons required to generate the \Fermi\ bubbles is harder than expected for electrons accelerated in supernova shocks in the disk, and such disk-produced electrons would be even softer after several kpc diffusion; the morphology of the bubble structure is also quite different to that of the lower-energy electrons traced by the Haslam 408 MHz map. Even setting aside the \WMAP\ haze, the \Fermi\ bubbles are unlikely to originate from excess $\pi^0$ emission, as (by construction) they are spatially distinct from the SFD dust map, their spectrum is much harder than that of the dust-correlated emission, and the \rosat\ data suggest that the bubbles are hot and underdense rather than overdense. The morphology of the \Fermi\ bubbles, and the overlap of the \Fermi\ bubbles and \WMAP\ haze, also strongly \emph{disfavor} the hypothesis that a significant fraction of the high energy gamma rays observed by \Fermi\ in the bubble region are photons directly produced by dark matter annihilation.



The \Fermi\ bubble structures were likely created by some large episode of energy injection in the GC, such as a past accretion event onto the central MBH, or a nuclear starburst in the last $\sim$10 Myr. We have discussed some general possibilities and considerations in this work, and found shortcomings in each scenario; it seems likely that either significant modifications to one of these ideas, or some combination of different mechanisms, will be necessary.

Jets originating from AGN activity can potentially accelerate CR electrons to high energies, and transport them rapidly away from the GC; the cooling time of electrons at $100-1000$ GeV is only $10^{5-6}$ years, so if the CRs are injected and accelerated only in the GC, a very fast bulk transport mechanism is required to convey them throughout the bubbles before they lose a significant fraction of their energy. However, filling the bubbles completely, with $n=10^{-2}$ cm$^{-3}$ gas, would require a mass injection of $\sim 10^8$ solar masses, so in any case it is more reasonable for the bulk of the material in the bubbles to be swept up and accelerated as the bubbles expand. Energetic shocks associated with jets can have high Mach number and thus efficiently accelerate CR electrons, producing hard spectra with $dN/dE \sim E^{-2}$, and the total energy required to heat the bubbles is also readily achievable by accretion events onto the central MBH.  

However, the north-south symmetry of the bubbles has no obvious explanation in the context of an AGN jet: there is no reason for one jet to be oriented perpendicular to the Galactic plane. The large width and rounded shape of the bubbles are also not typical of jets, which are generally much more collimated, although a precessing jet might help explain the wide opening angle of the bubbles. If the central MBH becomes active on a relatively short timescale, the \Fermi\ bubbles may be created by a number of past jets, which combine to give rise to the symmetric and uniform \Fermi\ bubbles.

An alternate source for the large required energy injection is a nuclear starburst. The wide opening angle of the bubbles is not a problem in this case; the bubble shape is similar to that observed in NGC 3079, and the X-ray features observed by \rosat\ are similar to those observed in other nearby starburst galaxies. However, no corresponding H$\alpha$ signal of the \Fermi\ bubbles is observed, in contrast to other known starburst galaxies: this problem might potentially be resolved if the H$\alpha$-emitting gas has cooled in the time since the starburst phase (gas hot enough to emit the X-rays observed by \rosat\ has a considerably longer cooling time). Also, generally gas filaments and clumps are observed in the X-rays in starburst galaxies, and it would seem that a relic of a past starburst should become more clumpy with gas clouds and filaments due to cooling of the gas. However, while no such structures are obvious in the \rosat\ maps, the signal-to-noise is insufficient to place strong constraints. 

The absence of any such filamentary structures inside the \Fermi\ bubbles, on the other hand, argues against a hadronic origin for the bubble gamma ray emission. Hadronic jets might accelerate protons to high energies, and the interactions of these protons with the ISM could then produce hard $\pi^0$ gammas and secondary $e^+ e^-$, which would scatter on the ISRF to produce more gamma rays. In this scenario, however, the gamma ray emission should trace the gas density, which we would not expect to be smooth and homogeneous.

Returning to the starburst scenario, the cosmic ray acceleration in this case would be due to shocks at the edge of the bipolar wind. However, the shocks expected in this scenario would be relatively weak and slow-moving, and thus may not be capable of generating a sufficiently hard electron spectrum to reproduce the signal. For example, in first-order Fermi acceleration, a shock Mach number of $\sim 3.3$ is needed to obtain an electron spectral index of $2.4$, as required for the synchrotron explanation of the \WMAP\ haze (see e.g. \cite{1991SSRv...58..259J}).

The \Fermi\ bubbles have sharp edges, also suggesting the presence of a shock at the bubble walls. If the CRs producing the gamma rays have a multi-kpc diffusion length (which is not expected to be the case for 1 TeV electrons, for example), then the edges can still be sharp if the bubble edge is moving outward faster than they can diffuse. If we assume the \Fermi\ bubbles are projected structures from three dimensional symmetric blobs towards the GC, the flat intensity profile of the bubbles requires the emissivity to rise at the bubble walls, but remain non-negligible in the bubble interior; the lack of spatial variation in the spectral index may also constrain models where the electrons diffuse long distances from an injection point. Magnetic reconnection in the interior of the bubbles, or some other mechanism such as dark matter annihilation, may help maintain a hard spectrum throughout the bubbles by accelerating existing lower-energy electrons or injecting electrons \emph{in situ}.

Dark matter annihilation or decay, while an effective mechanism for injecting hard electron CRs at high latitudes, cannot produce the features in the \rosat\ X-ray maps correlated with the bubbles, and would not be expected to result in sharp cutoffs in gamma-ray emission at the bubble edges. Dark matter annihilation or decay may be \emph{contributing} to the bubbles, or to gamma-ray emission in the inner Galaxy that is not well subtracted by either the bubble structure template or the models for known diffuse emission mechanisms; however, understanding the \Fermi\ bubbles will be a necessary step before extracting any such dark matter signal. 

The \emph{eROSITA}\footnote{\texttt{http://www.mpe.mpg.de/heg/www/Projects/EROSITA/main.html}} and \emph{Planck}\footnote{\texttt{http://www.rssd.esa.int/index.php?project=Planck}} experiments will provide improved measurements of the X-rays and microwaves, respectively, associated with the \Fermi\ bubbles, and so may help discriminate between these scenarios. \emph{eROSITA}, which is expected to launch in 2012, will provide the first imaging all-sky survey of mid-energy X-rays, studying the $0.2-12$ keV energy range with $\sim 100$ eV energy resolution and a PSF of 20''. The \emph{Planck} satellite, launched in 2009, will greatly improve the measurements of the \WMAP\ haze spectrum.  In addition, AMS-02\footnote{\texttt{http://www.ams02.org}} will launch in 2011, and may significantly advance our understanding of CR acceleration and propagation, and help to refine our interpretation of the \Fermi\ bubbles.


\vskip 0.15in {\bf \noindent Acknowledgments:} We acknowledge helpful
conversations with Avery Broderick, Greg Dobler, Martin Elvis, Jim Gunn, Jill Knapp, Maxim Markivitch, David Merritt, Simona Murgia, Norm Murray, Paul Nulsen, Aneta
Siemiginowska, David Spergel, Anatoly Spitkovsky, and Neal Weiner. DF and TS are partially supported by NASA grant
NNX10AD85G.  TS is partially supported by a Sir Keith Murdoch
Fellowship from the American Australian Association.  This research
made use of the NASA Astrophysics Data System (ADS) and the IDL
Astronomy User's Library at Goddard.\footnote{Available at
  \texttt{http://idlastro.gsfc.nasa.gov}}

\newpage
\bibliography{bubble}
\bibliographystyle{apj}

\end{document}